\documentclass[iop]{emulateapj}

\usepackage[utf8]{inputenc}
\usepackage{graphicx}
\usepackage{amsmath}
\usepackage{mathrsfs}
\usepackage{natbib}
\usepackage{array}
\usepackage{hyperref}

\graphicspath{{}}

\def\imagetop#1{\vtop{\null\hbox{#1}}}

\slugcomment{Corrections after referee report}

\shorttitle{Dust-vortex instability for well-coupled grains}
\shortauthors{C. Surville et al.}

\begin{document}

\title{Dust-vortex instability in the regime of well-coupled grains}

\author{Cl\'{e}ment Surville, Lucio Mayer}
\affil{Institute for Computational Science, University of Zurich, Winterthurerstrasse 190, 8057 Zurich, Switzerland}
\email{clement.surville@physik.uzh.ch}

\begin{abstract}
We present a novel study of dust-vortex evolution in global two-fluid disk simulations to find out if evolution toward high dust-to-gas ratios can occur in a regime of well-coupled grains with low Stokes numbers ($St=10^{-3}-{4\times 10^{-2}}$). We design a new implicit scheme in the code RoSSBi, to overcome the short timesteps occurring for small grain sizes. We discover that the linear capture phase occurs self-similarly for all grain sizes, with an intrinsic timescale (characterizing the vortex lifetime) scaling as $1/St$. After vortex dissipation, the formation of a {{global active dust ring}} is a generic outcome confirming our previous results obtained for larger grains. We propose a scenario in which, irrespective of grain size, multiple pathways can lead to local dust-to-gas ratios of order unity and above on relatively short timescales, $< 10^5$ yr, in the presence of a vortex, even with $St=10^{-3}$. When $St>10^{-2}$, the vortex is quickly dissipated by two-fluid instabilities, and large dust density enhancements form in the global dust ring. When $St<10^{-2}$, the vortex is resistant to destabilization. As a result, dust concentrations occur locally due to turbulence developing inside the vortex. Whatever the Stokes number, dust-to-gas ratios in the range $1-10$, a necessary condition to trigger a subsequent streaming instability, or even a direct gravitational instability of the dust clumps, appears to be an inevitable outcome. Although quantitative connections with other instabilities still need to be made, we argue that our results support a new scenario of vortex-driven planetesimal formation.

\end{abstract}


%%%%%%%%%%%%%%%%%%%%%%%%%%%%%%%%%%%%%%%%%%%%%%%%%%%%%
\section{ Introduction }
\label{Sect_Intro}

	The composition of early protoplanetary disks is dominated by a mixture of hydrogen and helium gas. An important, but small fraction of the disk mass, a few percent, is comprised of small dust grains with sizes ranging from micrometers to millimeters. These dust grains are responsible for the infrared excess observed in young disk emissions \citep[see, e.g. ][]{Gras-Velazquez2005, Meyer2008, Smith2008, Bryden2009}. This solid component is the building material for pebbles and planetesimals from which planets and other large-size objects may form. 

	Grain growth models try to explain how small grains become larger, e.g. by sticking, bouncing. However, existing models do not yet fully capture the coupled gas-dust dynamics in the disk. In particular, the drag force exerted on small grains produces a strong mixing with the gas, making it difficult for dust concentrations to arise. Alternatively, if particular gas structures such as vortices appear, it is possible that small grains accumulate in these locations, albeit on different timescales --- depending on the grain size. Due to numerical constraints, studies often focus on larger grains that are only weakly coupled to the gas, i.e. grains sizes with Stokes numbers close to unity \citep{Fu2014, Zhu2014a}. Examples of such numerical studies are those aimed at studying two-fluid instabilities which can trigger stronger local dust concentrations, such as the streaming instability (SI) \citep{Youdin2005, Youdin2007, Johansen2007, Jacquet2011, Kowalik2013, Simon2016}, instabilities resulting from the evolution and dissipation of anticyclonic vortices \citep{Fu2014, Surville2016} or, more recently, a broad class of such instabilities occurring in a resonant regime \citep{Lin2017, Hopkins2017, Squire2017} of which SI is an example. 
	
	From previous studies, it is unclear whether significant evolution of the dust density distribution can also occur when the grains are small and strongly coupled with the gas. Understanding the dynamics in this regime is important, for example, in validating whether the SI and the gravitiational instability can work in tandem as a viable pathway to planetesimal formation \citep{Youdin2011, Johansen2012, Simon2016, Schafer2017}. While the premise is attractive, the required conditions are restrictive. In particular the dust must be dominated by weakly-coupled particles --- namely with Stokes numbers of order unity so that the back reaction is strong --- and local dust-to-gas ratios of order unity must be already in place. Such restrictions apply in local and global simulations, in two and three dimensions \citep{Kowalik2013, Simon2016}. It has yet to be shown that such conditions can occur for sub-mm dust grains that are still strongly coupled to the gas.

	In this paper we study the dynamics of smaller grains, with Stokes number $St<10^{-2}$, which is a nearly unexplored territory at the moment. Our study is a follow-up to the investigation of \cite{Surville2016}, in which we considered the effect of vortices on dust evolution in global 2D disk simulations. Almost without exception, we found that the dust-to-gas ratio inside vortices would exceed unity and eventually spread out into turbulent, long-lived dust rings. These dust rings might offer the missing link to generate naturally conditions that are favorable to the SI. It is thus crucial to find out if the same multi-stage evolution of the dust component towards global structures with high dust-to-gas ratios can occur even when the initial grain size is in a regime of small Stokes number. This is not obvious since in our previous results the back-reaction of dust onto gas was shown to play a crucial role in driving the formation of turbulent dust rings after vortex dissipation. We expect the back reaction to become weaker as the coupling between the gas and dust becomes stronger at small Stokes numbers.
	
	{{The vortex instability that develops after the capture of the solids as well as the active dust rings that form eventually after vortex dissipation are source of diffusion of the dust grains of larger size. They limit the dust-to-gas ratio to $1-10$ in the eddies. However, as concerns the small sized well-coupled grains, turbulent diffusion may have an impact on the evolution of the dusty vortex.
	
	The origin of turbulence in the vortex is unclear, and for the moment, the major process could be the elliptical instability of incompressible vortices. This instability proposed in \cite{Lesur2009} concerns 3D vortices of  aspect ratio smaller than 4, with closed streamlines (a consequence of incompressibility). However, it is unclear if such vortices exist in a fully compressible flow. As shown in \cite{Surville2015}, 2D vortices of  aspect ratio smaller than 4 have open streamlines.  As a result, they may resist the elliptical instability in a 3D context, but it has to be confirmed. A step toward such a confirmation is present in \cite{Meheut2010}, where they found 3D vortices with open streamlines, and vertical motion inside the vortex. Such properties would reduce the efficiency of the instability, and keep the vortex flow laminar. 

	In this work, the large vortex with which dust is interacting has an aspect ratio close to 6, and will be quiet to hydrodynamical turbulence. We thus propose a study of this category of non-turbulent vortices with open streamlines. We will show how dust diffusion generated by the two-fluid interaction is efficient, and how it compares to previous works where turbulent diffusion of dust in elliptical vortices is at play. 

 }}

	The implications of such a study are important not only to investigate possible routes towards planetesimal formation but also because, if substructures in the dust component can still be triggered even in the regime of small grains, this would have immediate observational consequences. In other words, addressing this regime can be potentially important to interpret correctly the infrared observations of T Tauri and debris disks carried out by ALMA as well as by other instruments \citep{Perez2014, Partnership2015, Andrews2016}. 
	
	Our new focus on small dust grains poses some significant numerical challenges in a global disk simulation. Indeed numerical integration of the drag force interaction for small grains is demanding because it implies large source terms in the dynamical equations. As these source terms are proportional to the dust density but inversely to the Stokes number, they can develop very steep gradients in dense regions. Any type of second order explicit method would require a very small timestep and force the evolution to proceed on the natural timescale of the drag force, i.e $St/( \Omega_k)$. Therefore we developed a semi-implicit method to circumvent this problem. We show how efficient and accurate this is for the pressure-less fluid approximation. This method was used to accomplish the various simulations presented in this paper. 

	The paper is organized as follows. Section \ref{Sect_Friction} covers the details of the implicit numerical method used to accurately follow the dynamics of dust under the small Stokes number regime. Section \ref{Sect_Setup} describes the initial conditions of the simulations and the disk model. Section \ref{Sect_Results} presents the results of the numerical runs, and describes the major observations about the evolution of the dusty vortices, in particular the different main phases of evolution. These phases are detailed and analyzed during the Discussion, Section \ref{Sect_Discussion}, and a novel framework on the evolution paths of dusty disks as a function of grain size is presented. Finally, summary and perspectives are given in the Conclusions, Section \ref{Sect_Conclusion}.

%%%%%%%%%%%%%%%%%%%%%%%%%%%%%%%%%%%%%%%%%%%%%%%%%%%%%
\section{ Methods: Integration of the aerodynamical friction }
\label{Sect_Friction}

	In this first section, we present the main equations describing the flow of gas and dust in the disk. The dust component is modeled by a pressure-less fluid, which is particularly appropriate in the case of well-coupled grains, i.e. for small grains. The two fluids interact via aerodynamical friction, a force inversely proportional to the Stokes number of the dust grains, $St$. As a consequence, the friction process becomes much faster than the other components of the motion, like the Keplerian rotation and the wave propagation. Following friction with an explicit scheme like a Runge-Kutta integration becomes inaccurate and inefficient as long as the grain size are small.
	
	We propose an implicit method that removes the timestep restrictions imposed by the drag, thereby allowing us to default to the usual hydrodynamic time interval used in a single-phase simulation. 
	
%----------------------------------------------------
\subsection{ Equations of coupled fluids }

	In order to keep the conservative form of the Euler equations, we consider the momentum of the gas and particle fluids, $\vec{\mathcal{P}}_g=\sigma_g \vec{V}_g$ and $\vec{\mathcal{P}}_p=\sigma_p \vec{V}_p$, respectively. We warn the reader not to confuse $\vec{\mathcal{P}}$, which is an \emph{momentum in this section}, with any pressure. With these quantities, the conservation of movement of gas and particle fluids can be written as
%
%-----------------------
\begin{eqnarray}
\label{Equ_Main_Euler}
	\partial_t \vec{\mathcal{P}}_g & = & \vec{\mathcal{A}}_g + \Omega_k(r) St^{-1} \left( \vec{\mathcal{P}}_p - \frac{\sigma_p}{\sigma_g} \vec{\mathcal{P}}_g\right), \\
	\partial_t \vec{\mathcal{P}}_p & = & \vec{\mathcal{A}}_p - \Omega_k(r) St^{-1} \left( \vec{\mathcal{P}}_p - \frac{\sigma_p}{\sigma_g} \vec{\mathcal{P}}_g\right),
\end{eqnarray}
	with $\Omega_k(r)\propto r^{\beta_{\Omega}}$ the Keplerian frequency, so $\beta_{\Omega}=-3/2$.

	In writing this system of equations, we have combined advection terms, Keplerian and geometrical source terms in unique operators $\vec{\mathcal{A}}_g$ and $\vec{\mathcal{A}}_p$ for gas and particles, respectively. Until this point, we have not made any approximation. We now introduce the local dust-to-gas ratio, $\varepsilon=\sigma_p/\sigma_g$, and assume it is quasi-steady over the time period considered, typically the timestep of integration. Upon, defining the drag frequency, $\omega_p=\Omega_k(r) St^{-1}$, we obtain a single equation
%
%-----------------------
\begin{equation}
	\partial_t \left(\vec{\mathcal{P}}_p - \varepsilon \vec{\mathcal{P}}_g\right) =  \vec{\mathcal{A}}_p - \varepsilon \vec{\mathcal{A}}_g - \left(1+\varepsilon\right)\omega_p \left( \vec{\mathcal{P}}_p - \varepsilon \vec{\mathcal{P}}_g\right).
\end{equation}
Finally, we introduce the differential quantities $\Delta \vec{\mathcal{A}} = \vec{\mathcal{A}}_p - \varepsilon \vec{\mathcal{A}}_g$, and $\Delta \vec{\mathcal{P}}=\vec{\mathcal{P}}_p - \varepsilon \vec{\mathcal{P}}_g$, to obtain
%
%-----------------------
\begin{equation}
\label{Equ_Principal}
	\partial_t \Delta \vec{\mathcal{P}} =  \Delta \vec{\mathcal{A}} - \left(1+\varepsilon\right)\omega_p \Delta \vec{\mathcal{P}}.
\end{equation}

	Solving this equation for $\Delta \vec{\mathcal{P}}$, over the interval $[t, \: t+\Delta t]$, is possible assuming that $\Delta \vec{\mathcal{A}}$ is constant over the time interval. This is relevant for second-order time integration schemes, commonly found in Finite Volume methods. Its value is either the one at time $t$, for the first step of integration over $[t, \: t+\Delta t/2]$, or the one estimated at $t+\Delta t/2$ for the final step of integration over $[t, \: t+\Delta t]$. {{We can integrate over the timestep as a function of a local time $t'\in [0, \: \Delta t]$. The general solution of equation \ref{Equ_Principal} is combined with the initial condition that $\Delta \vec{\mathcal{P}}(t'=0)=\Delta \vec{\mathcal{P}}(t)$, at the beginning of the timestep. As a result we obtain}}
%
%-----------------------
\begin{equation}
\label{Eq_Sol_Delta_P_exact}
\begin{split}
	\Delta \vec{\mathcal{P}}(t') & =  \frac{\Delta \vec{\mathcal{A}}}{(1+\varepsilon)\omega_p} \left(1-\exp[-(1+\varepsilon)\omega_p t']\right) \\
	& + \Delta \vec{\mathcal{P}}(t) \exp[-(1+\varepsilon)\omega_p t'] ,
\end{split}
\end{equation}
which is exact as long as $\omega_p$ is constant over the time interval.

	There are two interesting regimes where this solution could be simplified, in order to decouple the drag force operator from the advection, which may be required by some numerical schemes based on Riemann solvers. {{ First consider the regime when $(1+\varepsilon)\omega_p t'$ is small.  In this case, the Taylor expansion of the exponential function in factor of $\Delta \vec{\mathcal{A}}$ shows that the term depending on the presence of dust, i.e. in $\omega_p$,  appears at the order in $t'^2$.  This term is neglected by the numerical scheme because the time integration scheme for advection is only second order, i.e. integrates only terms up to $t'$. Moreover, the expansion of the exponential in factor of $\Delta \vec{\mathcal{P}}(t)$ contains terms in $\omega_p$ already at the order in $t'$. Then the term in $\Delta \vec{\mathcal{A}}$ can be ignored and the solution can be simplified to }}
%
%-----------------------
\begin{equation}
\label{Equ_Sol_Delta_P}
	\Delta \vec{\mathcal{P}}(t') = \Delta \vec{\mathcal{P}}(t) \exp[-(1+\varepsilon)\omega_p t'],
\end{equation}
which is the solution of the differential equation without the term in $\Delta \vec{\mathcal{A}}$. Importantly, in this case, solving equation \ref{Equ_Principal} is possible by superposition of linear solutions. The integration of the terms depending on advection, $\vec{\mathcal{A}}$, is done by the numerical scheme, as usual.

	Secondly consider the regime when $(1+\varepsilon)\omega_p t'$ is large, i.e. for short friction times, large dust-to-gas ratios, or coarse resolutions ( $t'\sim \Delta t$ is large). {{ In this case the Equation \ref{Equ_Principal} can be simplified according to the following argument. The operators $\vec{\mathcal{A}}_g$ and $\vec{\mathcal{A}}_p$ are dominated by terms in $\Omega_k \sigma \vec{V}$, so in amplitude the term $\Delta \vec{\mathcal{A}}$ is comparable to $\Omega_k \Delta \vec{\mathcal{P}}$. As a consequence, in the right-hand terms of Equation \ref{Equ_Principal}, the dominant one is $- \left(1+\varepsilon\right)\omega_p \Delta \vec{\mathcal{P}}$ when $(1+\varepsilon) \omega_p \gg \Omega_k$. In this limit, the equation becomes
%-----------------------
\begin{equation}
\label{Equ_Short_friction}
	\partial_t \Delta \vec{\mathcal{P}} = - \left(1+\varepsilon\right)\omega_p \Delta \vec{\mathcal{P}}.
\end{equation}
giving the same solution as Equation \ref{Equ_Sol_Delta_P}.}} This regime corresponds to the short friction time approximation used to estimate the terminal velocity of the dust in many studies \citep{Johansen2005, Booth2015, Lin2017}.

	To conclude, the drag force source term in the equations of conservation of movement can be expressed as
%
%-----------------------
\begin{equation}
\label{Equ_Sol_Friction}
	\pm \omega_p \Delta \vec{\mathcal{P}}(t) \exp[-(1+\varepsilon)\omega_p t'],
\end{equation}
over the time interval of integration, with $+$ for the gas and $-$ for the particle fluids. We can interpret it as a modified Epstein drag law, that takes into account the two asymptotic regimes discussed previously. The intermediate regime, $(1+\varepsilon)\omega_p t' \sim 1$, may suffer from neglecting the coupling with $\Delta \vec{\mathcal{A}}$, but the deviation is small because the neglected term is at least one order smaller in time and $\Delta \vec{\mathcal{A}}(t)$ gets smaller and smaller as $(1+\varepsilon)\omega_p t'$ increases. The main advantage of this modified friction is the ability to model large and small dust grains without the need to modify the timestep or the numerical scheme.

%----------------------------------------------------
\subsection{ Source integral of the friction }

	The time evolution of the momentum equations is solved by integrating the system of equations \ref{Equ_Main_Euler} over the timestep, giving for the gas
%
%-----------------------
\begin{equation}
%\label{Equ_Main_Euler_Modified_g}
\begin{split}
	\vec{\mathcal{P}}_g(t+\Delta t) - \vec{\mathcal{P}}_g(t) = & \int_t^{t+\Delta t} \vec{\mathcal{A}}_g dt' \\
		& + \int_0^{\Delta t} \omega_p \Delta \vec{\mathcal{P}}(t) \exp[-(1+\varepsilon)\omega_p t'] dt',
\end{split}
\end{equation}
and for the particles
%
%-----------------------
\begin{equation}
%\label{Equ_Main_Euler_Modified_p}
\begin{split}
	\vec{\mathcal{P}}_p(t+\Delta t) - \vec{\mathcal{P}}_p(t) = & \int_t^{t+\Delta t} \vec{\mathcal{A}}_p dt' \\
		& - \int_0^{\Delta t} \omega_p \Delta \vec{\mathcal{P}}(t) \exp[-(1+\varepsilon)\omega_p t'] dt',
\end{split}
\end{equation}
where $\Delta \vec{\mathcal{P}}(t)=\vec{\mathcal{P}}_p(t) - \varepsilon \vec{\mathcal{P}}_g(t)$, and $\varepsilon = \sigma_p(t)/\sigma_g(t)$ are evaluated at time $t$.

	The integration of the advection and source terms in $\vec{\mathcal{A}}$ is done in the normal way specified by the numerical scheme used, while the modified friction source term is integrated analytically as follows:
%
%-----------------------
\begin{equation}
\label{Equ_Sol_Friction_Int}
\begin{split}
	\int_0^{\Delta t} \omega_p \Delta \vec{\mathcal{P}}(t) \exp[-(1+\varepsilon)\omega_p t'] dt' = & \frac{\Delta \vec{\mathcal{P}}(t)}{(1+\varepsilon)} \\
		\times & \left[ 1 - \exp[-(1+\varepsilon)\omega_p \Delta t]  \right].
\end{split}
\end{equation}
Decoupling the friction from the full Euler equations in this manner yields a second-order accurate implicit timestepping scheme.

%-----------------------
\begin{figure}[t]
%% Figure 1
	\begin{center}
	\includegraphics[height=6.cm]{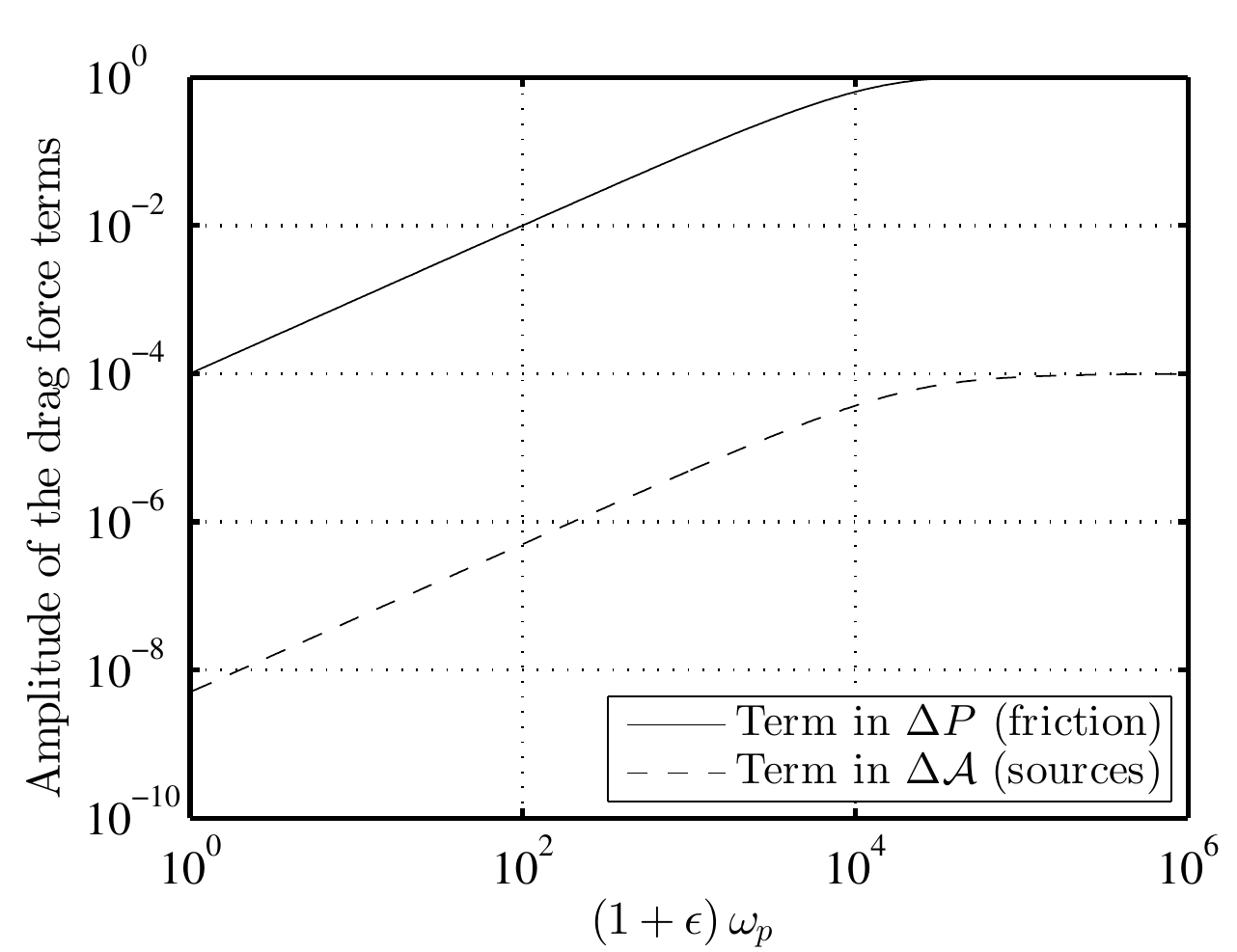}
	\caption{\label{Fig_Error_in_Delta_A} Estimation of the error induced by the implicit drag solution. Using a typical timestep such as $\Omega_0 \Delta t = 10^{-4}$, we show the amplitude of the two components of the drag force integral as a function of $\omega_p$, i.e.  $\Omega_k/St$. We observe that the term of friction in $\Delta \vec{\mathcal{P}}$ (solid line) is $10^4$ times larger than the contribution of the advection and source terms, $\Delta \vec{\mathcal{A}}$ (dashed line).  }
      \end{center}
\end{figure}

	Figure \ref{Fig_Error_in_Delta_A} compares the coefficients in front of each terms in the exact solution of Equation \ref{Eq_Sol_Delta_P_exact}, assuming $\Omega_0 \Delta t = 10^{-4}$, a typical value in the numerical setups presented in this study. The solid line corresponds to the term directly related to the drag force. The dashed line is the coefficient multiplying $\Delta \vec{\mathcal{A}}(t)$ that comes out from Equation \ref{Eq_Sol_Delta_P_exact}. In the regime of $(1+\varepsilon)\omega_p$ small (here we are limited to 1 as $St<1$ by assumption), we find that this amplitude is smaller than the one of $\Delta \vec{\mathcal{P}}$ by a factor ${\Delta t}$, as mentioned earlier. In fact this ratio of amplitudes is conserved up to the asymptotic case of short friction, $(1+\varepsilon)\omega_p >> 1$. Furthermore, the value of $\Delta \vec{\mathcal{A}}(t)$ converges to zero in the limit of fully coupled fluids, as the velocity difference between gas and dust will tend to zero.
	
	As a conclusion, using a modified friction source term as proposed in Equation \ref{Equ_Sol_Friction}, and integrating it implicitly for $t'$ over the timestep gives an accurate approximation of the drag force, as the coupling with advection and other operators is one order in time smaller. The implementation of this implicit scheme given by Equation \ref{Equ_Sol_Friction_Int} for the drag force into the code RoSSBi is done using the following operator splitting method:
%-----------------------
\begin{enumerate}
\item Euler variables are updated from time $t$ to $t+\Delta t/2$ with the implicit drag force, 
\item advection and other sources are integrated from $t$ to $t+\Delta t$ from the estimates of Euler variables provided by the previous step
\item Euler variables are finally updated from $t+\Delta t/2$ to $t+\Delta t$ with the implicit drag force computed from the estimates of Euler variables provided by the previous step
\end{enumerate}

	This way of splitting the operators is also second order in time, so it does not degrade the accuracy of the advection scheme. This kind of semi-implicit scheme has been implemented in various methods based on particle approach (Lagrangian or SPH) for treating dust-gas interaction \citep{Booth2015, Yang2016}, but we propose a new implementation for multi-fluid methods. This updated version of the code RoSSBi is thus very efficient in following the dynamics of dusty flows for any grain size, and/or of local dust-to-gas ratios. It enabled us to investigate the evolution of a dusty vortex in presence of grains much smaller than in our previous study.

%%%%%%%%%%%%%%%%%%%%%%%%%%%%%%%%%%%%%%%%%%%%%%%%%%%%%
\section{ Numerical simulations setup }
\label{Sect_Setup}

	In this section, we present the setups of the numerical simulation runs performed to investigate the evolution of a vortex in presence of small dust grains. We use the same context as in \cite{Surville2016} and follow the evolution of a vortex with parameters
%-----------------------
\begin{equation}
(Ro, \: \chi_r, \: \chi_\theta)=(-0.13, \: 0.1, \: 6.5),
\end{equation}
for which we have observed that the formation of a dust ring was possible for grains of $St>4 \times 10^{-2}$.

	The disk profiles are the same for all the simulations, with background gas density and temperature: ${\sigma_0(r) \propto \left(r/r_0\right)^{\beta_\sigma}}$ and ${T_0(r) \propto \left(r/r_0\right)^{\beta_T}}$. The power law exponents are fixed to standard values, $\beta_\sigma = -1$ and $\beta_T = -0.5$. The isothermal disk scale height at the reference radius $r_0=7.5$ au is $H_0(r_0) = 0.05 \; r_0$. The only parameter that differs from \cite{Surville2016} is the density slope which was $\beta_\sigma = -1.5$. We chose to use a shallower power-law index for the surface density in this study in order to slow the migration speed of the vortices through the disk. According to the dust capture model developed in \cite{Surville2016}, the capture timescale of the dust, $\tau_{1/2}$, scales as $St^{-1}$. So we follow the vortex evolution typically over several thousands of disk rotations. The observed migration of the vortex must be slow enough to keep it inside the disk domain over this period.

	The dust population is initially distributed with a density of $\epsilon \sigma_0(r)$, setting the initial dust-to-gas ratio to $\epsilon = 10^{-2}$. This value is the same for all the simulations; only the grain size $r_s$ varies in this study to explore different Stokes numbers. Recalling that the Stokes number equals
%-----------------------
\begin{equation}
\label{Equ_Stokes_number}
	St = \frac{\pi}{2} \rho_s r_s \sigma_g^{-1} \: ,
\end{equation}
using an internal grain density of $\rho_s=3 \: \text{g.cm}^{-3}$, we can relate the Stokes number at the reference radius $r_0$ to a grain size. To this end, we use the gas surface density normalization which is ${\sigma_0(r_0) = 83 \: \text{g.cm}^{-2}}$ in this study. We consider four different Stokes numbers in this paper, ${St = [1, 4, 10, 40] \times 10^{-3}}$, which is almost two orders of magnitude smaller than the values used in \cite{Surville2016} and other studies.

\begin{figure}[t]
%% Figure 2
	\begin{center}
	\begin{tabular}{c}
	\scriptsize{Stokes number} \\
	\includegraphics[height=5.7cm, trim=0mm 0mm 0mm 3mm, clip=true]{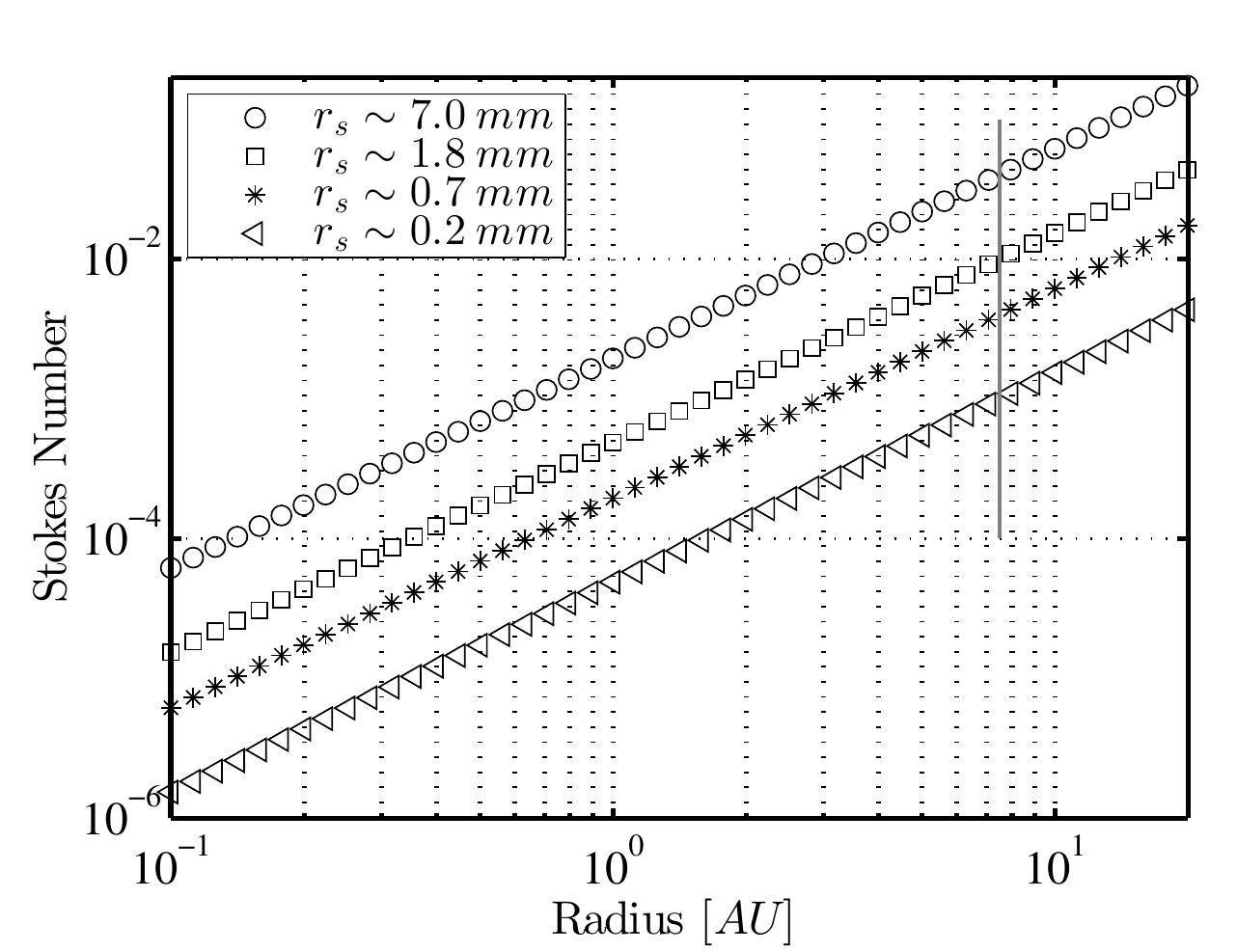} \\ \\
	\scriptsize{Grain size for $\rho_s=3\: g.cm^{-3}$} \\
	\includegraphics[height=5.7cm, trim=0mm 0mm 0mm 3mm, clip=true]{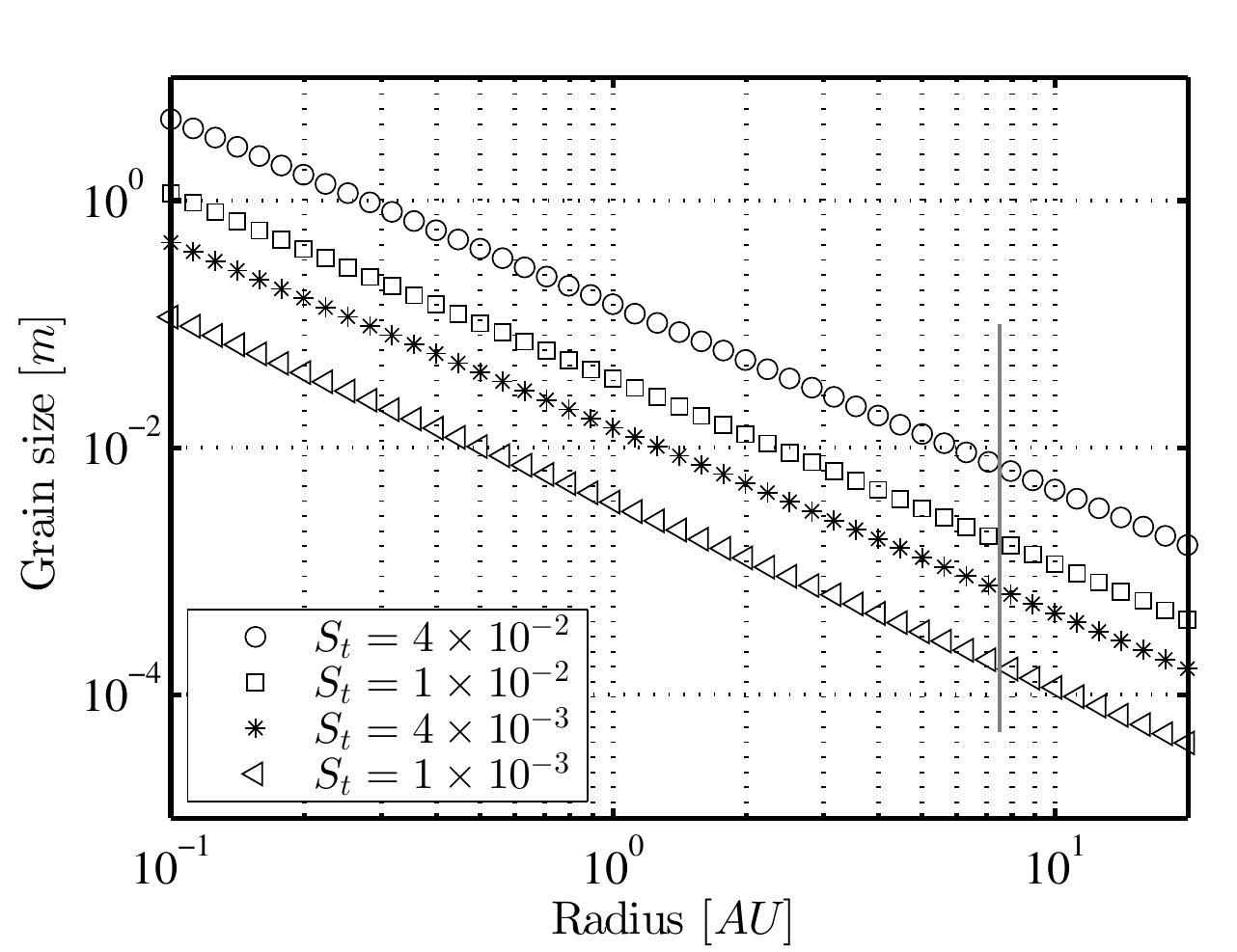}
	\end{tabular}
	\caption{\label{Fig_Stokes_in_disk} Top: Variations of the Stokes number in the disk for different grain sizes (symbols). Bottom: Variations of the grain size in the disk for different Stokes numbers (symbols). For both plots, we compute the values based on the equation \ref{Equ_Stokes_number} and assuming the disk model used in the simulations and described in Section \ref{Sect_Setup}. The vertical gray line identifies the reference radius used in our simulations.}
      \end{center}
\end{figure}

	As shown in Equation \ref{Equ_Stokes_number}, the relation between the Stokes number and the grain size also depends on the location in the disk through the local gas surface density $\sigma_g$. This dependence is presented Figure \ref{Fig_Stokes_in_disk}, for the disk profile used in this study, which is similar to the MMSN model. Firstly, the top panel shows that for a given grain population of radius $r_s$, the corresponding Stokes number increases with distance from the central star. Thus, grains orbiting at $1$ au will experience Stokes numbers $\sim 50$  times smaller than similarly sized grains orbiting at the reference radius of $r_0 = 7.5$ au. Even a cm-sized population will have a Stokes number smaller than $10^{-2}$ (commonly used value in numerical studies) in the disk region from $0.1$ to $2$ au, which promotes the necessity to resolve the drag interaction precisely in the regime of small Stokes numbers.

%-----------------------

Secondly, the bottom panel shows which grain sizes correspond to the Stokes numbers used in this study, as function of the orbital distance. It is striking that even a Stokes number as small as $4\times 10^{-3}$ (star symbol) would represent the dynamics of grains of a few centimeters (so-called pebbles) at $1$ au. We recall that cm-sized grains are critical in the pebble accretion scenario, which is argued to be an efficient way to grow the rocky cores of massive planets such as gas giants \citep{Lambrechts2014, Chatterjee2014, Levison2015}. The Euler equations being invariant as function of the gas density normalization (as long as the disk gravity is neglected), the results presented in this study could be applied at different positions in the disk, with respect to the scalings shown in these plots. We thus give insight to the dynamics of different grain sizes in different regions of the disk.

	We performed global 2D simulations with the code RoSSBi \citep{Surville2016}, which is now updated with the scheme presented Section \ref{Sect_Friction} to handle the small Stokes number regime. We tested our implementation of the new implicit algorithm in this paper against a previous case run performed in \cite{Surville2016} with ${St=4 \times 10^{-2}}$. The results matched to within the numerical noise level of the simulation, showing that the new implementation of the friction is robust even for non-critical Stokes numbers. After performing convergence tests with resolutions of $(N_r, \: N_\theta) = (2048, \: 4096)$, we concluded that a resolution of $(N_r, \: N_\theta) = (1024, \: 2048)$ is enough to resolve the dynamics of the systems we consider.
	 {{One run with ${St=4 \times 10^{-3}}$ was performed with twice the resolution, in order to follow with more details the evolution of the dusty vortex.}}

%%%%%%%%%%%%%%%%%%%%%%%%%%%%%%%%%%%%%%%%%%%%%%%%%%%%%
\section{ Results of the simulations }
\label{Sect_Results}

%----------------------------------------------------
\subsection{ Results of the main runs }
\label{Sect_Results_1}

%-----------------------
\begin{figure*}
%% Figure 3
	\begin{center}
	\begin{tabular}{ccp{15mm}}
	\scriptsize{Rossby number} & \scriptsize{Dust density}  & \\
	\imagetop{\includegraphics[height=5.5cm, trim=2mm 0mm 0mm 3mm, clip=true]{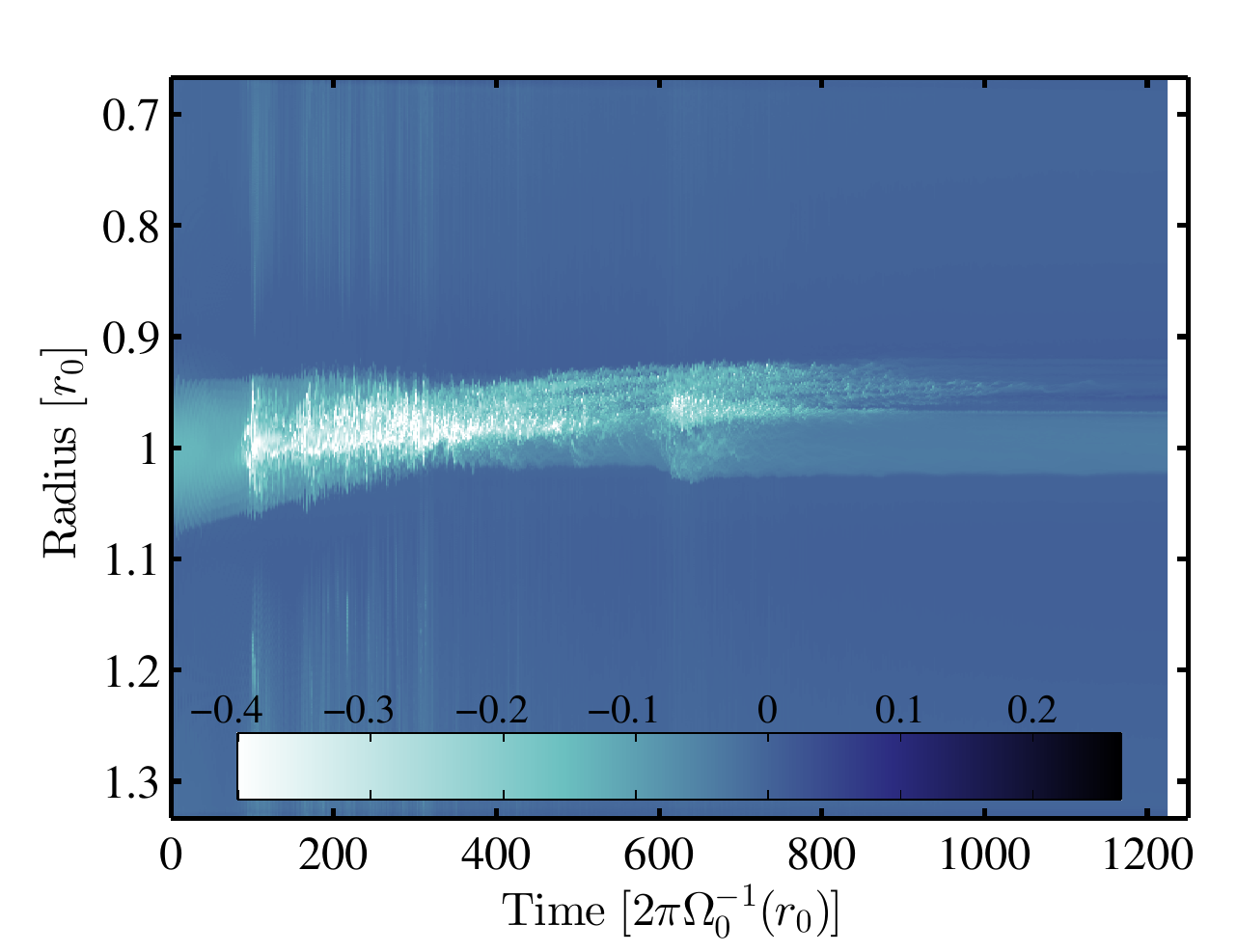}} &
	\imagetop{\includegraphics[height=5.5cm, trim=2mm 0mm 0mm 3mm, clip=true]{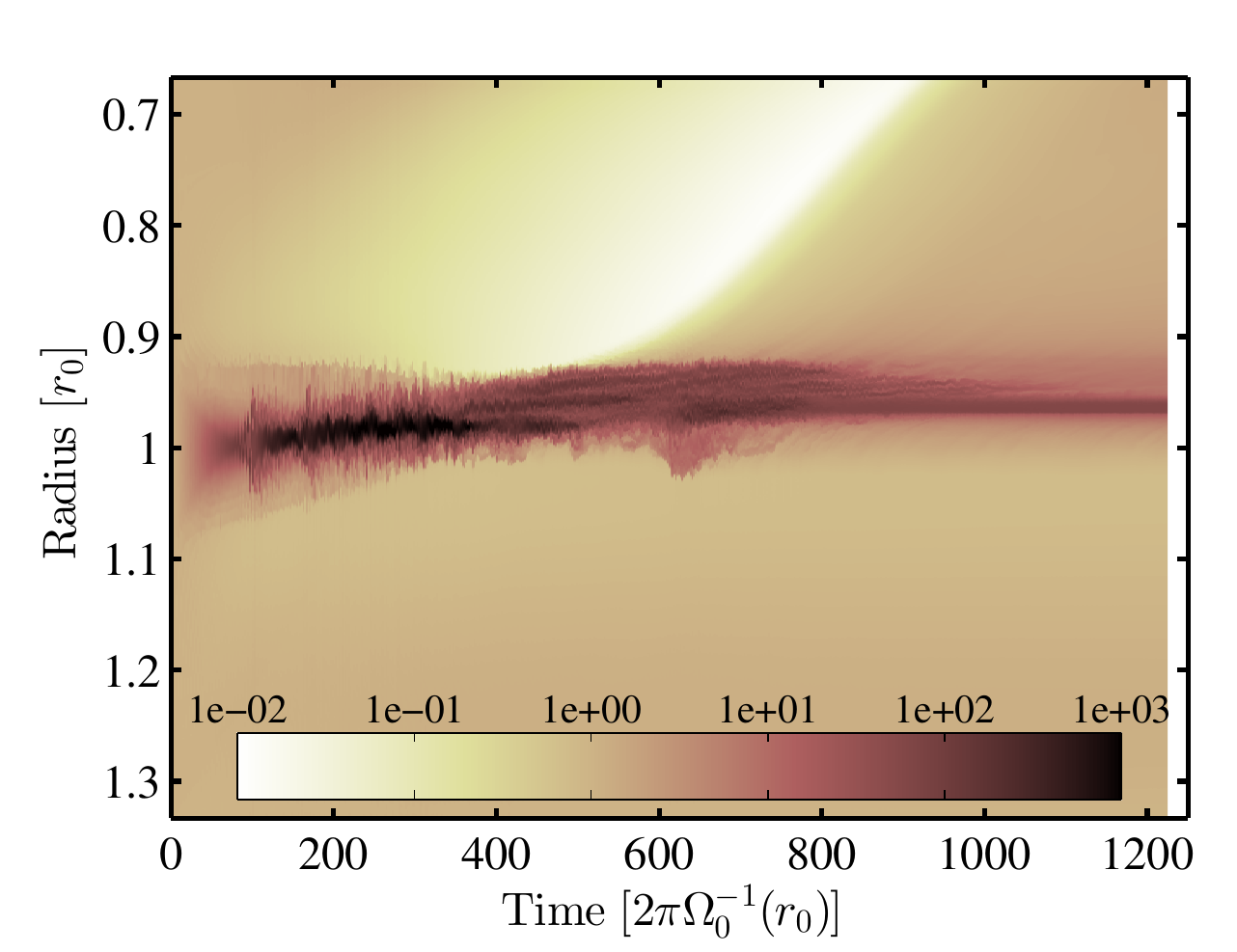}} &
	\scriptsize{$\:$ \newline \newline $St=4\times 10^{-2}$} \\

	\imagetop{\includegraphics[height=5.5cm, trim=2mm 0mm 0mm 3mm, clip=true]{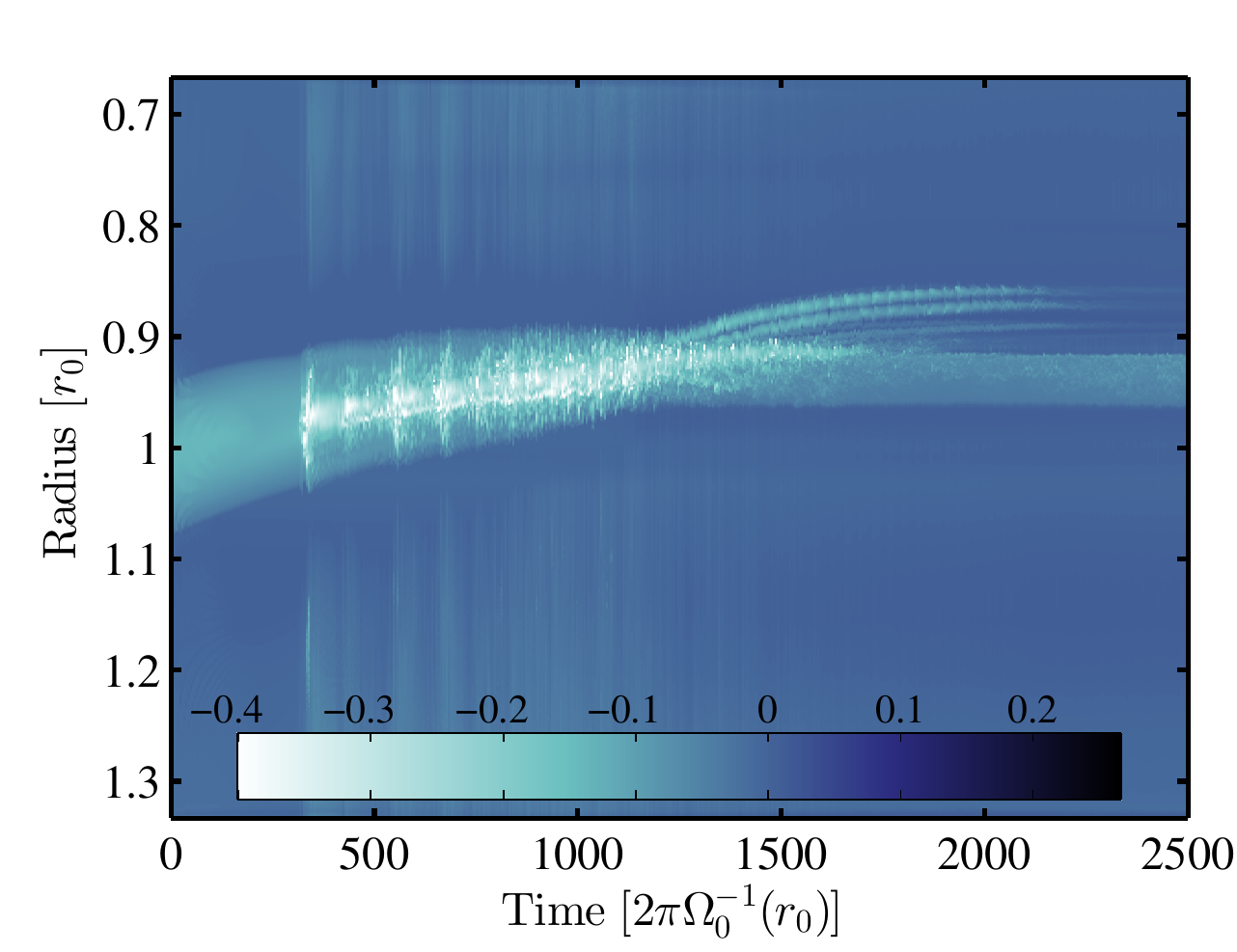}} &
	\imagetop{\includegraphics[height=5.5cm, trim=2mm 0mm 0mm 3mm, clip=true]{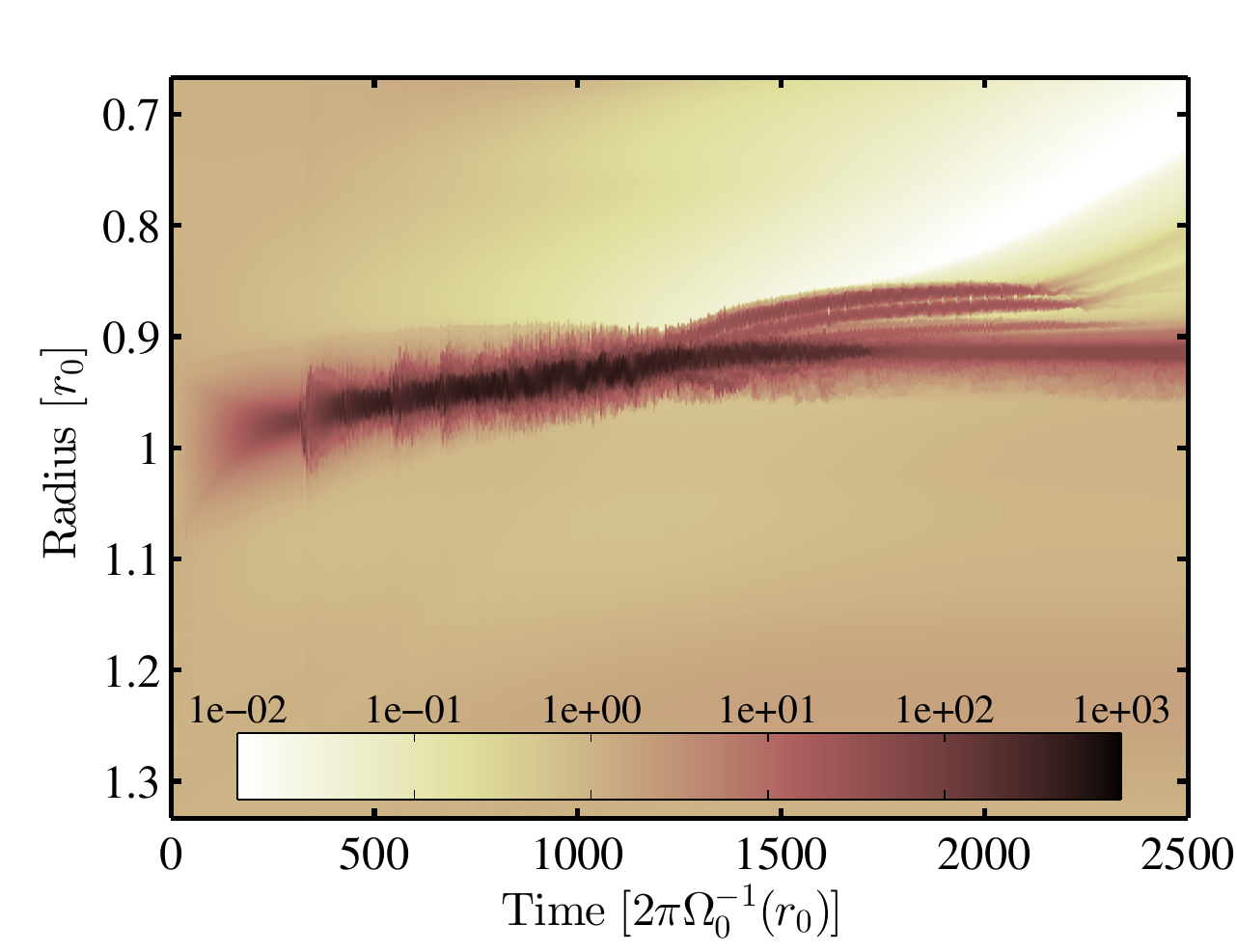}} &
	\scriptsize{$\:$ \newline \newline $St=1\times 10^{-2}$} \\

	\imagetop{\includegraphics[height=5.5cm, trim=2mm 0mm 0mm 3mm, clip=true]{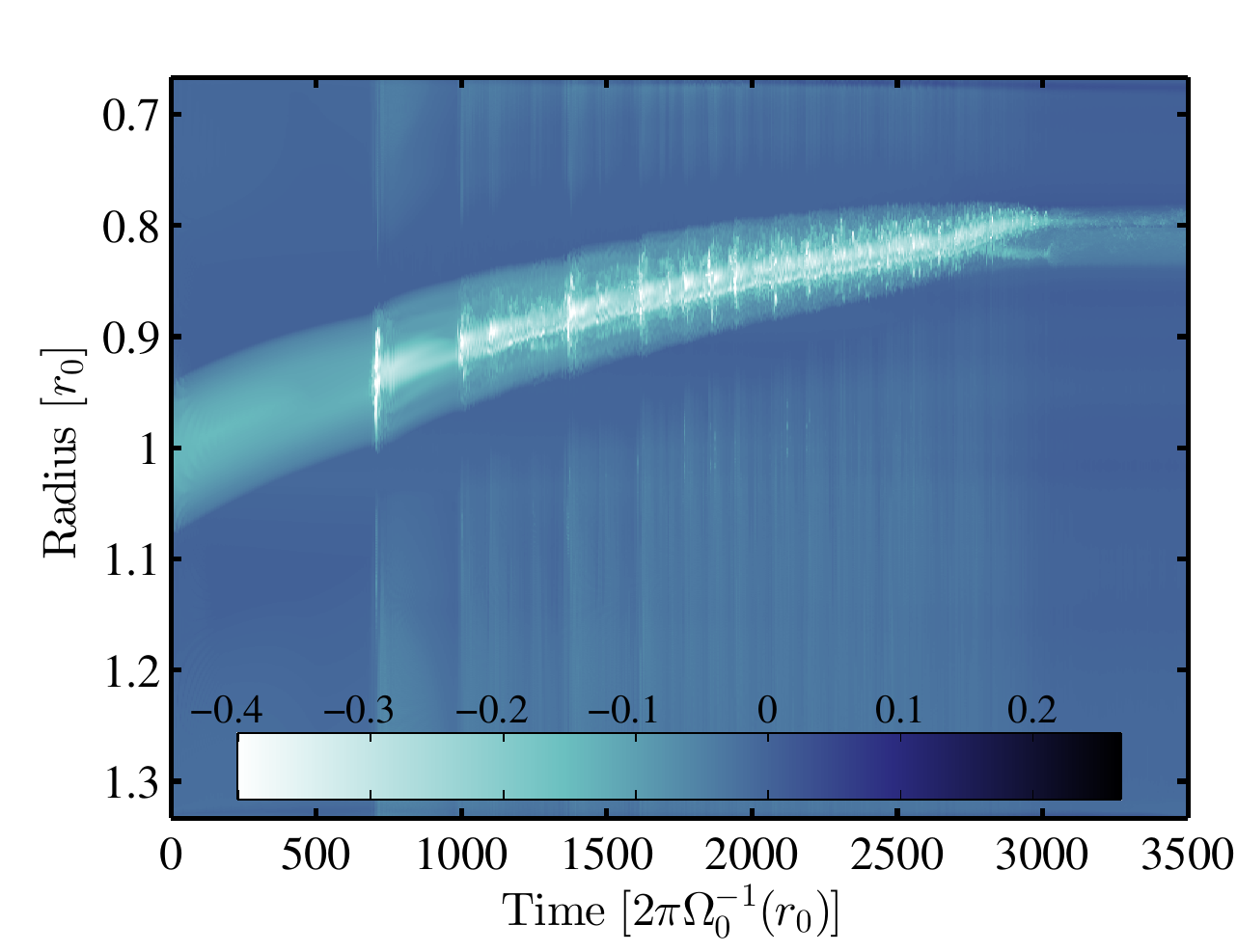}} &
	\imagetop{\includegraphics[height=5.5cm, trim=2mm 0mm 0mm 3mm, clip=true]{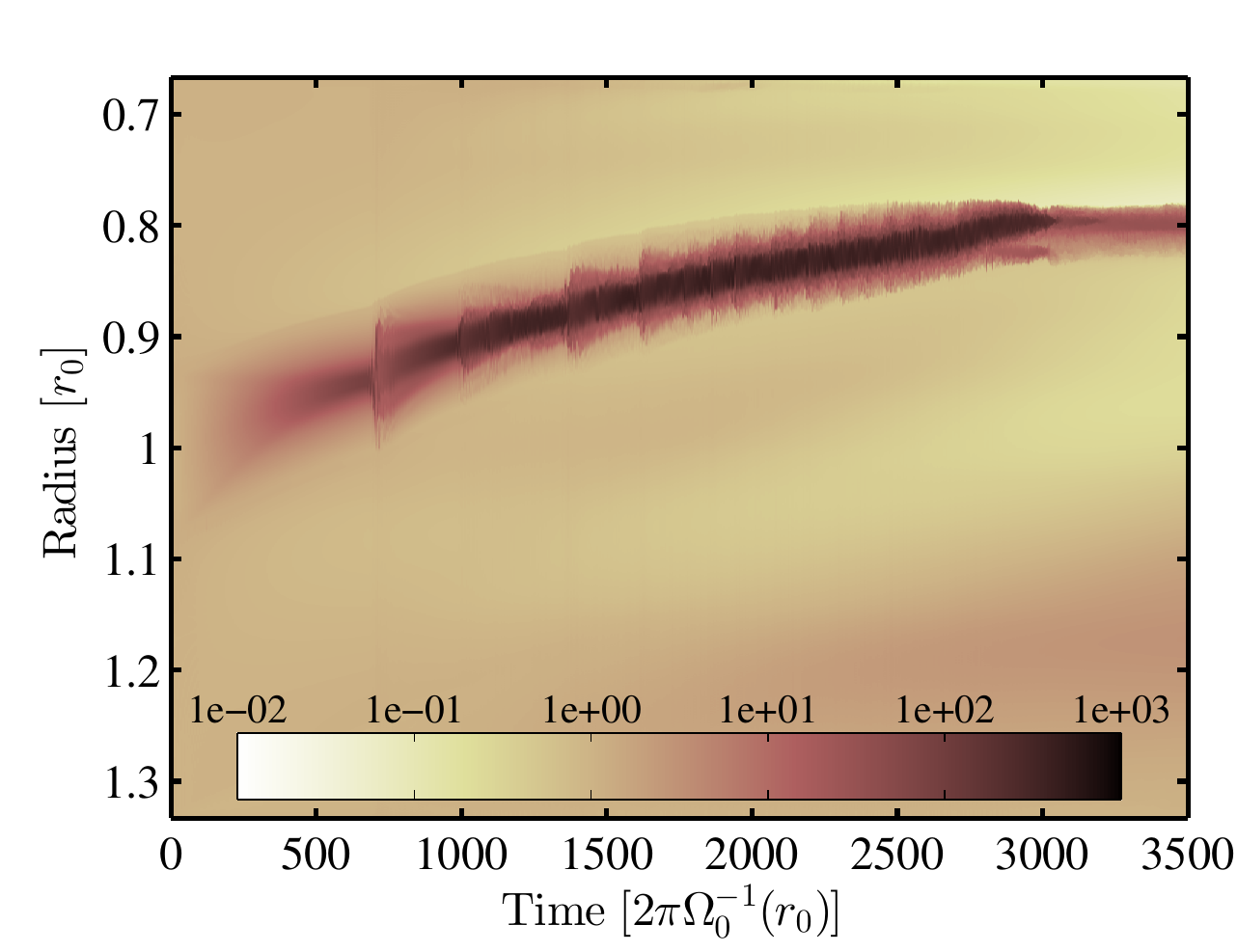}} &
	\scriptsize{$\:$ \newline \newline $St=4\times 10^{-3}$} \\

	\imagetop{\includegraphics[height=5.5cm, trim=2mm 0mm 0mm 3mm, clip=true]{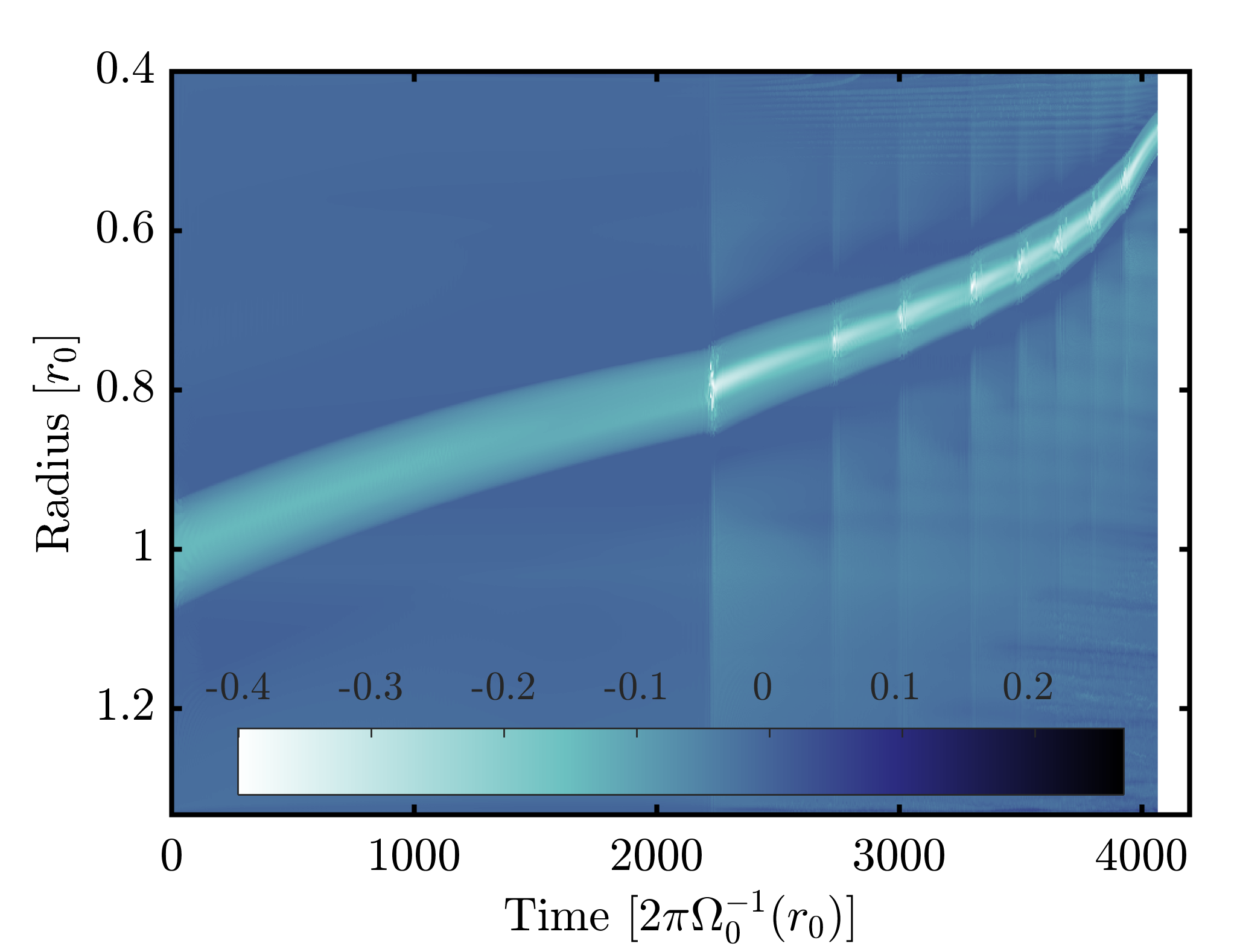}} &
	\imagetop{\includegraphics[height=5.5cm, trim=2mm 0mm 0mm 3mm, clip=true]{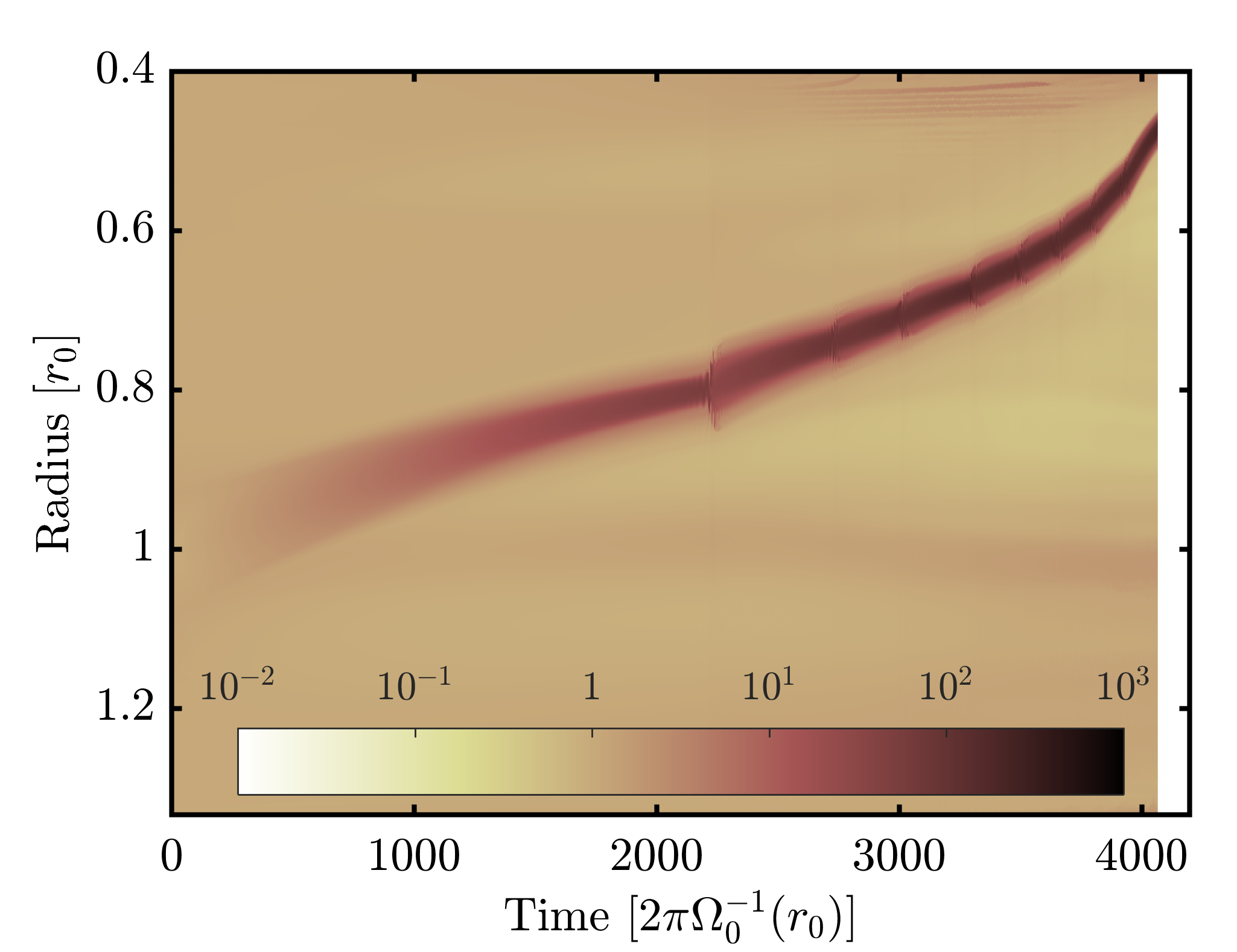}} &
	\scriptsize{$\:$ \newline \newline $St=1\times 10^{-3}$} \\

	\end{tabular}
	\caption{\label{Fig_Time_evolution} Results of the main runs. Long term evolution of the azimutal minimum of the Rossby number (left), and of the maximum of the dust density (right) for different values of the Stokes number. From top to bottom: ${St=4\times 10^{-2}}$, ${St=1\times 10^{-2}}$, ${St=4\times 10^{-3}}$, and ${St=1\times 10^{-3}}$ respectively. The dust density is relative to the background disk. }
      \end{center}
\end{figure*}

	We report and describe here the results of the four main simulations, investigating the evolution of vortices in the presence of dust populations with Stokes numbers ${St = [1, 4, 10, 40] \times 10^{-3}}$. We cover a range of values larger than an order of magnitude. 
%	
	%We will show that the evolution across three different stages (linear capture, unstable vortex phase, dust ring formation) is general down to Stokes of ${4 \times 10^{-3}}$, so over a decade in Stokes number variation. However the timescale from beginning to the third phase is heavily dependent on the Stokes number, while the maximum dust over density that can be reached, both locally in the vortex and in the ring, is of the same order of magnitude.
%
	In order to present the time evolution of the disk in a concise manner, we show for each run the time evolution of: {\it{(i)}} the azimuthal minimum of the Rossby number $Ro=||\vec{\nabla} \times [\vec{V}-\vec{V_0}(r)]||/[2 \Omega_k(r)]$, with $\vec{V_0}(r)$ the disk background velocity, {\it{(ii)}} the azimuthal maximum of the normalized dust density $\sigma_p/[\epsilon\sigma_0(r)]$. These extrema highlight the concentrations obtained in the vortex, and allows to present Figure \ref{Fig_Time_evolution} the whole evolution of the four runs.

	A case with ${St=4 \times 10^{-2}}$ and $\epsilon=10^{-2}$ was presented in \cite{Surville2016} but was only followed for 500 disk rotations. In this study, we evolve the case with ${St=4 \times 10^{-2}}$ (first row) during 1200 disk rotations until we observe a significant numerical dissipation of the structures. We observe three phases of evolution, in agreement with the results of \cite{Surville2016}. First, a phase of dust capture into the vortex occurs during the first hundred disk rotations, which can be seen in the dust density map (right) as a conic shape, revealing the inflow of dust toward the vortex center. Then the vortex instability develops at $t=100$ rotations, generating some strong vorticity throughout the vortex and further increasing the maximal dust density. The regime of unstable dusty vortex continues for almost $200$ rotations, until time $t=380$ rotations. During this period, dusty eddies are generated/destroyed inside the vortex, but never reorganize into a smooth structure. As a results, the dust density has a radially extended maximum, covering the majority of the vortex surface, as opposed to the strong confinement in the vortex center observed prior to the onset of the instability. Finally, the vortex is stretched along the whole azimuthal direction, and a turbulent dust ring forms. It consists in several dusty eddies and filaments evolving at different orbits, which explains the wide radial extent of the dust density maximum, which is almost $0.1 \: r_0$. After $t=800$ disk rotations, the numerical diffusion dissipates the eddies into a quasi-axisymmetric, quiet dust ring. In this near steady state, the maximum dust density drops down to around $50$ times the dust background.

	When we reduce the Stokes number to ${St=10^{-2}}$ (second row), we obtain a similar sequence of events. The dust capture into the vortex happens on a longer timescale, approximately $300$ disk rotations, before the vortex instability develops. In \cite{Surville2016}, we derived an analytical formulation of the capture time, $\tau_{1/2}$, which was approximately proportional to $1/St$ for the well-coupled grain regime, explaining a longer period of capture. Compared to the previous case, the Stokes number is reduced by a factor of 4, however the time of the end of the capture is only longer by a factor of 3. We will explain this discrepancy in the next section. After the vortex instability, the disk enters a long phase when the unstable dusty vortex survives and resists destruction. Similarly to the case with ${St=4 \times 10^{-2}}$, the structure remains turbulent and the vortex splinters and eventually dissolves. This period lasts almost $900$ disk rotations, until $t=1200$ rotations. After that time, the disk enters the dust ring phase, which lasts $500$ rotations, with robust dust eddies and filaments surviving. In particular, two persistent small dusty vortices detach from the main region of formation, and can be seen on the plots after $t=1200$ rotations in the region $0.8<r<0.9 \: r_0$. After $t=2200$ disk rotations, the structures are dissipated by the numerical diffusion.

	On the third row, we present the results obtained with ${St=4\times 10^{-3}}$. In this case, the capture of dust into the vortex takes about $700$ disk rotations before the vortex becomes unstable. However, a significant difference with the previous cases is visible. Just after the instability, the vortex becomes again quiet between $t=700$ and $t=1000$ rotations approximately. One can see on the dust density map (on the right) that a capture is again at work, revealed by the characteristic conic shape. A series of three additional vortex instabilities occurs during the period between $t=700$ and $t=1700$ disk rotations. This new phase of evolution of the dusty vortex is visible only for small well-coupled grains, and will be discussed in the next section. After that phase, the vortex is unstable, and persists in the disk until $t=2800$ rotations, with a high dust loading. Inside the vortex, the dust density is enhanced by more than $500$ times, and the local dust-to-gas ratio is in the range $5-8$. Finally, the dusty vortex is destroyed and a dust ring forms, and evolves during the last $500$ rotations of the simulation. We could not follow the evolution much longer, but based on the dissipation observed during the last hundred orbits, the dust ring may persist for a long time after the end of the run.

%%%
	The last setup of the main runs (bottom row) is performed with ${St=10^{-3}}$. {{The disk domain is more extended, with ${0.4<r<1.5 \: r_0}$ to provide even more room for radial migration. In order to keep a comparable numerical resolution with the other runs, we used $(N_r, \: N_{\theta})=(2048, \: 2048)$ grid cells. Assuming a grain density in the range of $\rho_s=1-3\: g.cm^{-3}$, the grain size corresponding to this Stokes number ranges in $r_s=0.2-0.5 \: mm$. Such small dust grains are well-coupled, and expected to have a negligible deviation from the gas streamlines. However, this small discrepancy is still enough to deviate the dust from the gas motion if the evolution is followed for a sufficiently long time. Our results show that, even for such a small Stokes number, the capture of the solids into the vortex is efficient.}}
	
	The dust is strongly accumulated to a critical density of $50$ times the background, and the vortex instability develops at $t=2100$ disk rotations. A second period of dust capture follows, from $t=2100$ to $t=2800$ disk rotations, and ends by a second event of vortex instability. This process happens again three times until $t=3600$ disk rotations, and is responsible for a significant increase of the dust density inside the vortex. After that phase, the vortex suffers an unstable evolution as observed in the other runs, but its high vorticity fastens quickly its radial migration. Even with this extended disk run, the vortex is not destroyed after $4000$ disk orbits of evolution, and we stop the simulation before observing the formation of a dust ring. However, at the end of the run, the vortex is very turbulent and is much more elongated than at earlier stage of evolution. It is an evidence of destabilization and possible vortex destruction, but this conclusion is not fully confirmed by the numerical simulation.

	Vortex migration is particularly visible in the ${St=4\times 10^{-3}}$, and ${St=1\times 10^{-3}}$ runs. Even if the axes of the figures are different, due to different disk size, in the two runs the vortex migration rate of $-1\times 10^{-4} \: r_0$ per disk rotation is similar. It is not affected by the grain size, {{indeed it depends mostly on the local disk conditions and on the vorticity of the vortex \citep{Paardekooper2010, Surville2012, Surville2013}}}. In fact, migration of the vortex occurs in the four runs; however the vortex is destabilized for the larger grains before it can migrate very far. For the smaller grains, because the capture timescale is much longer, the migration of the vortex is more apparent. In fact, for the ${St = 1 \times 10^{-3}}$ case, the vortex even reached the inner boundary before being destroyed into a dust ring.

%----------------------------------------------------
\subsection{ Results of the high resolution run }
\label{Sect_Results_2}

%-----------------------
\begin{figure}[t]
%% Figure 4
	\begin{center}
	\begin{tabular}{c}
	\scriptsize{Rossby number} \\
	\includegraphics[height=6cm, trim=2mm 0mm 0mm 3mm, clip=true]{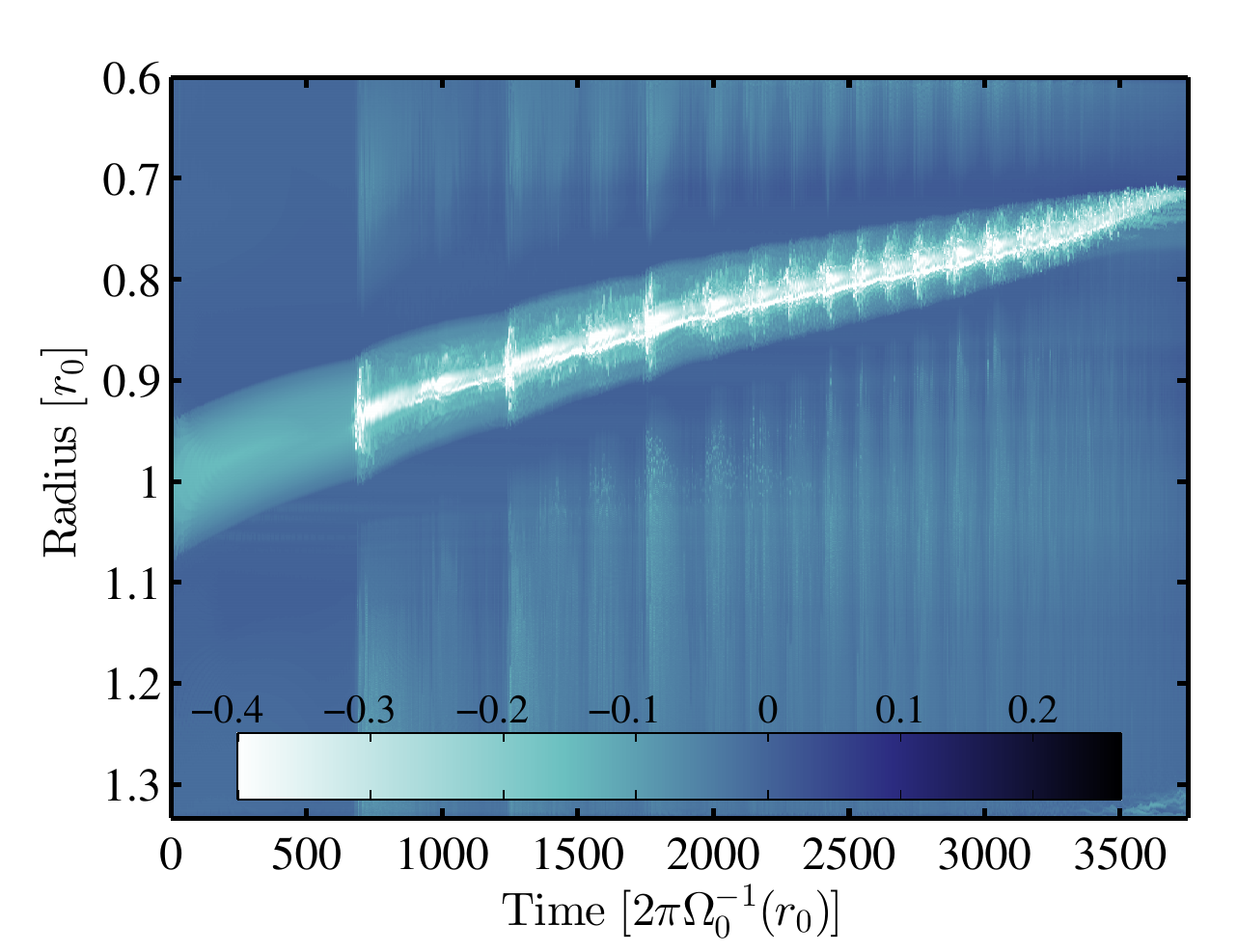} \\ \\
	\scriptsize{Dust density}  \\
	\includegraphics[height=6cm, trim=2mm 0mm 0mm 3mm, clip=true]{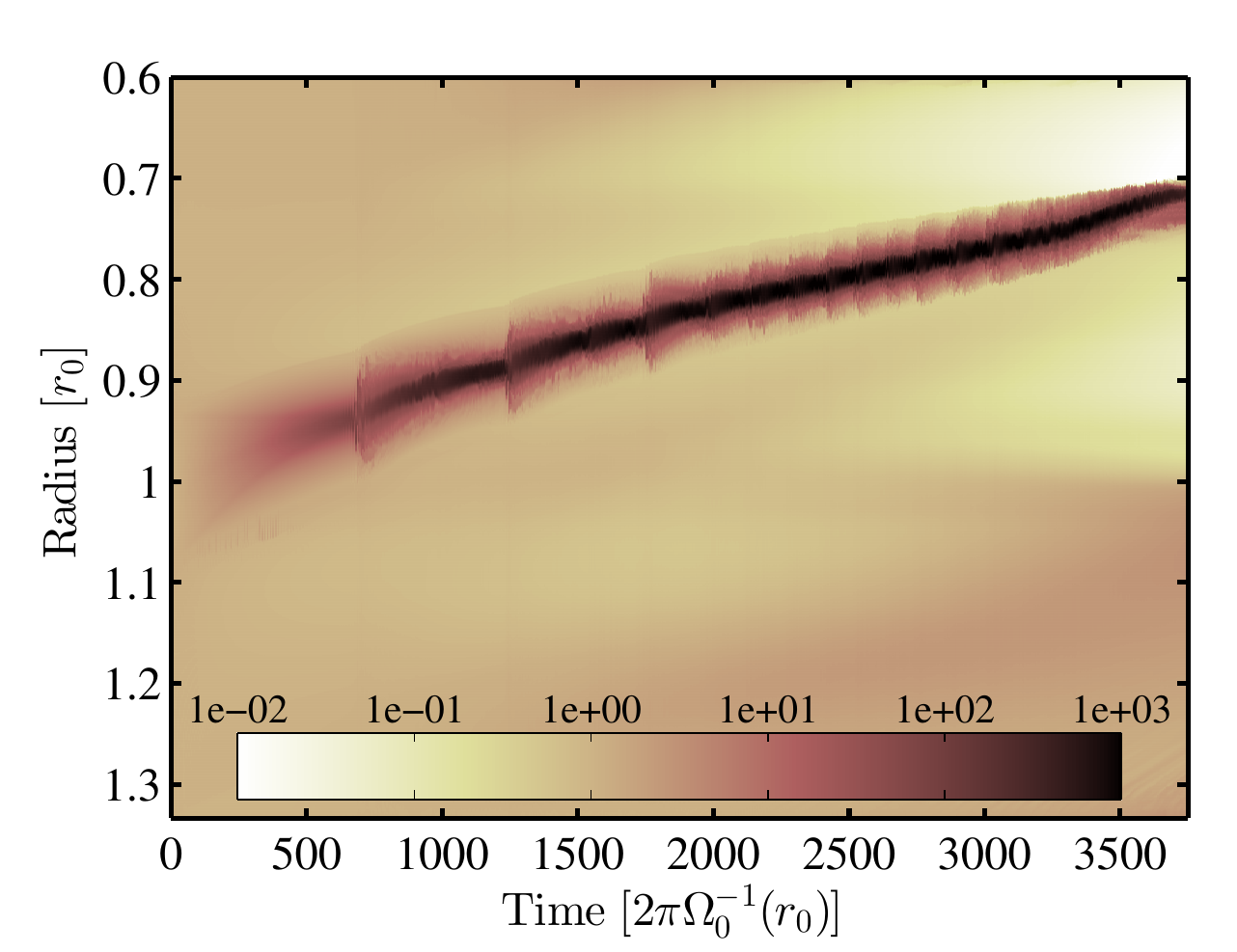} 
	\end{tabular}
	\caption{\label{Fig_Time_evolution_2k_4k} High resolution run for ${St=4\times 10^{-3}}$, using a numerical resolution of $(N_r, \: N_{\theta}) = (2048, \: 4096)$. We show the long term evolution of the azimutal minimum of the Rossby number (top), and of the maximum of the dust density (bottom). }
      \end{center}
\end{figure}

%-----------------------
\begin{figure}[t]
%% Figure 5
	\begin{center}
	\begin{tabular}{c}
	\scriptsize{Rossby number} \\
	\includegraphics[height=6cm, trim=2mm 0mm 0mm 3mm, clip=true]{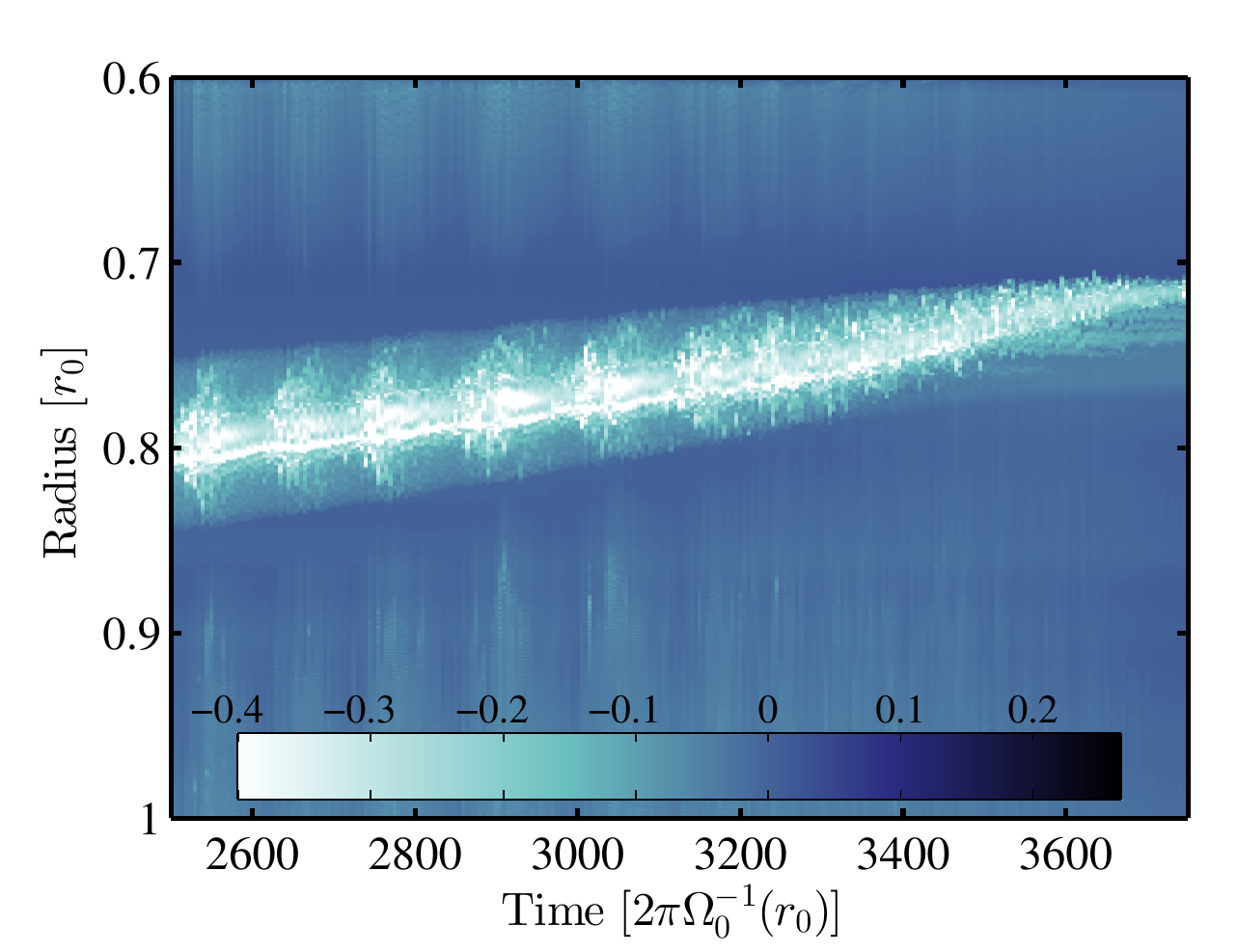} \\ \\
	\scriptsize{Dust density}  \\
	\includegraphics[height=6cm, trim=2mm 0mm 0mm 3mm, clip=true]{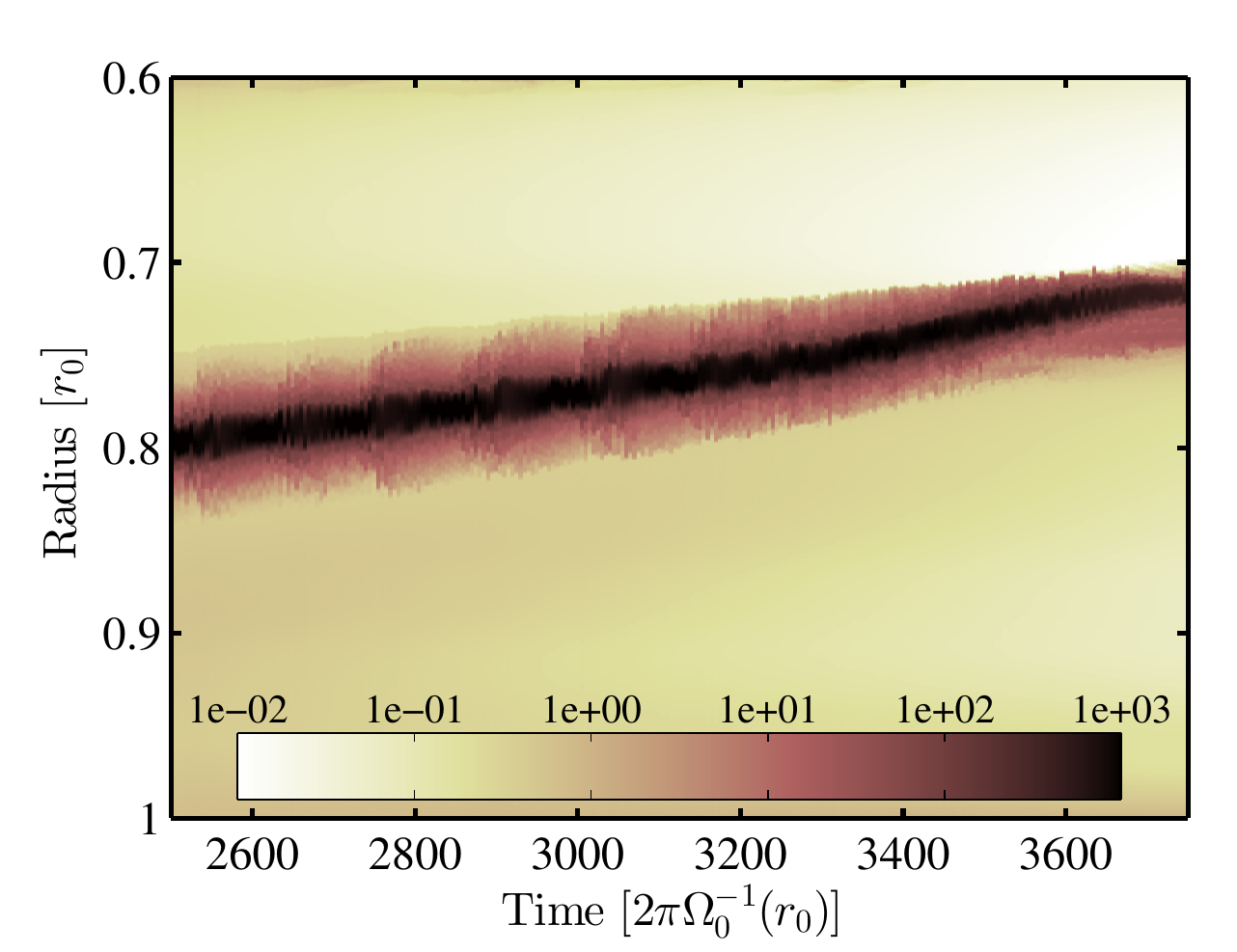} 
	\end{tabular}
	\caption{\label{Fig_Time_evolution_2k_4k_zoom} Ring formation phase for the high resolution run. We show the same results as in Figure \ref{Fig_Time_evolution_2k_4k} but we focus on the last thousand orbits and we have zoomed in radially. }
      \end{center}
\end{figure}

	The high resolution run is devoted to study with higher accuracy the evolution of the dust ring obtained in the ${St=4 \times 10^{-3}}$ case. The first motivation is that dust rings are composed of small scale eddies and filaments, which require high resolution to be well resolved. The second motivation is that, by using a resolution of $(N_r, \: N_{\theta})=(2048, \: 4096)$, one can compare this additional run with the results of \cite{Surville2016} where the Stokes numbers are larger. We have anticipated the possible radial migration of the vortex that was observed in the main run, and used a disk inner boundary at ${r=0.6 \: r_0}$ rather than $2/3$. All the other parameters are the same as in the main run (except resolution). The analysis of the high resolution additional run is presented Figure \ref{Fig_Time_evolution_2k_4k}.

	The evolution begins with the dust capture, the duration of which is almost identical to the results of the main run. The effect of the resolution is minor during this phase, implying that the main runs have enough resolution to capture the process accurately. The vortex instability develops at $t=700$ disk rotations, and is followed by a second phase of capture. However, this second phase is almost twice as long as in the main run, and the dust is more confined inside the vortex. {{ A second event of instability starts at $t=1250$ disk rotations, and results in a vortex with a turbulent core, the size of which is significantly wider than after the first event of instability. Then the vortex calms down and dust is again accumulated in its core. At $t=1700$ rotations, another vortex instability occurs followed by a short period of dust capture. After these three major instability events, affecting the vortex from its center to its border, the vortex cannot calm down or resist the turbulent flow sustained in its core. Over $\sim2000$ disk rotations, the nonlinear state of the two-fluid instability saturates and a frequent repetition of weaker instabilities at the vortex core is observed. The distinction between the different phases is more evident in the high resolution results than in the main run. The vortex migration is also more prominent in the high resolution run, probably because the vortex flow has less numerical dissipation.

	The formation of a dust ring was observed in the high resolution run at $t=3400$ disk rotations. It is significantly later than in the main run ($t=2800$ rot.) because the phase of sustained turbulence lasts longer in this additional run. Figure \ref{Fig_Time_evolution_2k_4k_zoom} zooms in the last thousand disk rotations of the simulation, when dust ring formation occurs. After $t=3400$ rotations, the vortex suffers destruction and dusty eddies are spread in the disk. It is visible on the Rossby number plot as horizontal stripes created at $0.70<r<0.75 \: r_0$. The dust density enhancement is kept higher than $50$ times in this region, in particular in a narrow region close to $r=0.71 \:r_0$ where it is almost $10^3$ larger than the disk background.}}

\bigskip

	In this section we have described the numerical results obtained for different grain sizes in the context of interaction with a large scale vortex. 
	In summary, we observe the same phases of vortex evolution at smaller grain sizes that we observed previously in \cite{Surville2016}, i.e. the initial capture of dust, the vortex instability, {{the unstable evolution when an active dusty flow is sustained inside the vortex}}, and the eventual dust ring formation. However, the duration of capture in the vortex is shorter for larger grains, and the repetition of instabilities of the vortex was discovered only for small grains, i.e. $St<4 \times 10^{-3}$.

%%%%%%%%%%%%%%%%%%%%%%%%%%%%%%%%%%%%%%%%%%%%%%%%%%%%%
\section{ Discussion }
\label{Sect_Discussion}

	This section is devoted to provide more insights on the different processes described in the previous section, and discuss the consequences on planet formation scenarios.

%----------------------------------------------------
\subsection{ Dust distribution during the capture }
\label{Sect_Linear_capture}

	In all the simulations performed with different grain sizes, the capture of solids inside the vortex was observed during the first phase of evolution. As discussed during the previous section, the estimated timescale of capture shows that it takes longer for smaller grains to accumulate to the critical point where the vortex is destabilized. This result was discovered in \cite{Surville2016} for $0.04<St<0.36$, and is extended down to $St=10^{-3}$ in this study. We will compare here the capture observed in the simulations with the analytical model, to infer the robustness of the model and to confirm the physics behind this capture.

	The analytical model of linear capture developed in \cite{Surville2016} is based on the coupled evolution of the gas and dust fluids. We will not detail here the calculation, we refer to Section 3 of this reference, but we recall the main equations of the coupled system:
%-----------------------
\begin{equation}
\label{Sys_evo_1}
\begin{aligned}
	& \partial_T \sigma_p^*  = 4 \pi A |Ro| \sigma_p^* \: , \\
	& \partial_T |Ro|   = - 4 \pi A \tilde{\epsilon} |Ro| \sigma_p^* \: .
\end{aligned}
\end{equation}
	
	Here $\sigma_p^*$ is the maximum of the dust density at the vortex center normalized to ${\epsilon \sigma_0(r)}$, $|Ro|$ is the absolute value of the Rossby number, ${T=2 \pi \Omega_0^{-1}}$ is the local period at the vortex orbit, and ${\tilde{\epsilon}= \left(1+\beta_\Omega/2\right) \epsilon \sigma_0(r)/ \sigma_g}$ is a modified dust-to-gas ratio at the vortex center. The parameter ${A= 2 St/(1+{St}^2)}$ represents the degree of coupling between the two fluids.

	From this coupled system of equations, we can compute the time evolution of $\sigma_p^*(T)$ as:
%-----------------------
\begin{equation}
\label{Equ_evo_dust_density}
	\sigma_p^*(T) = 1 + |Ro| \tilde{\epsilon}^{-1} \times \frac{1- exp(-4\pi A I_0 T)}{1 + |Ro| \tilde{\epsilon}^{-1} exp(-4\pi A I_0 T)} \:,
\end{equation}
	with $I_0 = \tilde{\epsilon} + |Ro|$ estimated at the beginning of the simulation, i.e. at ${t=0}$.

%-----------------------
\begin{figure}[t]
%% Figure 6
	\begin{center}
	\includegraphics[height=6.cm]{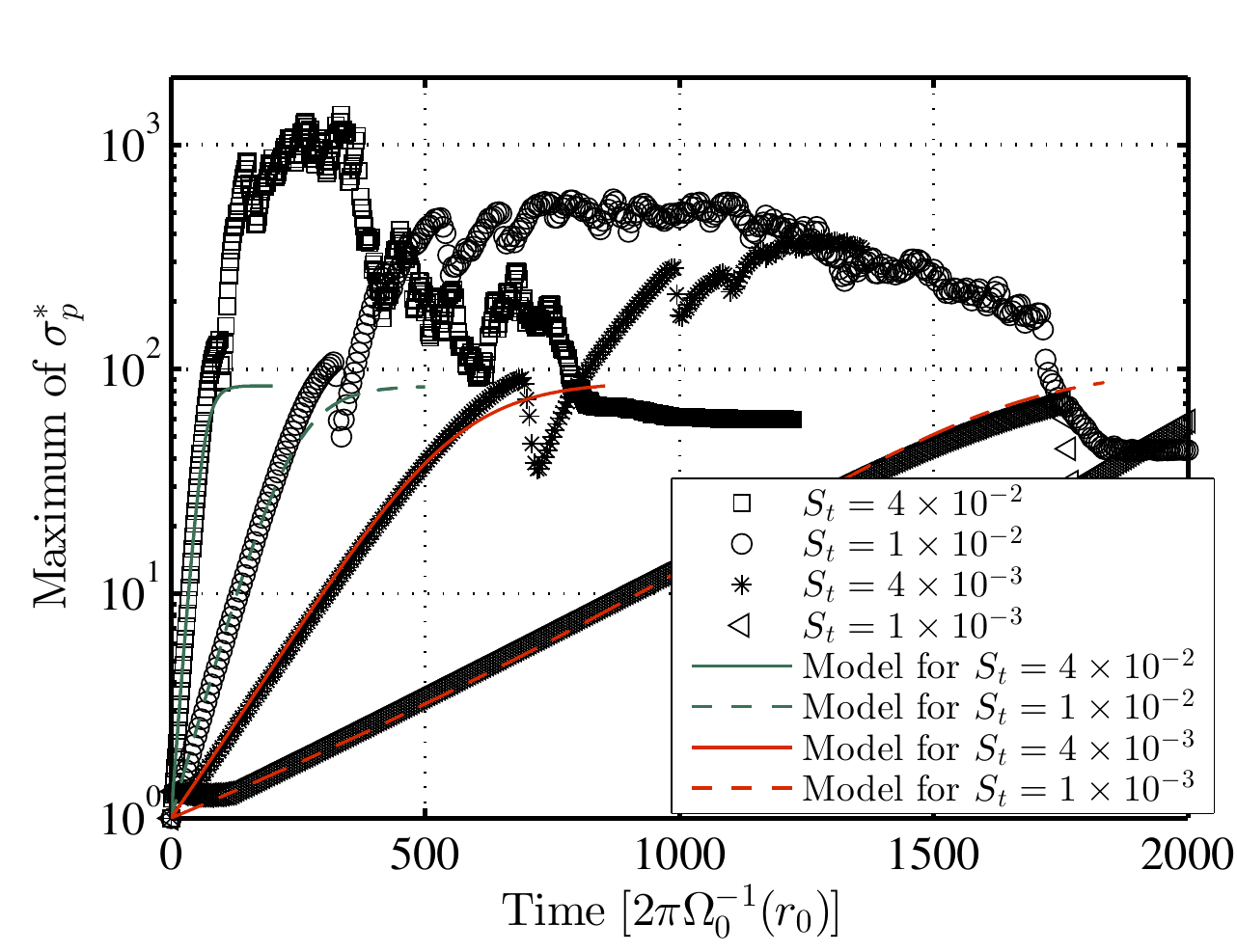}
	\caption{\label{Fig_Linear_capture} Linear dust capture into the vortex. We show the evolution of the dust maximum density, relatively to the disk background, for different grain sizes (symbols). The lines represent the evolution predicted from the dust capture model of \cite{Surville2016}. A very good agreement is obtained when taking into account the vortex migration, which modifies the local time reference (as $\Omega_k(r)$ reduces with $r$).}
      \end{center}
\end{figure}

	The main difference between the cases of well-coupled grains and larger ones relies on a significant vortex migration. This effect is due to the long duration of the capture before the vortex becomes unstable. Thus the local orbital period of the vortex $T$ in the model may change as long as the dust is accumulated. However, this change is slow, and can be estimated as $T(r_v)=2 \pi \Omega_0^{-1}(r_v)$ with the vortex orbit $r_v={r_v|}_{t=0} + v_{mig} T$. The measured migration velocity of this vortex is $v_{mig}= -1\times 10^{-4} \: r_0/T(r_0)$. In practice, we have measured the orbit of the vortex as a function of time for the different runs, and used these values for $r_v$. The change of $T(r_v)$ being slow, the integration of the system (\ref{Sys_evo_1}) with a varying $T$ can be approximated to a good accuracy by only replacing $T$ by $T(r_v)$ in the solution Equation (\ref{Equ_evo_dust_density}).

	The comparison of this modified model with the main runs is presented Figure \ref{Fig_Linear_capture}. The symbols represent the four different Stokes numbers used in the main runs. For each one, the result of the model is plotted with a colored line. It is striking that the increase of the maximum of dust density at the vortex center follows an exponential growth as predicted, even for Stokes numbers as small as $10^{-3}$. The growth rates of the maximum dust density are correctly estimated by the modified model for all the tested grain sizes, which confirms its validity under a wide range of grain sizes and vortex profiles.
	
	We can be confident that the drag interaction between gas and dust is responsible for the accumulation of solids in anticyclonic regions, as described by our model. The correction of the local orbital period, necessary to fit the numerical results, has significant implications. It shows that the exponential growth of dust density depends on local conditions of the dusty vortex, and on the grain size. We show evidence that there is no critical grain size (or Stokes number) for which this growth disappears. In fact even well-coupled dust grains undergo the capture due to the drag. As predicted by our capture model, the growth rate asymptotically tends to zero as the Stokes number decreases, due to proportionality in this regime.

	Despite this analytical result, there must, be in reality, a threshold in Stokes number below which this capture would be too long to be relevant. Considering for example that, at a few au, the giant planets cores are formed by core-accretion within million years, the planetesimal formation must occur on faster timescales. As a consequence, the solid material contained in small dust grains must assemble even faster into pebbles and planetesimals. This implies there is an upper limit on the timescale in which this scenario of capture and subsequent collapse of high concentrations of small grains is useful. Estimating the capture timescale from the results of the analytical model of \cite{Surville2016} (see Equation 38 of this reference), and limiting it to $10^5$ orbits at $5$ au, gives a minimum Stokes number in the range $10^{-5}-10^{-4}$. There is one final limiting factor --- the survival of large-scale vortices over such a long period. One can reasonably reduce by a factor of ten this estimate, and argue that grains of Stokes numbers smaller than $10^{-4}$ cannot contribute to the disk evolution in the scenario of vortex driven planetesimal formation.

%-----------------------
\begin{figure*}
%% Figure 7
	\begin{center}
	\begin{tabular}{ccp{15mm}}
	\scriptsize{Rossby number} & \scriptsize{Dust density}  & \\
	\imagetop{\includegraphics[height=5.5cm, trim=2mm 0mm 0mm 3mm, clip=true]{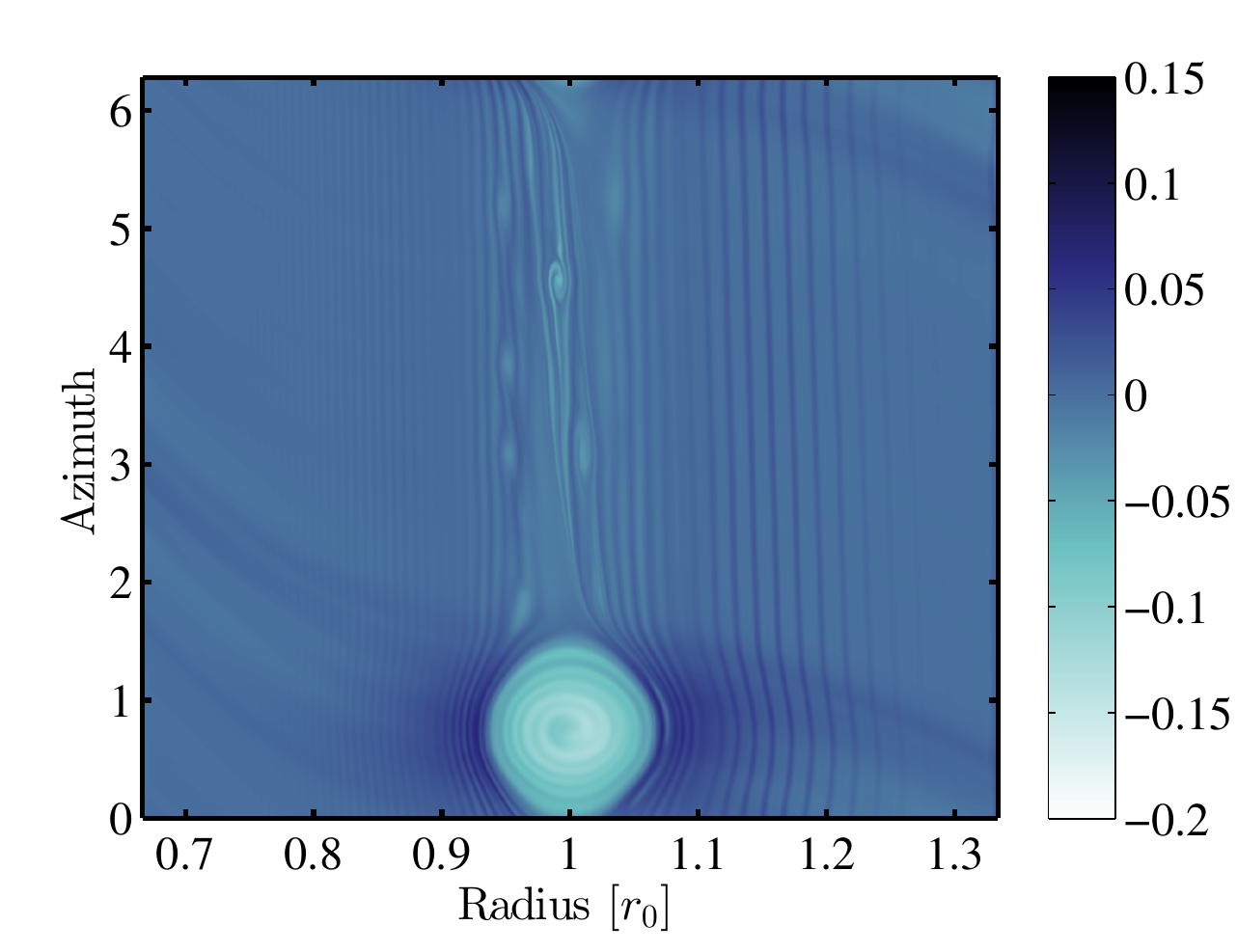}} &
	\imagetop{\includegraphics[height=5.5cm, trim=2mm 0mm 0mm 3mm, clip=true]{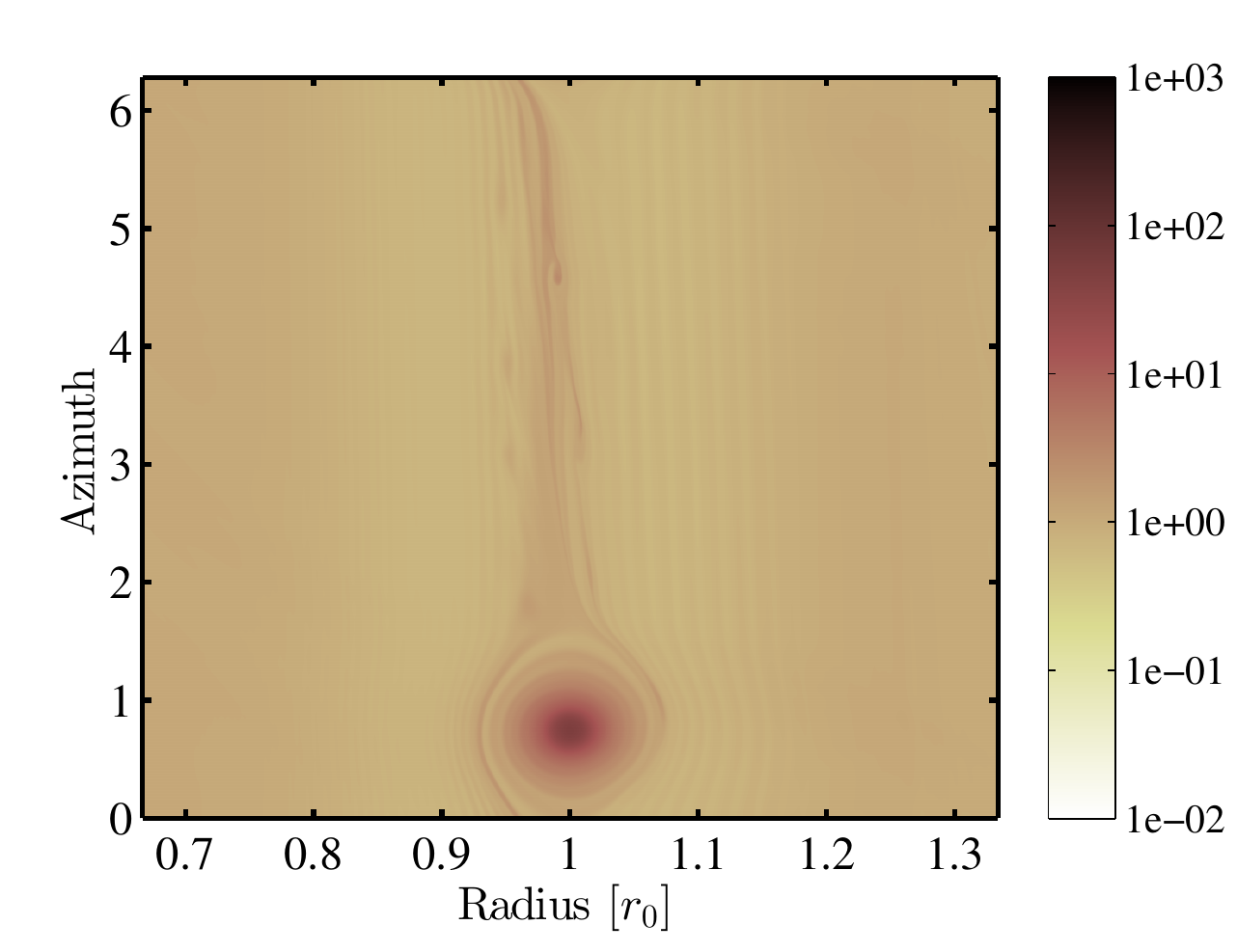}} &
	{\scriptsize{ $\:$ \newline \newline ${St = 4\times 10^{-2}}$ \newline $t=60 \: rot$}} \\

	\imagetop{\includegraphics[height=5.5cm, trim=2mm 0mm 0mm 3mm, clip=true]{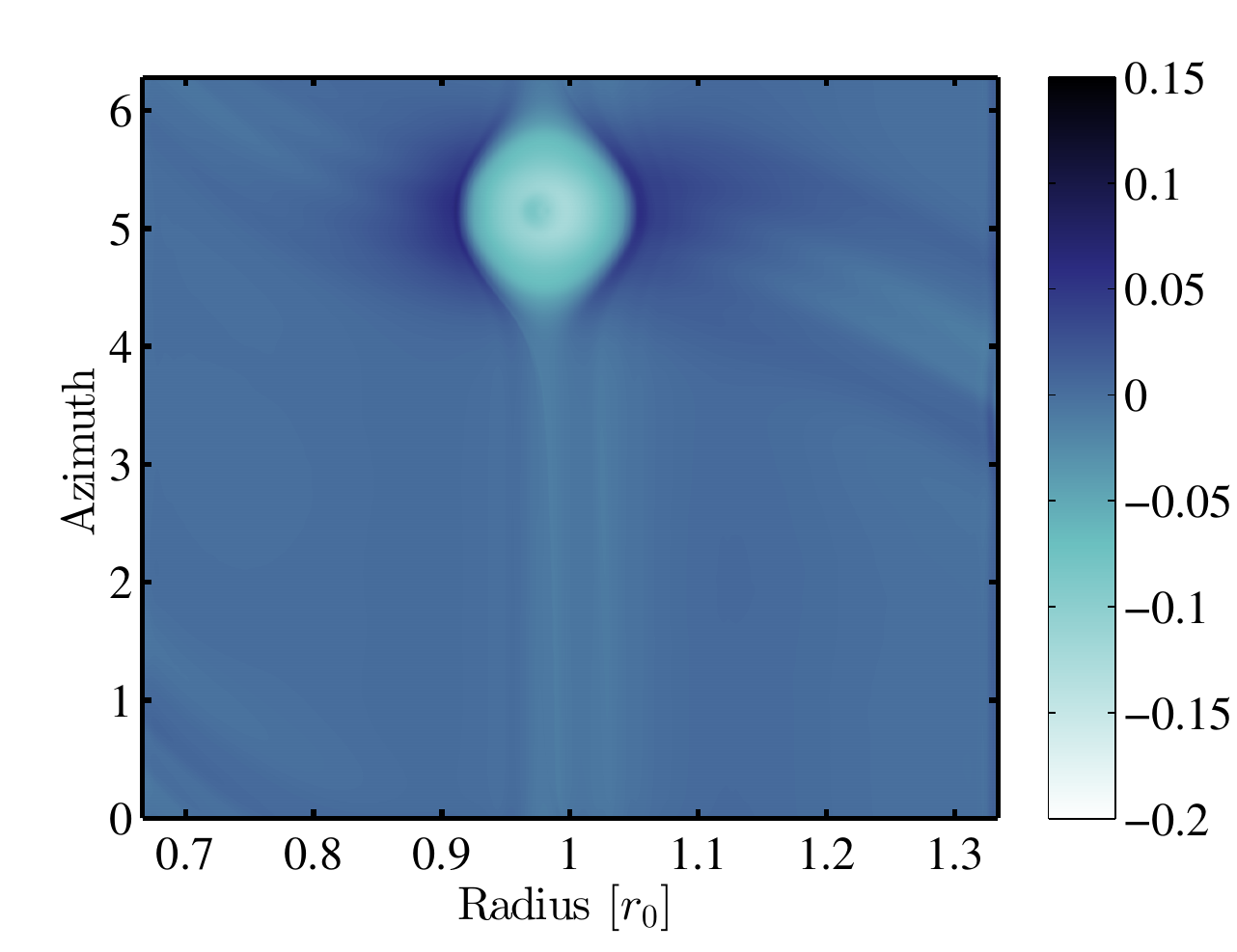}} &
	\imagetop{\includegraphics[height=5.5cm, trim=2mm 0mm 0mm 3mm, clip=true]{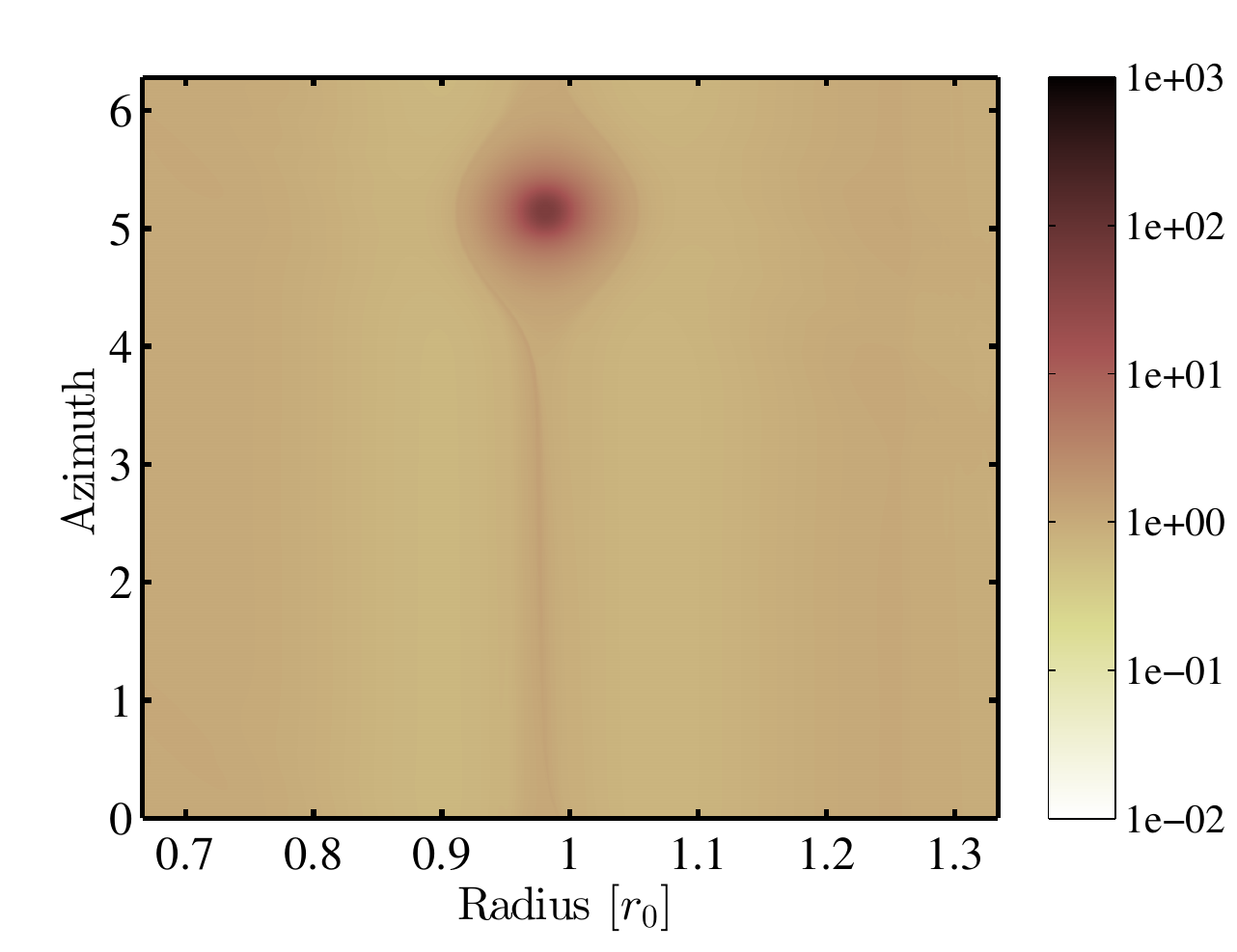}} &
	\scriptsize{ $\:$ \newline \newline ${St = 1\times 10^{-2}}$ \newline $t=235 \: rot$} \\

	\imagetop{\includegraphics[height=5.5cm, trim=2mm 0mm 0mm 3mm, clip=true]{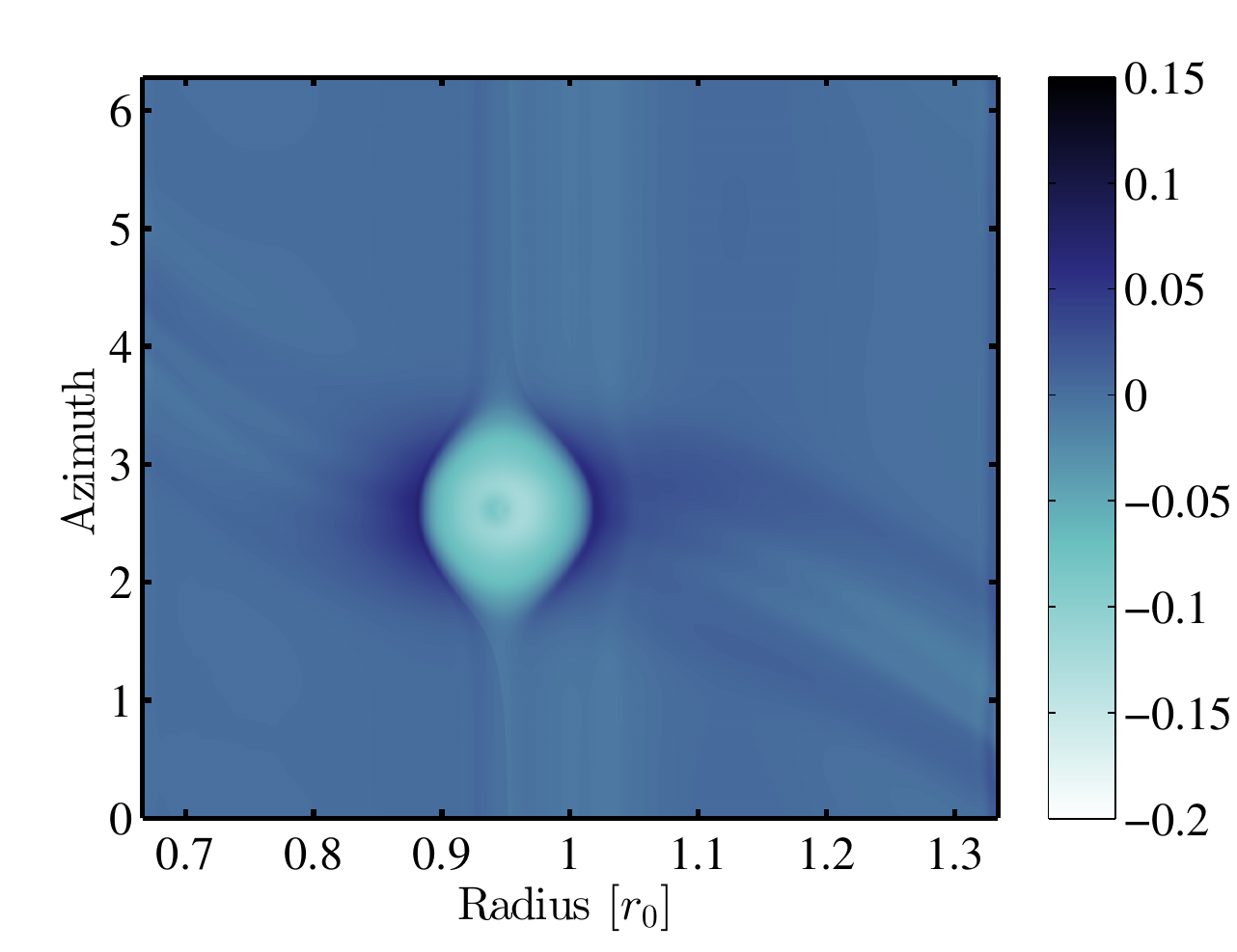}} &
	\imagetop{\includegraphics[height=5.5cm, trim=2mm 0mm 0mm 3mm, clip=true]{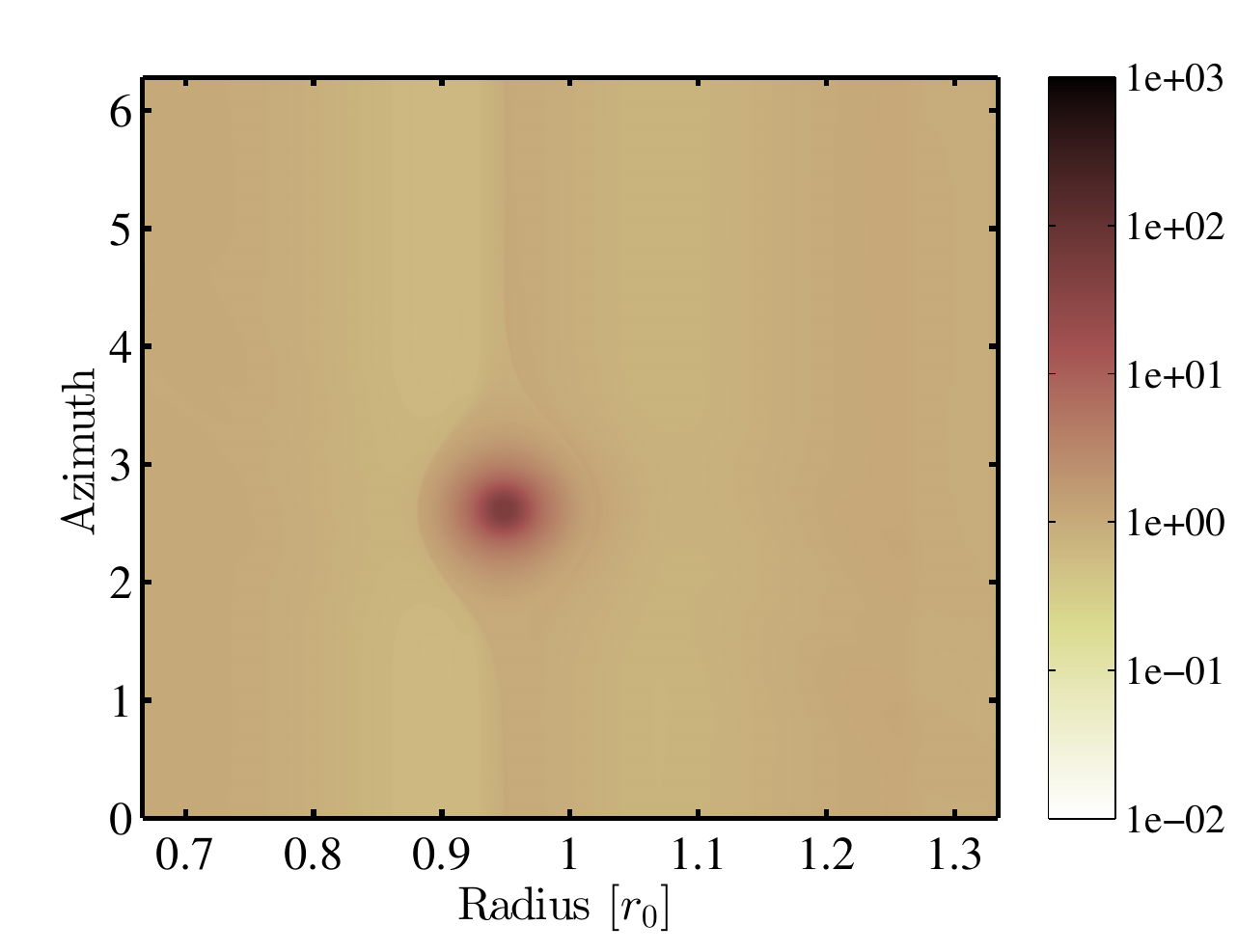}} &
	\scriptsize{ $\:$ \newline \newline ${St = 4\times 10^{-3}}$ \newline $t=545 \: rot$} \\

	\imagetop{\includegraphics[height=5.5cm, trim=2mm 0mm 0mm 3mm, clip=true]{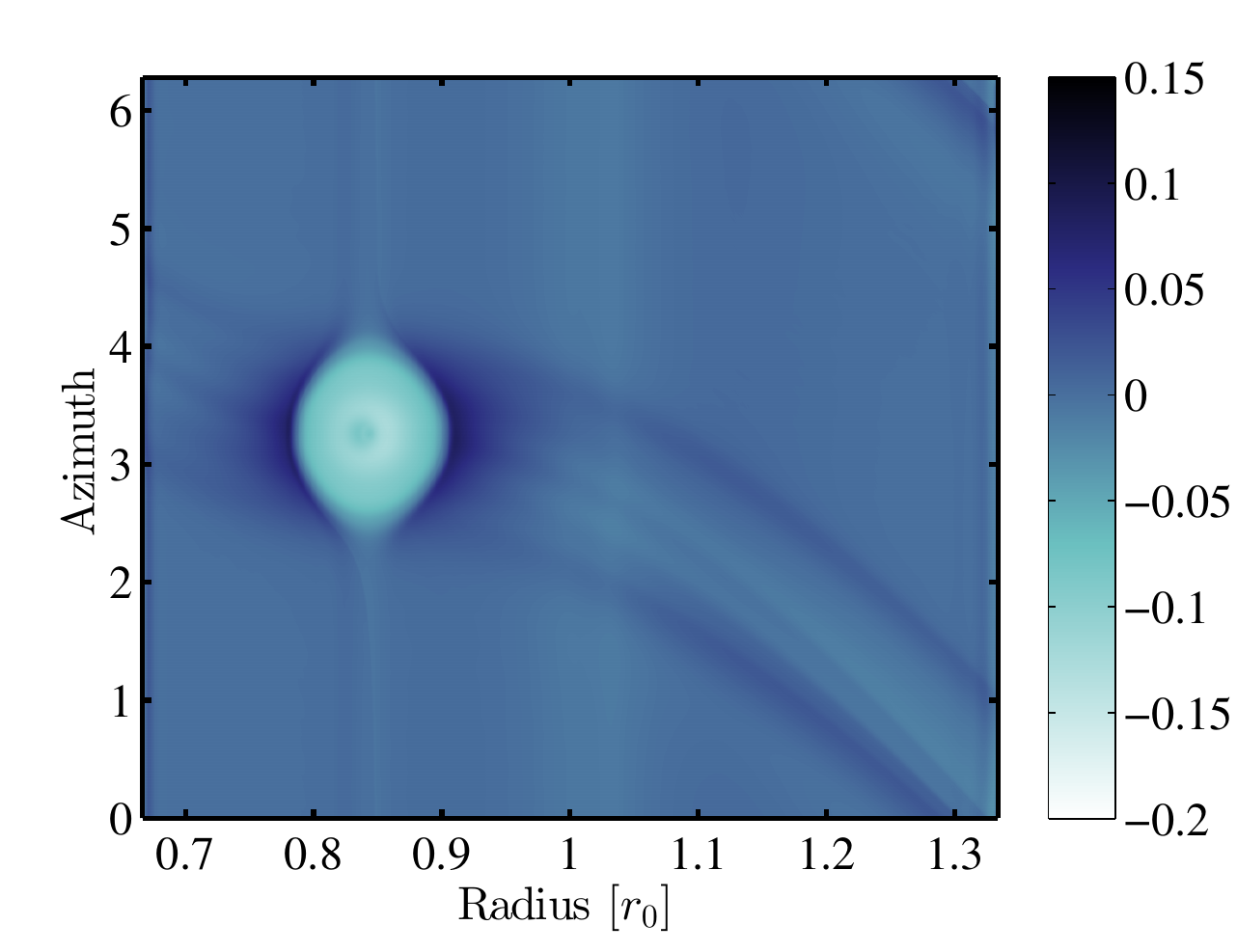}} &
	\imagetop{\includegraphics[height=5.5cm, trim=2mm 0mm 0mm 3mm, clip=true]{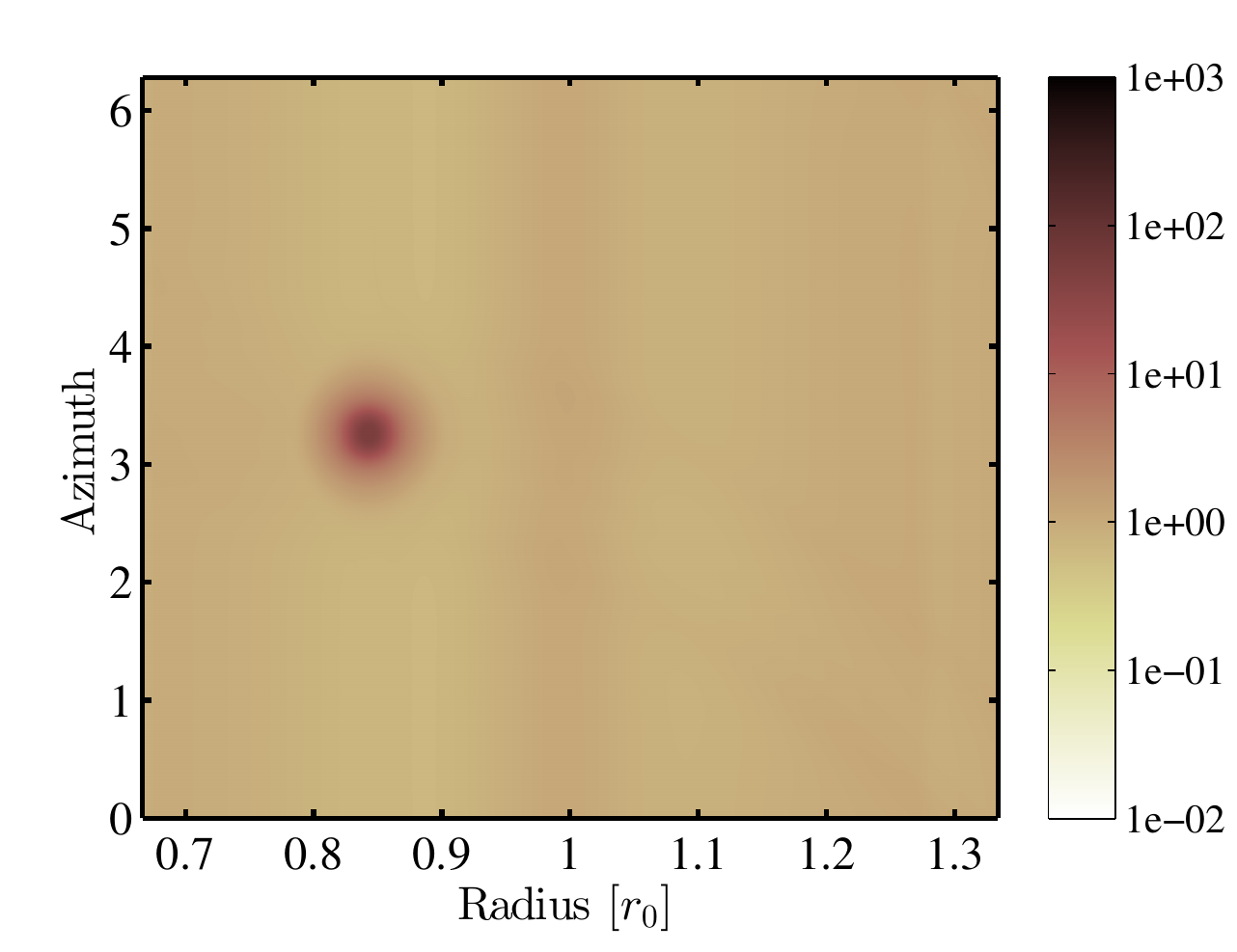}} &
	\scriptsize{ $\:$ \newline \newline ${St = 1\times 10^{-3}}$ \newline $t=1540 \: rot$}
	\end{tabular}
	\caption{\label{Fig_Capture_correspond} Disk profiles at equivalent stages of evolution, during the capture phase, when $\sigma_p^*=50$ at the vortex center. From top to bottom: Snapshots at ${t=61, \: 235, \: 545,}$ and $1540$ disk rotations for ${St= 4 \times 10^{-2},}\: 10^{-2}, \: {4 \times 10^{-3},}$ and $10^{-3}$, respectively. On the left are reported the $(r, \: \theta)$ maps of the Rossby number and on the right the maps of the dust density, in a logarithmic color scale. }
      \end{center}
\end{figure*}

	It is also important that we consider the duration of capture. The analytical model predicts that the maximum density of dust achieved at the end of capture is the same for all the Stokes numbers, because this value depends only on the Rossby number of the vortex and on the initial dust-to-gas ratio. It is confirmed by the numerical results, as we measure a dust density enhancement in the range $80-120$ for the four grain sizes before the vortex instability. The analytical model of capture predicts a maximum dust enhancement of:
%-----------------------
\begin{equation}
	\sigma_p^*(T \to \infty) = 1+ \frac{|Ro|}{\tilde{\epsilon}} \sim 85,
\end{equation}
using the parameters of the vortex. The agreement is thus satisfying over decades of grain sizes, even when the vortex migration is significant. 

	The findings above suggest that the evolution of dusty vortices are self similar, during the phase of capture. Comparing a snapshot of the structure of the disk for different grain sizes requires that it corresponds to the same time during the process of capture. Defining this time correspondence is possible for example by choosing a value of dust enhancement. Here, if we choose $\sigma_p^* = 50$, the right times for equivalent state of evolution are ${t=61, \: 235, \: 545,}$ and $1540$ disk rotations for ${St= 4 \times 10^{-2},}\: 10^{-2}, \: {4 \times 10^{-3},}$ and $10^{-3}$, respectively.

	In Figure \ref{Fig_Capture_correspond}, we show the color maps of the Rossby number and of the dust density for the four main runs, at these 'equivalent' evolution times. It is striking to see how similar the distributions of dust are inside the vortex, but also the similarity of the vorticity profiles. Only the case with $St=4\times 10^{-2}$ shows noticeable differences, with some structures spiraling inside the vortex. Indeed, the vortex is still relaxing from the initial conditions and capturing the dust at the same time. These spirals are also visible in a run with only, and so not due to the presence of dust.

	In comparing these snapshots, it is very difficult to distinguish between the different grain sizes. The dust grain distribution inside the vortex follows approximately the same streamlines as the gas. However, as function of the distance from the vortex center, the density of dust is the same regardless the Stokes number.

{{Our results differ from some recent papers showing that the dust density profile inside a vortex depends on the grain size \citep{Lyra2013, Barge2017, Sierra2017}. The analytical models behind these studies assume a turbulent diffusion of dust, due to an unresolved (3D) turbulence, which allows for a steady state inside the vortex. This vortex turbulence may be produced by the saturation of the elliptical instability acting in incompressible vortices \citep{Lesur2009}. Thus, these studies concern a particular category of vortices. As a result, the dependence on the grain size is effective when the accumulation of dust in the vortex is large.}}

	{{In our study, the compressible vortex has open streamlines, and is unaffected by this instability (similarly to the 3D vortices found in \cite{Meheut2010}). This class of vortices evolves differently because the dust diffusion rises from the instability of the vortex triggered by the two-fluid interaction, at the end of the capture phase.}}

	{{The comparison between these two classes of vortices suggests that}}, if additional turbulence exists, the changes of gas vorticity and other profiles inside the vortex core -- due to turbulence -- will happen on short timescales, probably shorter than the time is takes for the drag to act on the dust. Then, the time averages over the drag timescale will be the quantities that describe the capture. We can reasonably assume that turbulence may not drastically change the capture law. Moreover, in both approaches, the concentration of solids in the vortex should be slower for smaller gains, and saturate either due to artificial diffusion or due to the two-fluid instability.

	The timescale of capture is thus fondamental for the comparing the evolution of different grain size. For example, the numerical simulations done in \citep{Barge2017, Sierra2017} compare the dust distribution inside a vortex for different grain sizes. This comparison is done only after 100 disk rotations, which biases the results. The smallest grains do not have time to concentrate enough to reach a steady state or the vortex instability, while the largest grains will pass beyond the vortex instability. In contrast, our results suggest that the vortex evolution during the phase of accumulation is the same for all grain sizes --- just with a different timescale --- until a saturation is reached, {{either because of the change of the vorticity profile in our model of compressible vortices, or by the effect of turbulent diffusion in incompressible turbulent vortices.}}

	Finally, additional processes like considering a wide distribution of grain sizes in the disk and the possibility for grains to grow or fragment may change the density profiles in the vortex. The observability of the well-coupled grains in vortices can be explored by producing simulated images, for example of what would be seen by ALMA. The multi-wavelength maps of the vortex region could help to disentangle the grain size of the dust \citep{Gonzalez2012}. This extension of our study to the observability will be addressed in a future publication.

%----------------------------------------------------
\subsection{ Effects of the grain size on the vortex instability }
\label{Sect_Grain_Size}

%-----------------------
\begin{figure*}
%% Figure 8
	\begin{center}
	\begin{tabular}{cc}
	\scriptsize{$St=4\times 10^{-2}$} & \scriptsize{$St=1\times 10^{-2}$}  \\
	\scriptsize{$t=84 \: rot$} & \scriptsize{$t=310 \: rot$}  \\
	\includegraphics[height=6.cm, trim=2mm 0mm 0mm 3mm, clip=true]{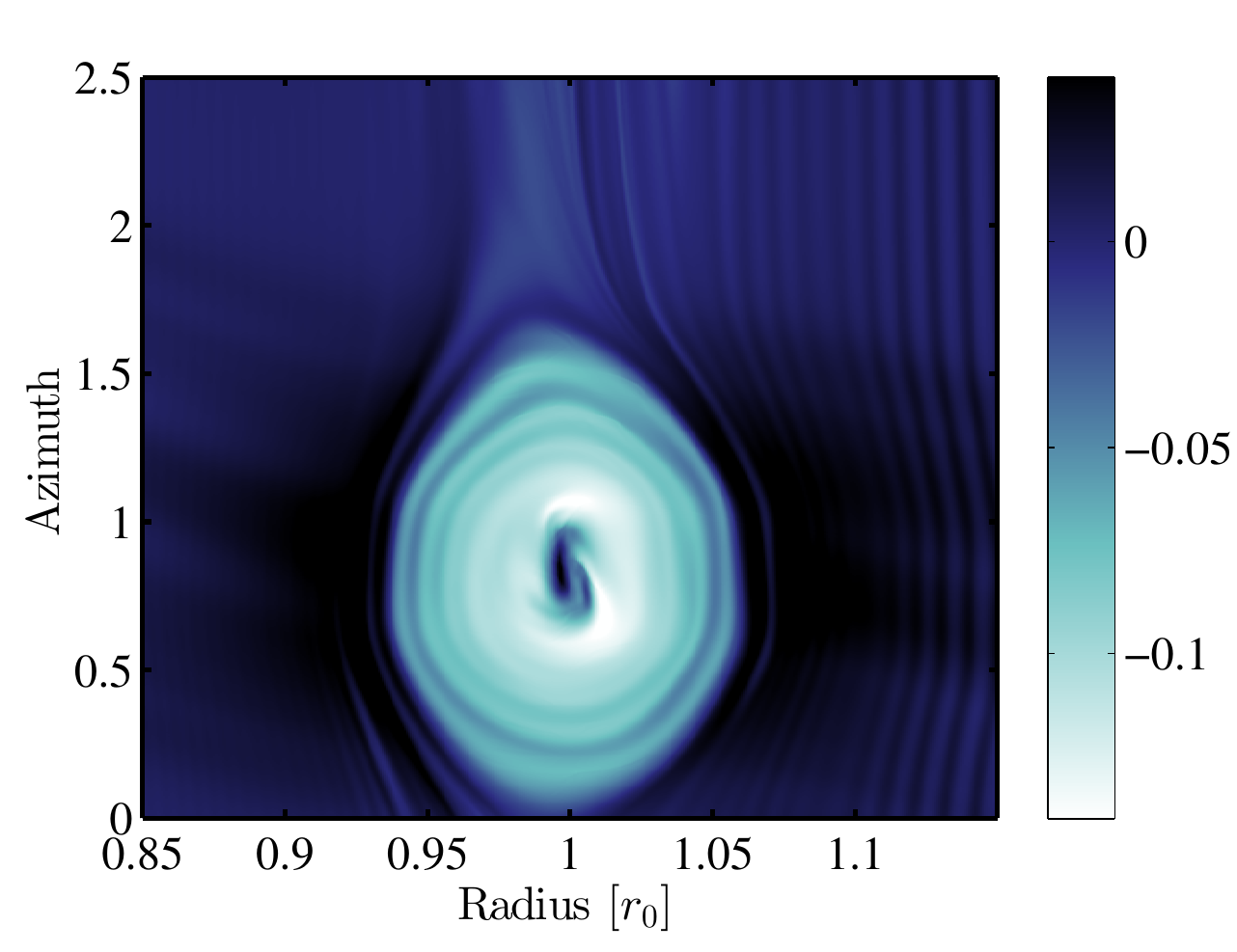} &
	\includegraphics[height=6.cm, trim=2mm 0mm 0mm 3mm, clip=true]{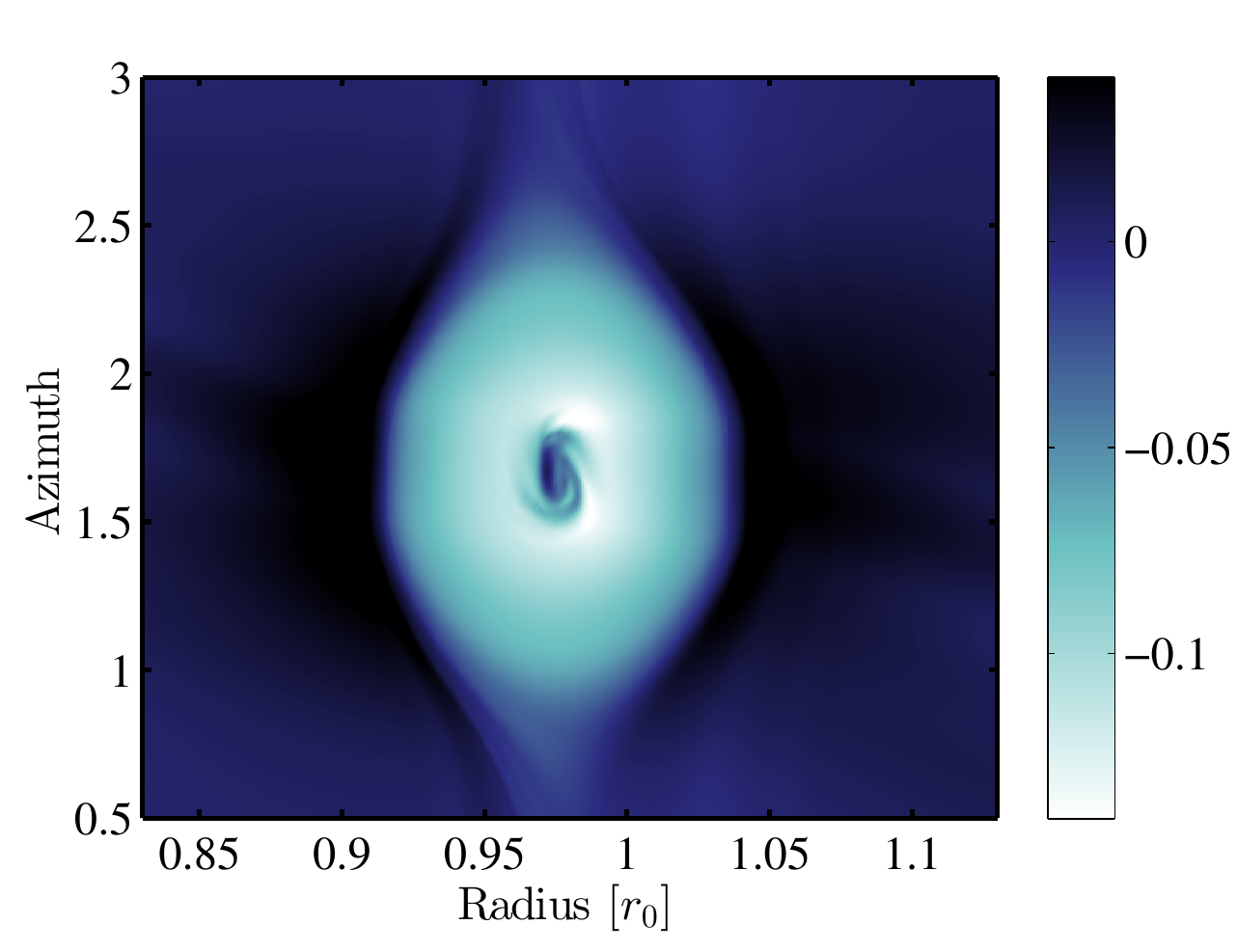} \\ \\

	\scriptsize{$St=4\times 10^{-3}$} & \scriptsize{$St=1\times 10^{-3}$}  \\
	\scriptsize{$t=685 \: rot$} & \scriptsize{$t=1740 \: rot$}  \\
	\includegraphics[height=6.0cm, trim=2mm 0mm 0mm 3mm, clip=true]{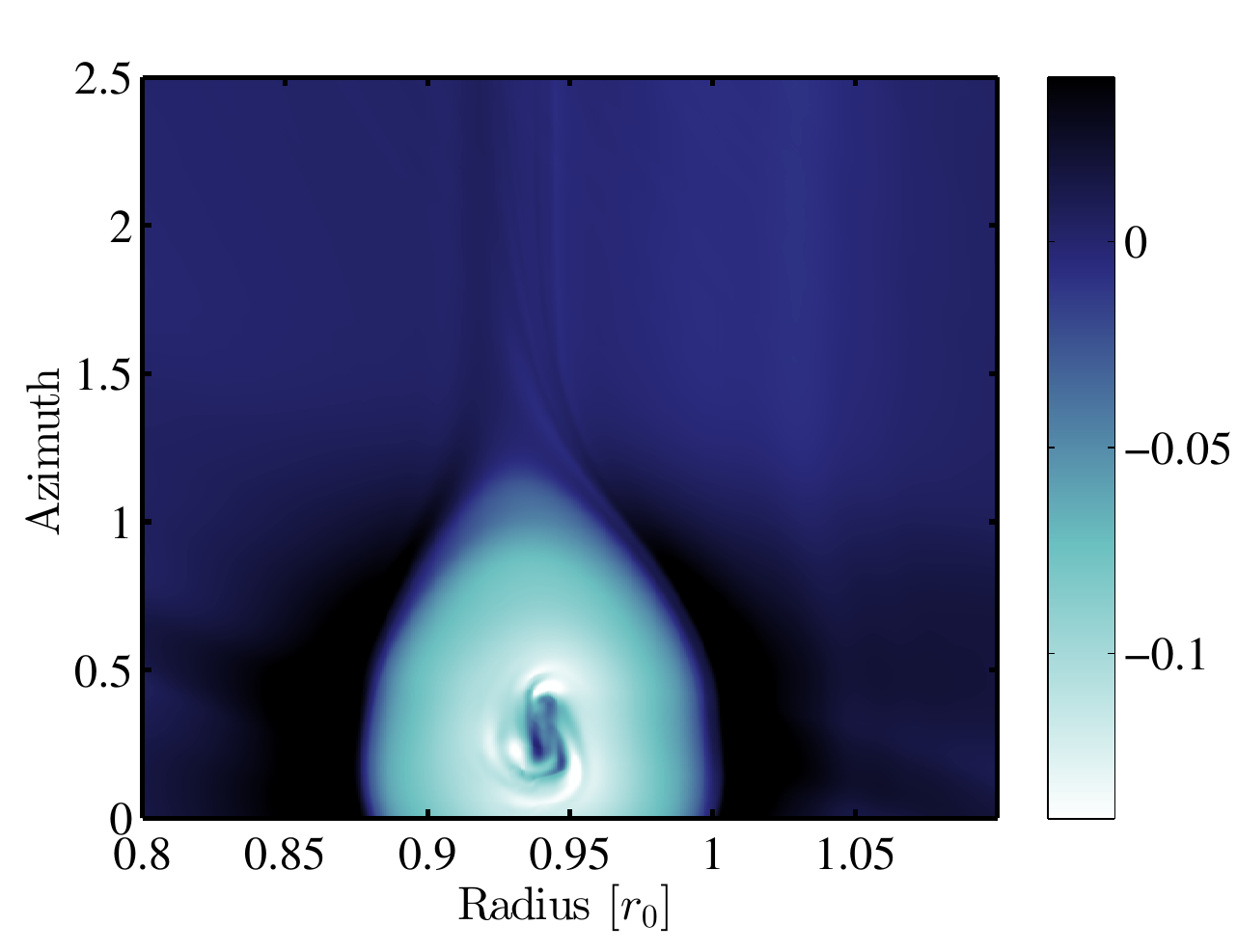} &
	\includegraphics[height=6.0cm, trim=2mm 0mm 0mm 3mm, clip=true]{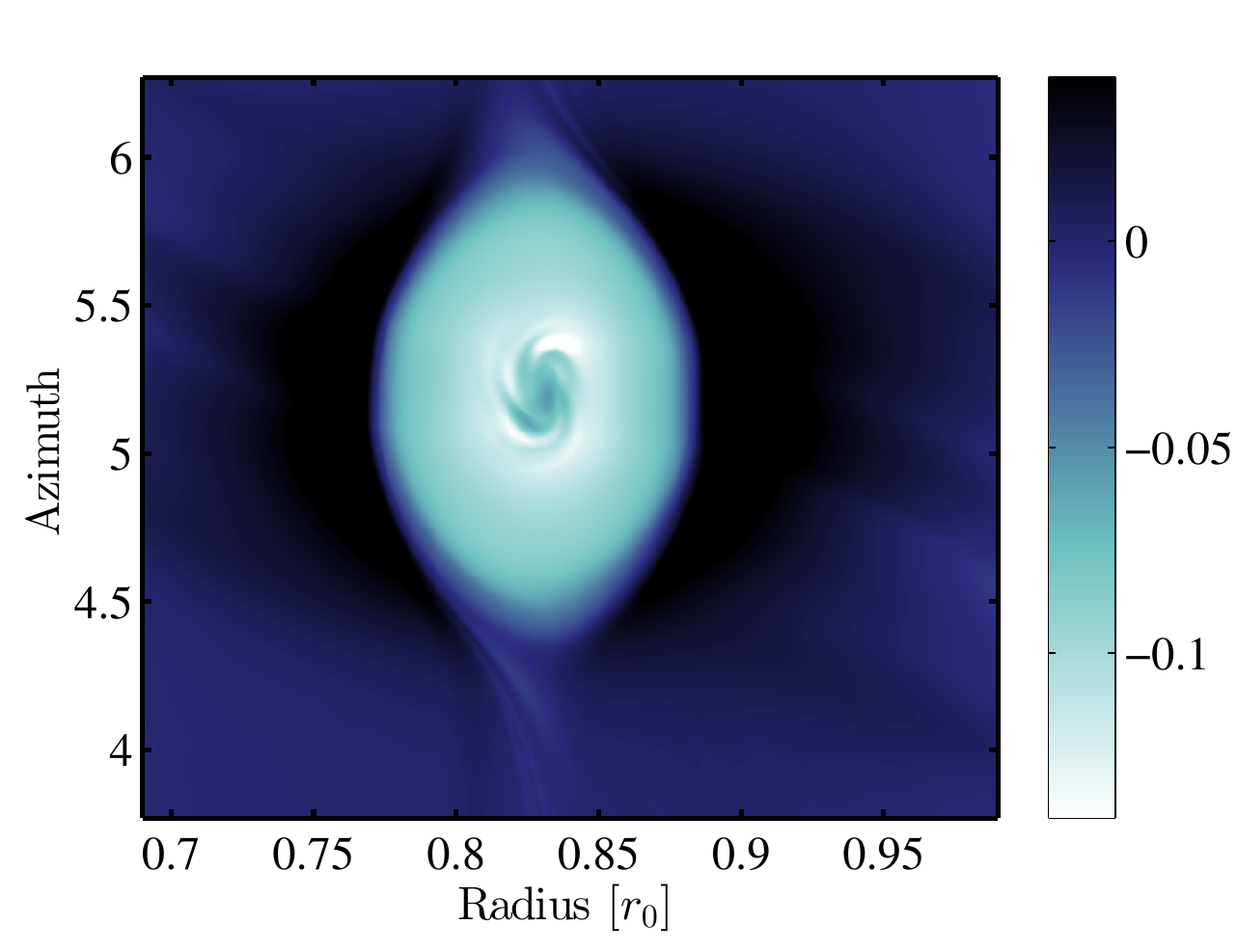} \\ 

	\end{tabular}
	\caption{\label{Fig_Threasold_insta_global} Saturation of the first modes during the vortex instability. We show a zoom in of the Rossby number maps for different Stokes numbers at the time when the first most recognizable modes are established. When decreasing the Stokes number (from left to right and top to bottom), higher modes saturate: $m\sim 2$, $m\sim 2-3$, $m\sim 3-4$ and $m\sim 4-5$ for the main runs with ${St=4\times 10^{-2}}$, ${St=1\times 10^{-2}}$, ${St=4\times 10^{-3}}$, and ${St=1\times 10^{-3}}$, respectively. }
      \end{center}
\end{figure*}

	In all the simulations done with different grain sizes, from $St=0.35$ to $St=0.04$ in our previous study \cite{Surville2016}, and here from ${St=4\times10^{-2}}$ to ${St=10^{-3}}$, the vortex becomes unstable after the period of dust capture. The duration of the instability is very limited, of the order of $50$ disk rotations. Its dependence on the Stokes number of the dust fluid is difficult to estimate, and varying $St$ over two orders of magnitude has no significant influence on the duration of the vortex instability. This observation indicates that the conditions of the flow triggering the instability are very similar in the different cases, and are likely linked to the gas profile in the vortex at the end of the process of dust capture. The similarities of the flow profiles for different grain sizes when compared at equivalent times of evolution, as discussed in the previous section (see Figure \ref{Fig_Capture_correspond}), corroborate this argument.

	We compared the vorticity profiles of the main runs, at the time when the instability of the vortex starts. These times are chosen to correspond to the equivalent states of evolution for the different grain sizes, ${t=84, \: 310, \: 685,}$ and $1740$ disk rotations for ${St= 4 \times 10^{-2},}\: 10^{-2}, \: {4 \times 10^{-3},}$ and $10^{-3}$, respectively. The profiles of the Rossby numbers are shown Figure \ref{Fig_Threasold_insta_global}, and are focused on the vortex region.

	To first order, the vorticity profiles look similar, with a near constant vorticity throughout the vortex (${Ro = -0.13}$), and a core of nearly zero vorticity. This region surrounding the core exhibits spiral structures, which reveal the development of the vortex instability. Upon closer inspection, we see that the modal structure varies with the Stokes number of the dust. For the two largest grains, the mode structure is similar, with a mixture of modes $m=2$ and $m=3$ in a local frame centered on the vortex center. As pointed out previously, the spirals visible in the ${St=4 \times 10^{-2}}$ case over the whole vortex are due to the vortex relaxation, and not to the interaction with dust. Thus, even if the grain size is reduced by a factor four between these two runs, the vortex profiles are surprisingly similar.
	
%-----------------------
\begin{figure*}
%% Figure 9
	\begin{center}
	\begin{tabular}{cc}
	\scriptsize{$(N_r, \: N_\theta) = (1024, \: 2048)$} & \scriptsize{$(N_r, \: N_\theta) = (2048, \: 4096)$} \\
	\scriptsize{$t=685 \: rot$} & \scriptsize{$t=675 \: rot$}  \\
	\includegraphics[height=6cm, trim=0mm 0mm 0mm 3mm, clip=true]{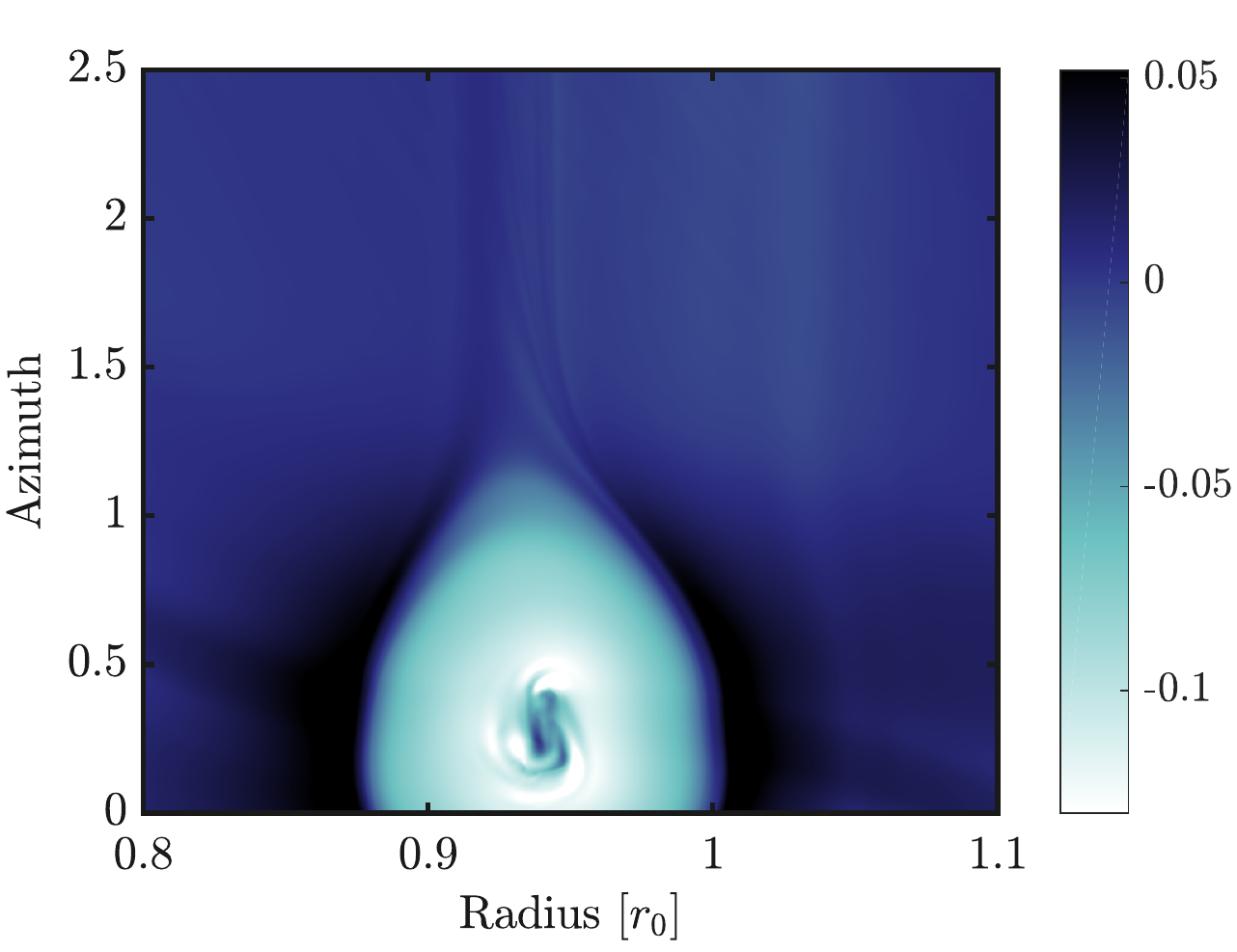} &  
	\includegraphics[height=6cm, trim=0mm 0mm 0mm 3mm, clip=true]{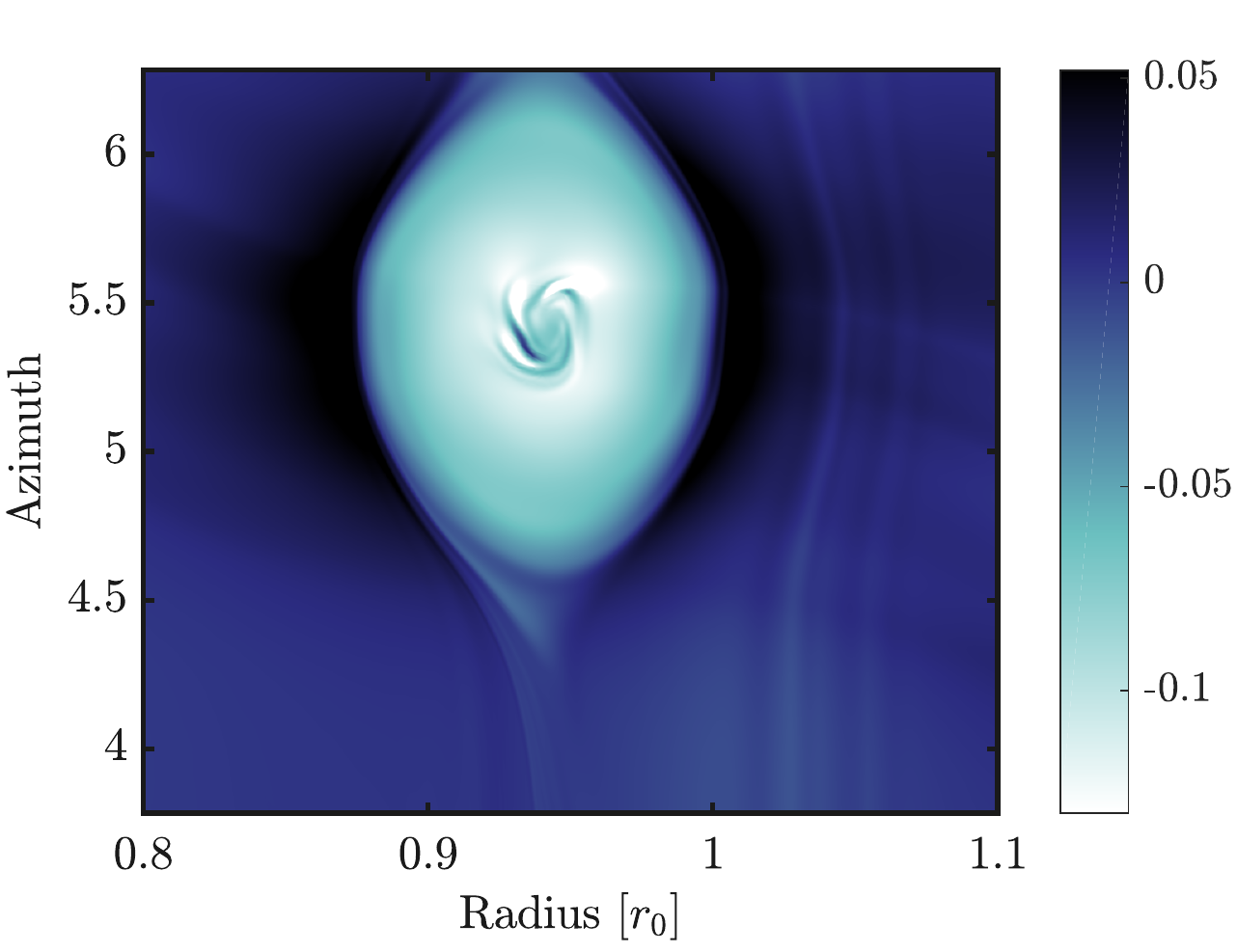} \\ \\

	\scriptsize{$t=690 \: rot$} & \scriptsize{$t=680 \: rot$}  \\
	\includegraphics[height=6cm, trim=0mm 0mm 0mm 3mm, clip=true]{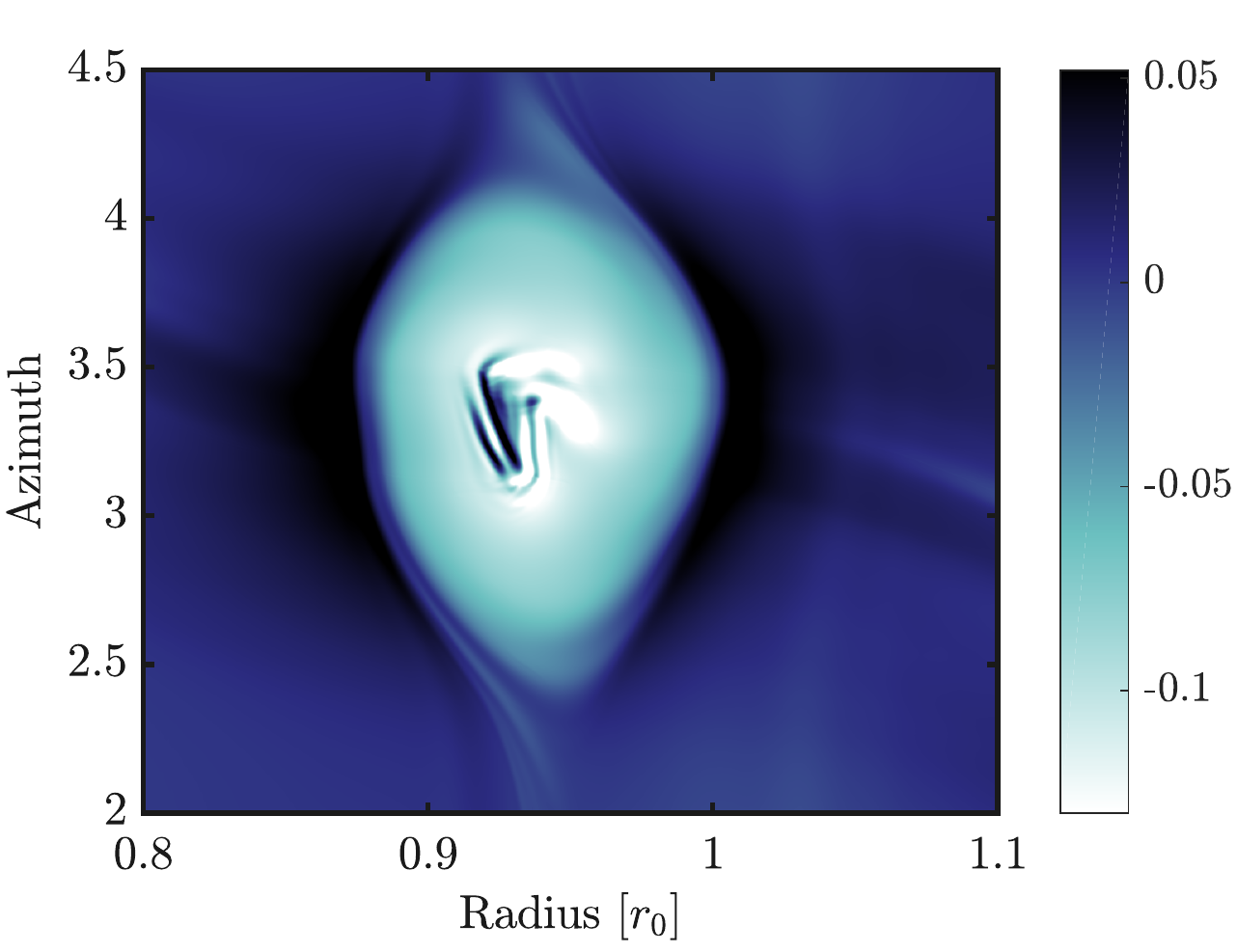} &  
	\includegraphics[height=6cm, trim=0mm 0mm 0mm 3mm, clip=true]{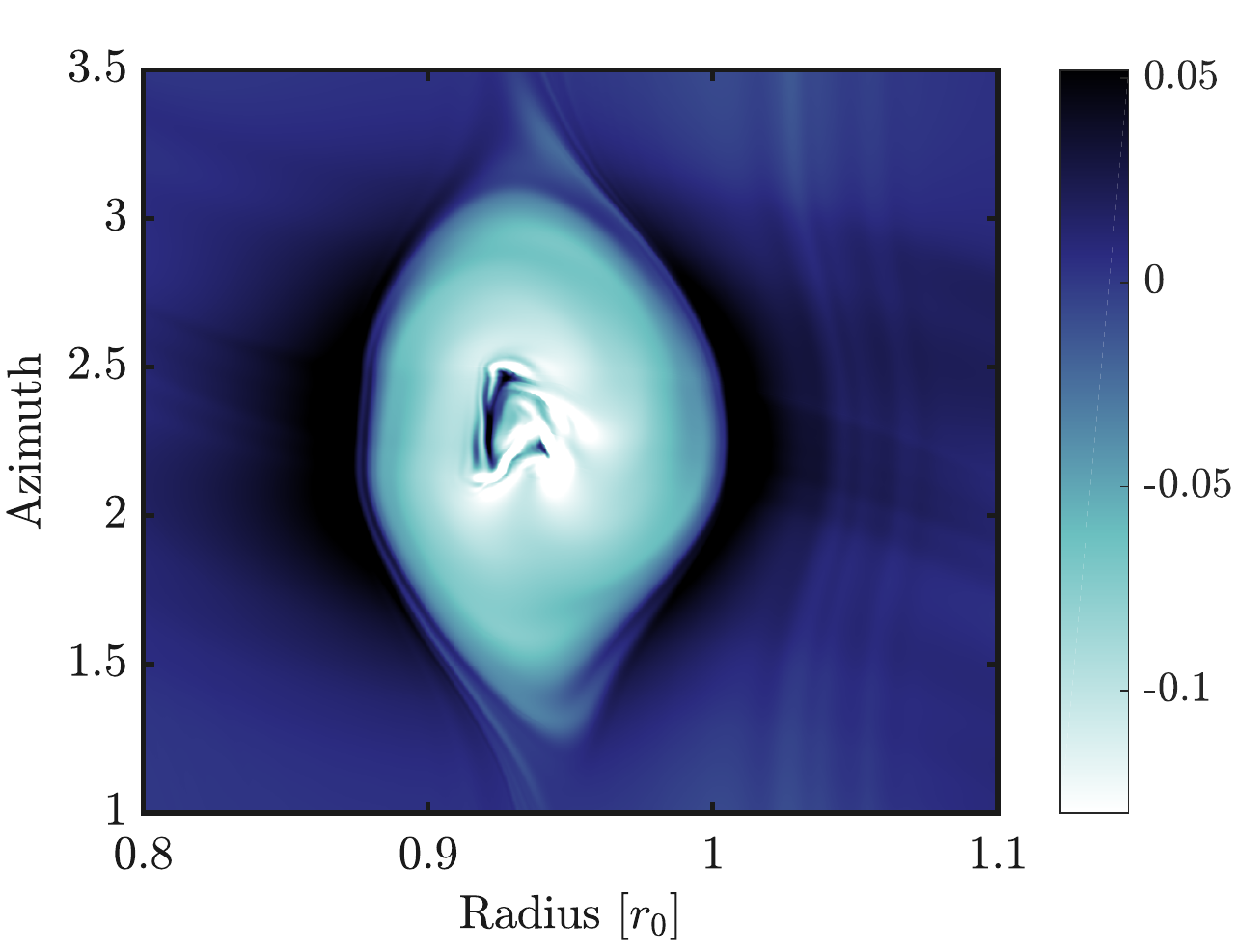}

	\end{tabular}
	\caption{\label{Fig_Threasold_insta_2k_4k}  Effect of the resolution on the instability, with ${St=4 \times 10^{-3}}$. We show a zoom in of the Rossby number maps of the vortex, when the first recognizable modes are established. Left: Results of the main run, $(N_r, \: N_\theta) = (1024, \: 2048)$, at $t= 685 \: rot$ (top) and $t= 690 \: rot$ (bottom). The flow close to the vortex center shows a mixture of $m=3$ and $m=4$ modes, locally, which saturate forming this typical triangular shape. Right: Results of the high resolution run, $(N_r, \: N_\theta) = (2048, \: 4096)$, at $t= 675 \: rot$ (top) and $t= 680 \: rot$ (bottom). The flow close to the vortex center shows more clearly a higher mode structure ($m=4-5$), which saturate forming this typical square shape. }
      \end{center}
\end{figure*}

	The effect of the dust Stokes number is more evident for the two smaller grains (bottom row). In the case with ${St=4 \times 10^{-3}}$, the flow close to the vortex center contains a mixture of $m=3$ and $m=4$ modes, which is nevertheless not an impressive discrepancy with the largest grains runs, with regard to the magnitude of the Stokes numbers. The last case with ${St=10^{-3}}$ is the one with the highest modes, $m=4$ and $m=5$, with also a different value of the Rossby number at the vortex center, differing from zero. The results published in \cite{Surville2016}, obtained with the same vortex model and with $St=0.17$, also exhibit the appearance of modes $m=3$ and $m=4$ modes at the beginning of the vortex instability (see Figure 14 of this reference).

	As a consequence, varying the Stokes number from ${St=0.17}$ down to ${St=10^{-3}}$, i.e. over two orders of magnitude, does not modify the modal structure of the vortex instability, as well as the growth rates because the duration of instability is qualitatively similar. The vortex instability has no significant dependence on the Stokes number. As a last remark, we mention that the study of a dusty Rossby waves instability in \cite{Surville2016} revealed that the azimuthal modes ${m=3,\: 4}$, and $5$ were the most unstable with comparable growth rates (see Figure 16 of this reference), with $St=0.1$. {{The fact that similar modes grow inside the vortex during the instability reflects the nature of the instability. It is due to the shear created in the gas by the drag back-reaction, and drags the dust while growing.}}

	If well-coupled grains are able to trigger the vortex instability after the exponential dust capture phase in a similar way as larger grains, the development of the instability may be sensitive to the numerical resolution. In fact, if this instability is related to the gas profile of vorticity, the typical length scale must be resolved by the numerical scheme. We explore this convergence problem by comparing the same case with ${St=4 \times 10^{-3}}$ at $(N_r, \: N_\theta) = (1024, \: 2048)$ (main run) and at $(N_r, \: N_\theta) = (2048, \: 4096)$ (high resolution run). The maps of Rossby number of these two cases at a similar time of evolution are presented Figure \ref{Fig_Threasold_insta_2k_4k}. The snapshots are chosen in both cases when the first established modes of instability are visible inside the vortex, and an later state (5 disk orbits after) show the saturation of the modes.

	Whereas the medium resolution result (main run, left panels) exhibits a mixture of modes $m=3$ and $m=4$ inside the vortex, the high resolution result (additional run, right panels) shows a higher mode structure, $m=4-5$. Even if the change is by only one mode number, it is an significant discrepancy, and suggests that higher resolutions are required to obtain the full convergence of the process. However, the evolution over five orbits (bottom row) shows that the two resolutions evolve finally in a similar manner. At that stage, it is difficult to distinguish the difference of the mode numbers.

It was not possible to conduct all the cases performed in this study at such a high resolution because of the duration necessary to cover the full evolution of the dusty vortex, i.e. up to $4200$ disk rotations. The degree of convergence obtained for the main runs is nevertheless satisfactory, and sufficient to provide trustful insights on the vortex instability and on the full process of evolution. Moreover, the effect of the grain size on the instability should not be affected by the used resolution, and if the growing modes could be higher than the ones observed in the main runs, the comparison between Stokes number values will give similar conclusions at higher resolution, i.e. the mode structure is almost not influenced by the value of $St$.

%----------------------------------------------------
\subsection{ Dust ring evolution for different grain sizes }
\label{Sect_Dusty_Ring}

%-----------------------
\begin{figure*}
%% Figure 10
	\begin{center}
	\begin{tabular}{ccp{15mm}}
	\scriptsize{Rossby number} & \scriptsize{Dust density}  & \\
	\imagetop{\includegraphics[height=5.5cm, trim=2mm 0mm 0mm 3mm, clip=true]{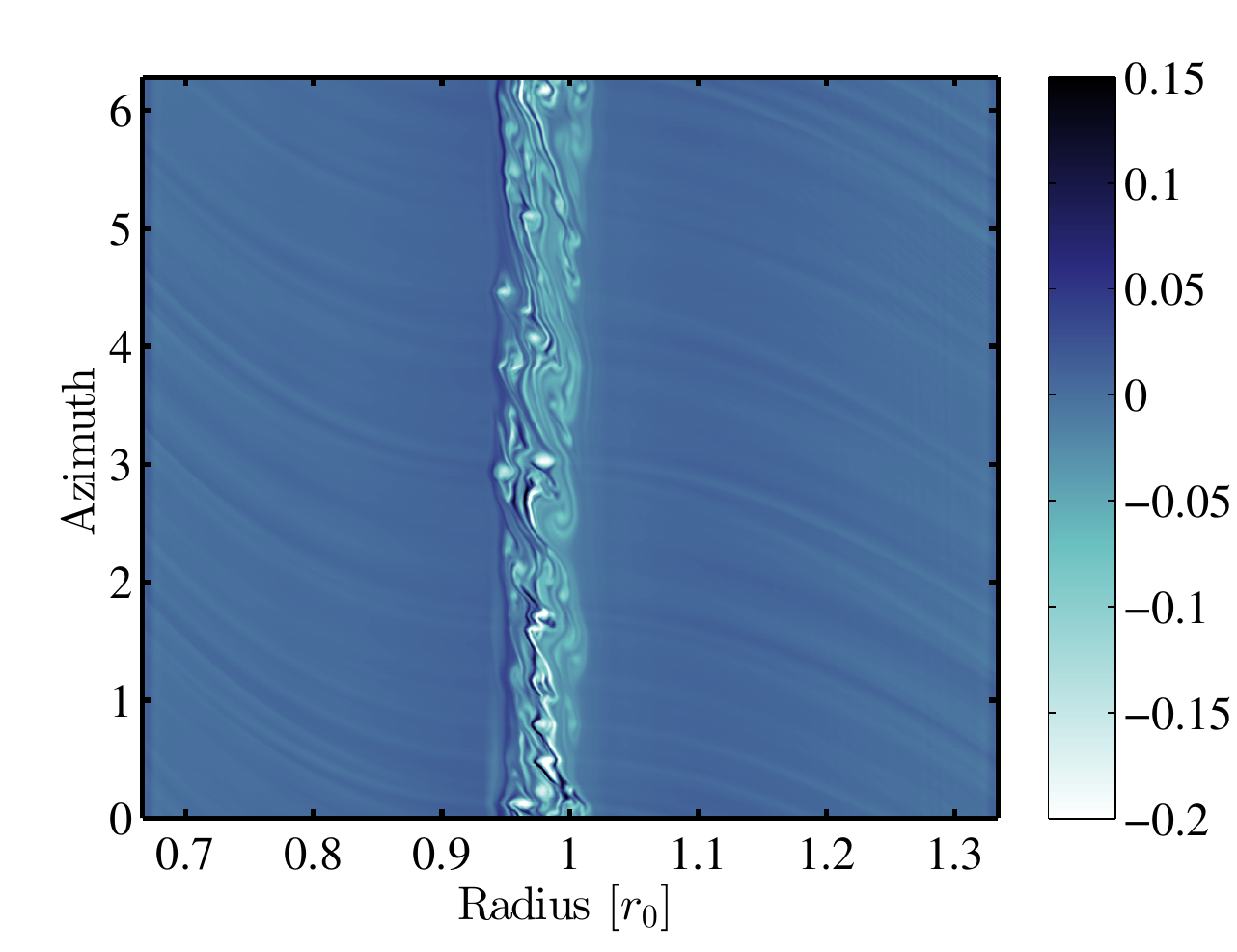}} &
	\imagetop{\includegraphics[height=5.5cm, trim=2mm 0mm 0mm 3mm, clip=true]{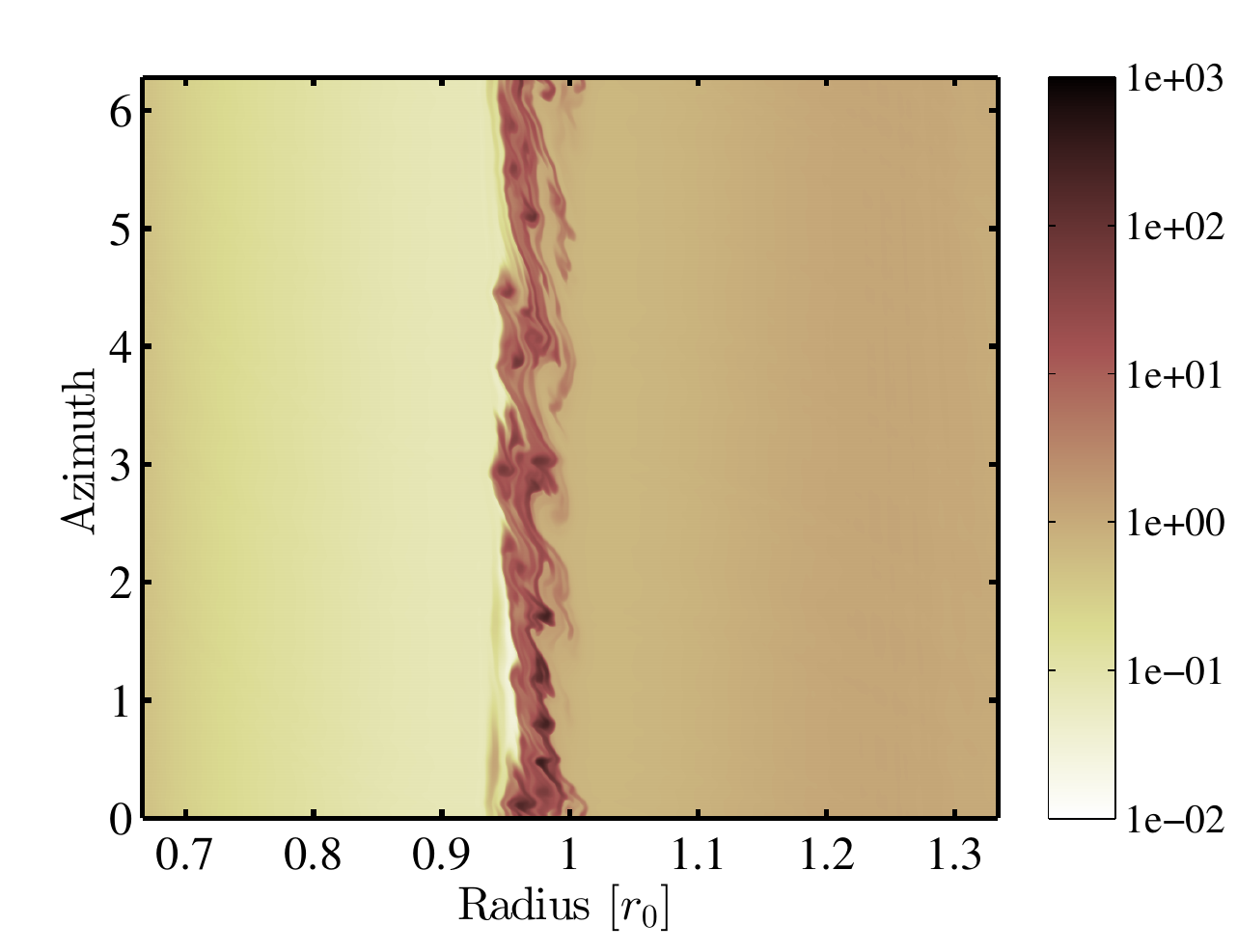}} &
	\scriptsize{ $\:$ \newline \newline ${St=4\times 10^{-2}}$ \newline $t=400 \: rot$} \\

	\imagetop{\includegraphics[height=5.5cm, trim=2mm 0mm 0mm 3mm, clip=true]{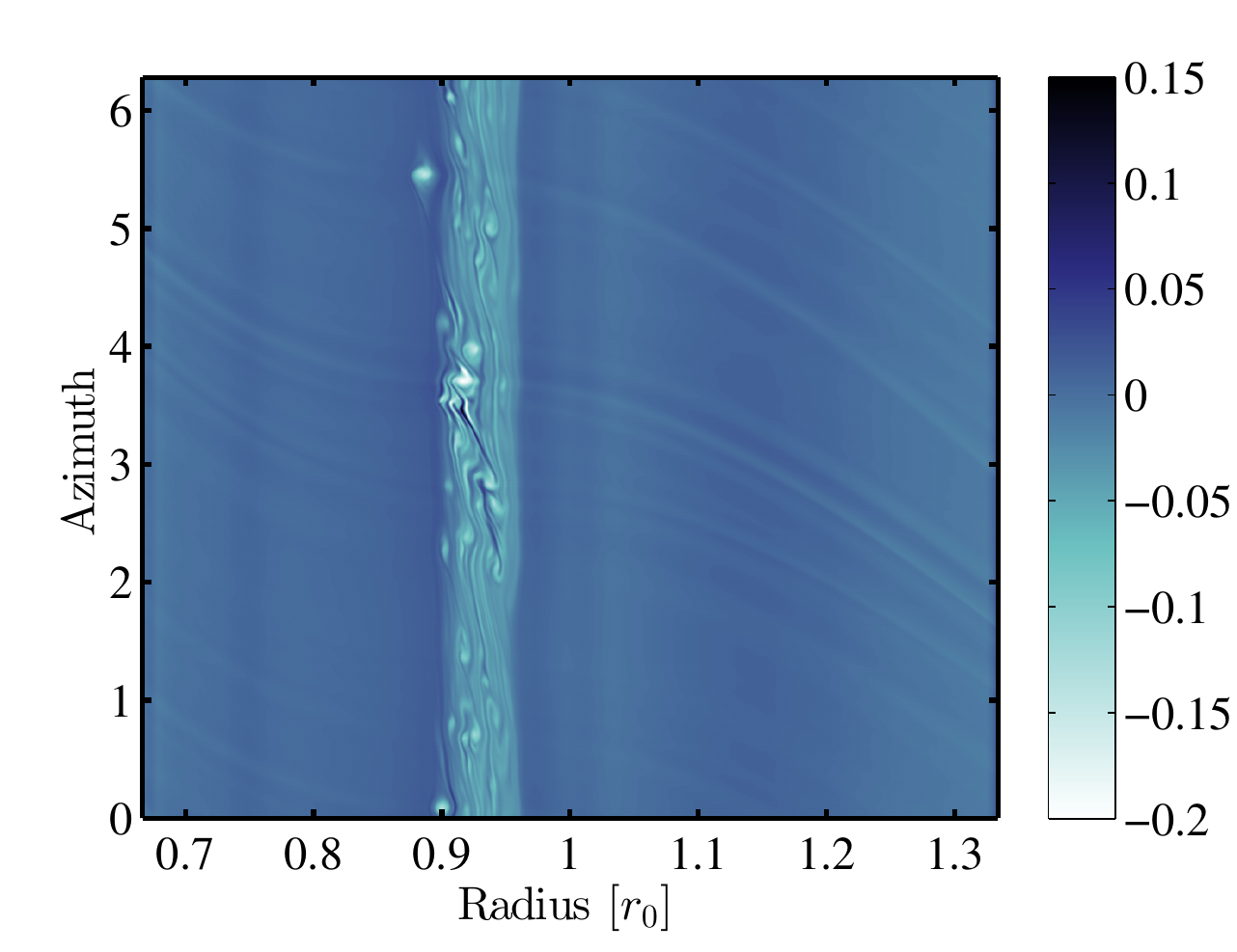}} &
	\imagetop{\includegraphics[height=5.5cm, trim=2mm 0mm 0mm 3mm, clip=true]{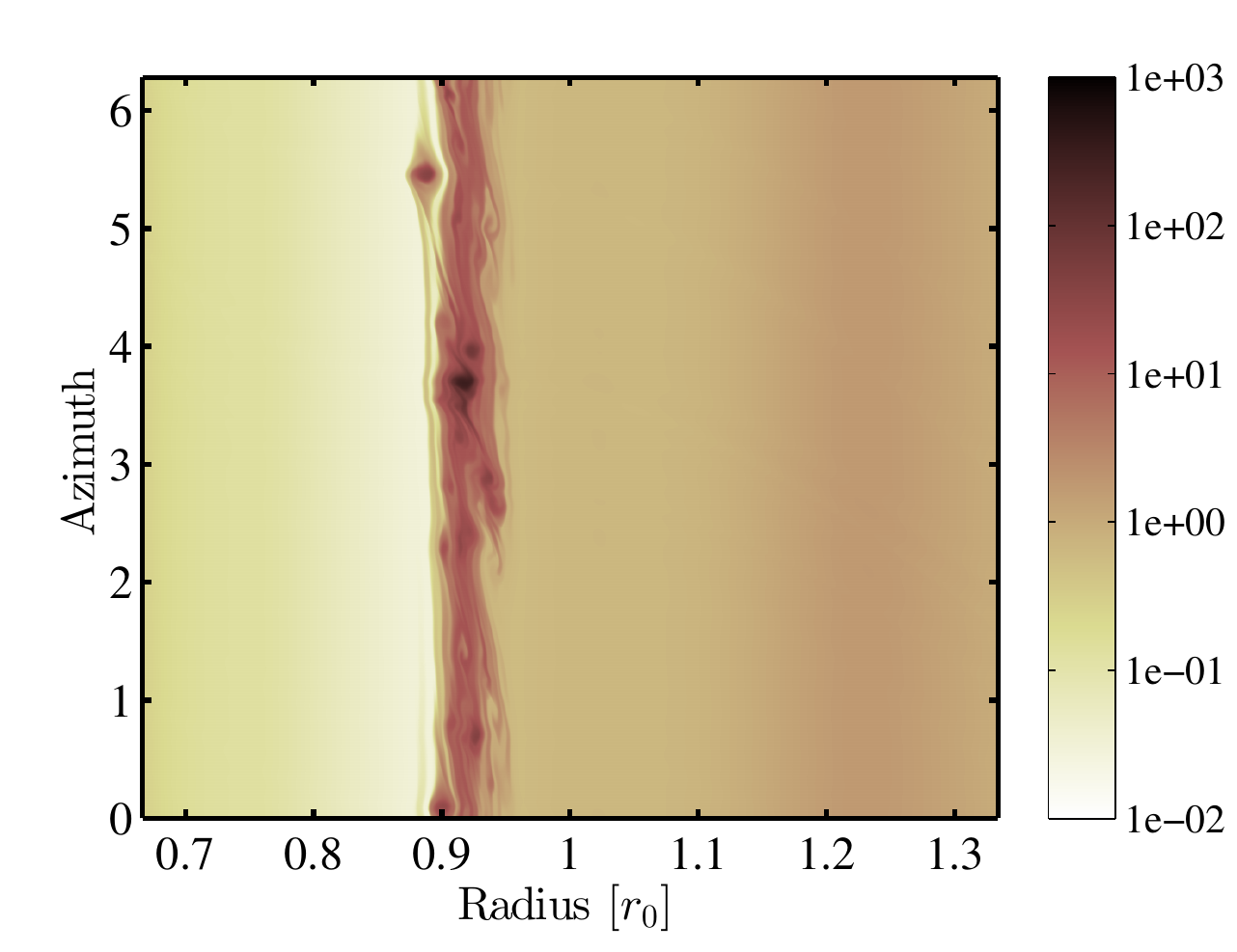}} &
	\scriptsize{ $\:$ \newline \newline ${St=1\times 10^{-2}}$ \newline $t=1400 \: rot$} \\

	\imagetop{\includegraphics[height=5.5cm, trim=2mm 0mm 0mm 3mm, clip=true]{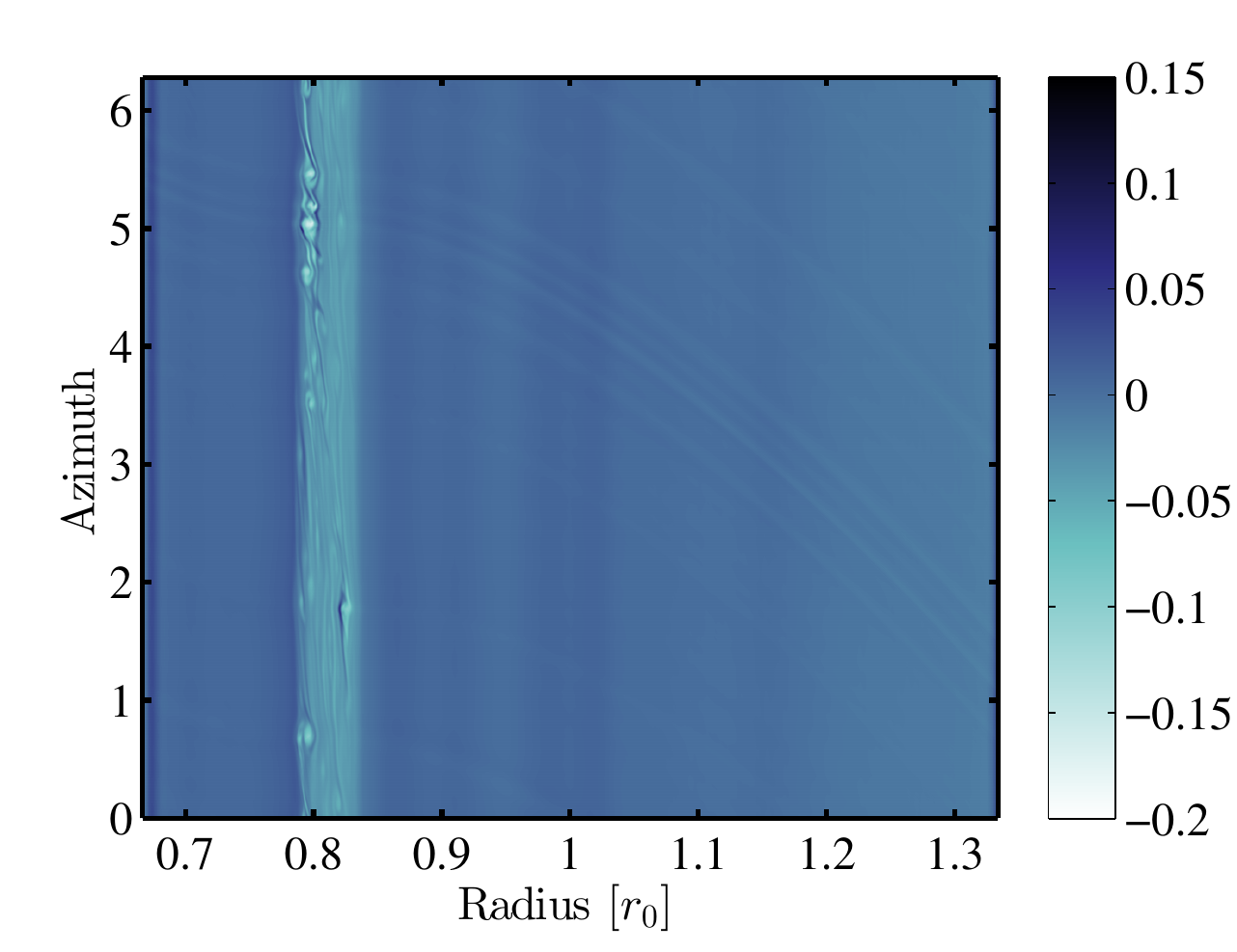}} &
	\imagetop{\includegraphics[height=5.5cm, trim=2mm 0mm 0mm 3mm, clip=true]{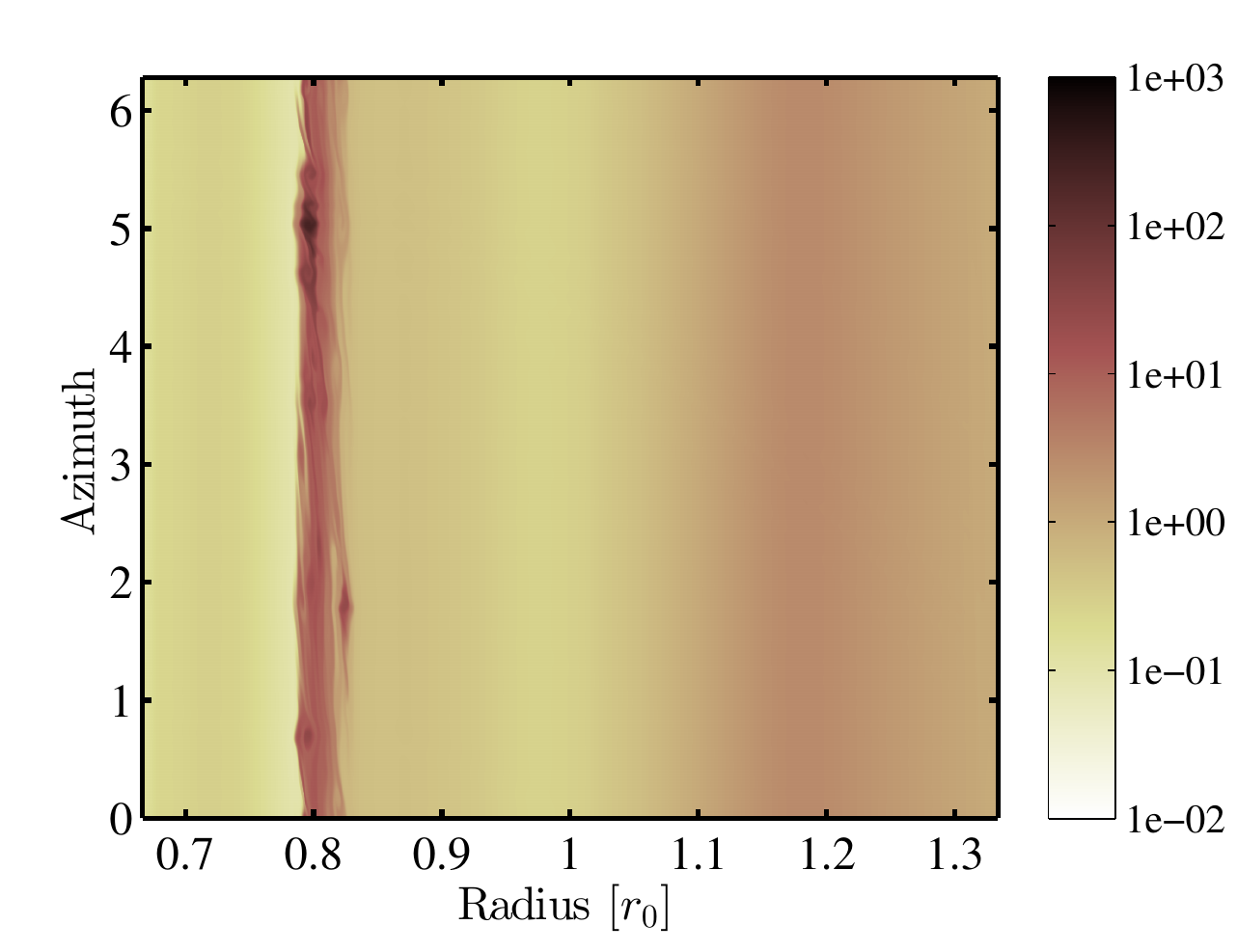}} &
	\scriptsize{ $\:$ \newline \newline ${St=4\times 10^{-3}}$ \newline $t=3000 \: rot$}

	\end{tabular}
	\caption{\label{Fig_Ring_global} Comparison of the structure of the dusty ring, for different grain sizes. We show snapshots of the main runs after the formation of the dust ring. From top to bottom: The grain sizes correspond to ${St=4\times 10^{-2}}$, ${St=1\times 10^{-2}}$, and ${St=4\times 10^{-3}}$, respectively. The last main run is not included as no dust ring is formed at the end of the simulation. On the left are reported the $(r, \: \theta)$ maps of the Rossby number and on the right the maps of the dust density, in a logarithmic color scale. }
      \end{center}
\end{figure*}

%-----------------------
\begin{figure*}
%% Figure 11
	\begin{center}
	\begin{tabular}{ccp{15mm}}
	\scriptsize{Rossby number} & \scriptsize{Dust density}  & \\
	\imagetop{\includegraphics[height=5.5cm, trim=2mm 0mm 0mm 3mm, clip=true]{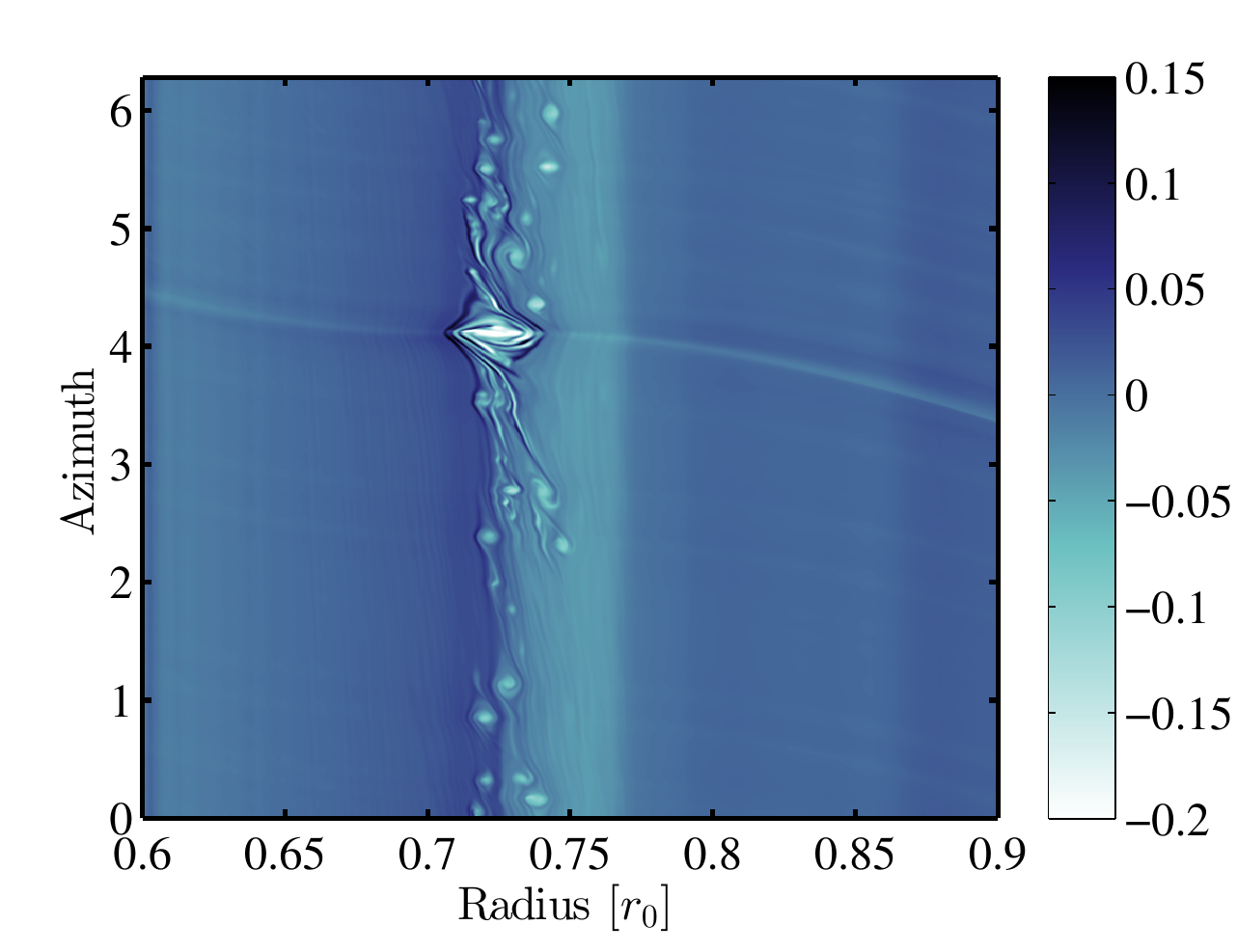}} &
	\imagetop{\includegraphics[height=5.5cm, trim=2mm 0mm 0mm 3mm, clip=true]{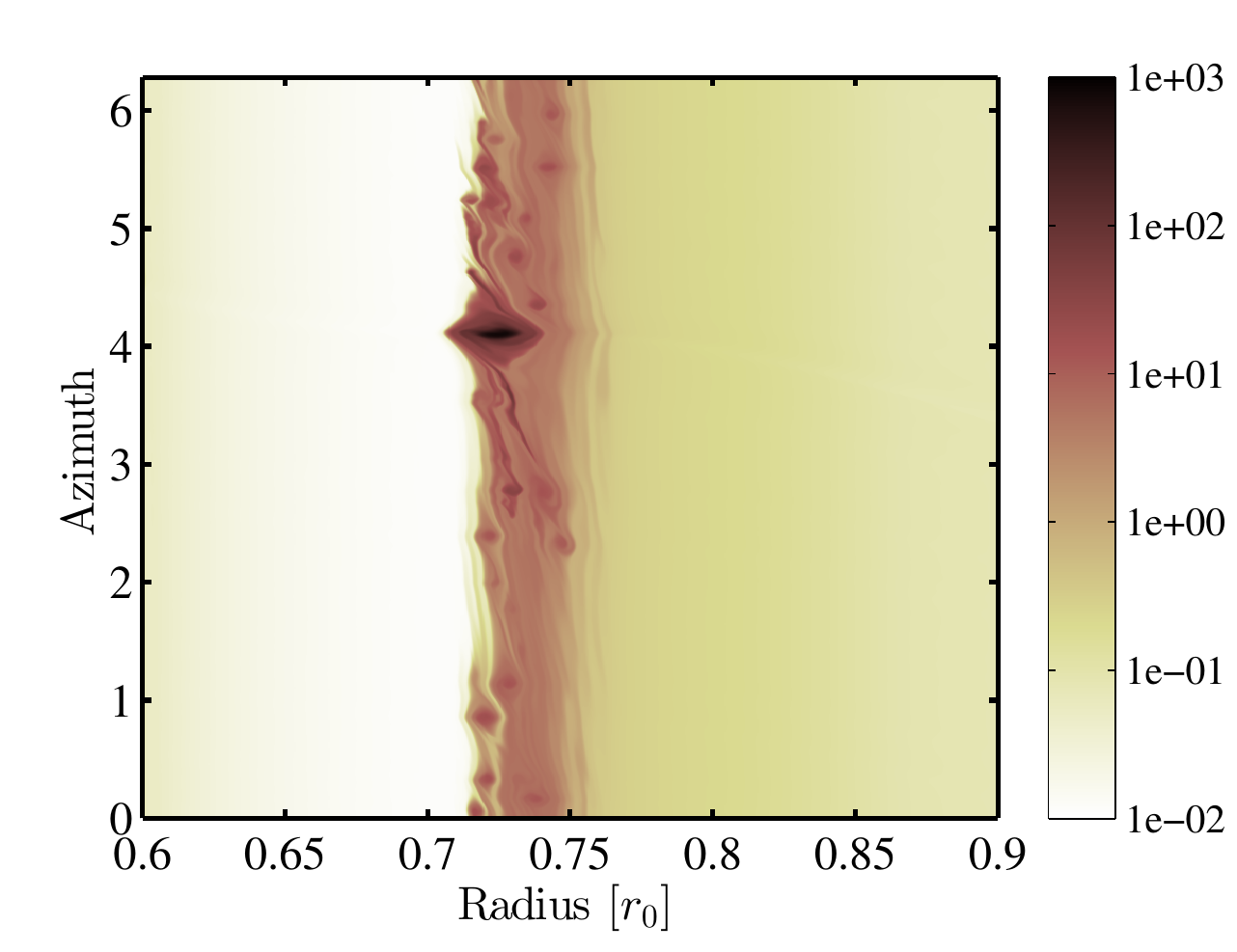}} &
	\scriptsize{ $\:$ \newline \newline $t=3600 \: rot$} \\

	\imagetop{\includegraphics[height=5.5cm, trim=2mm 0mm 0mm 3mm, clip=true]{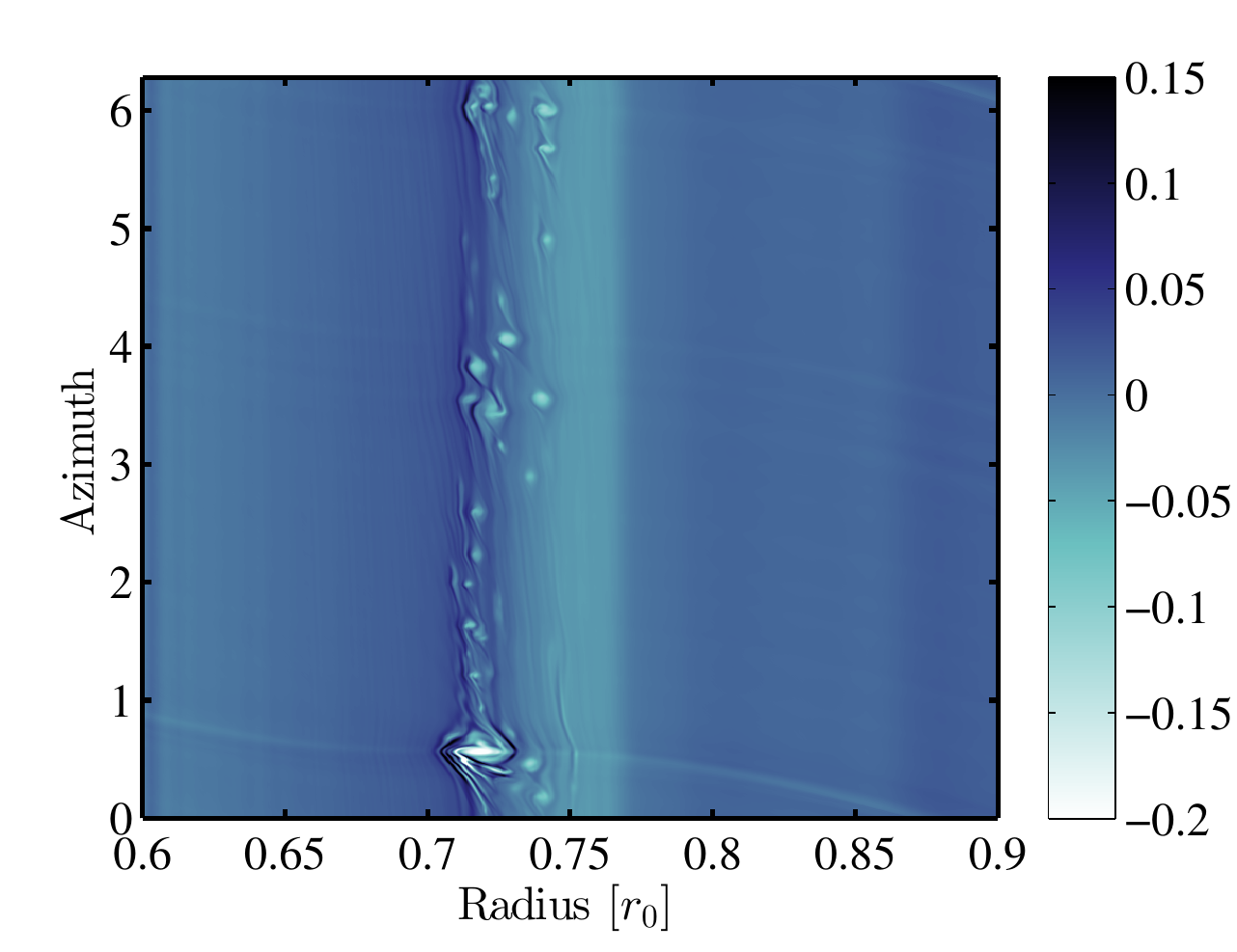}} &
	\imagetop{\includegraphics[height=5.5cm, trim=2mm 0mm 0mm 3mm, clip=true]{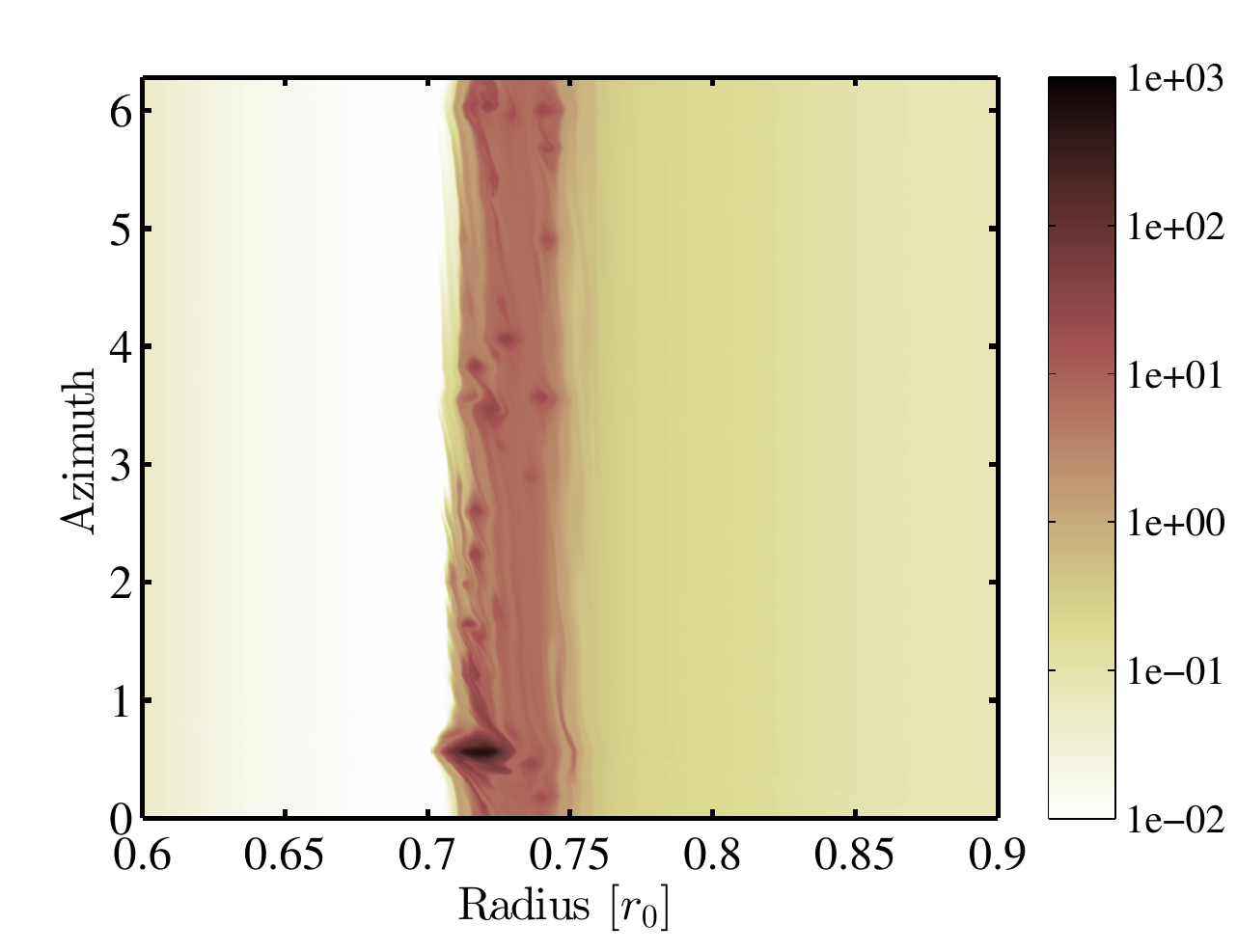}} &
	\scriptsize{ $\:$ \newline \newline $t=3675 \: rot$} \\

	\imagetop{\includegraphics[height=5.5cm, trim=2mm 0mm 0mm 3mm, clip=true]{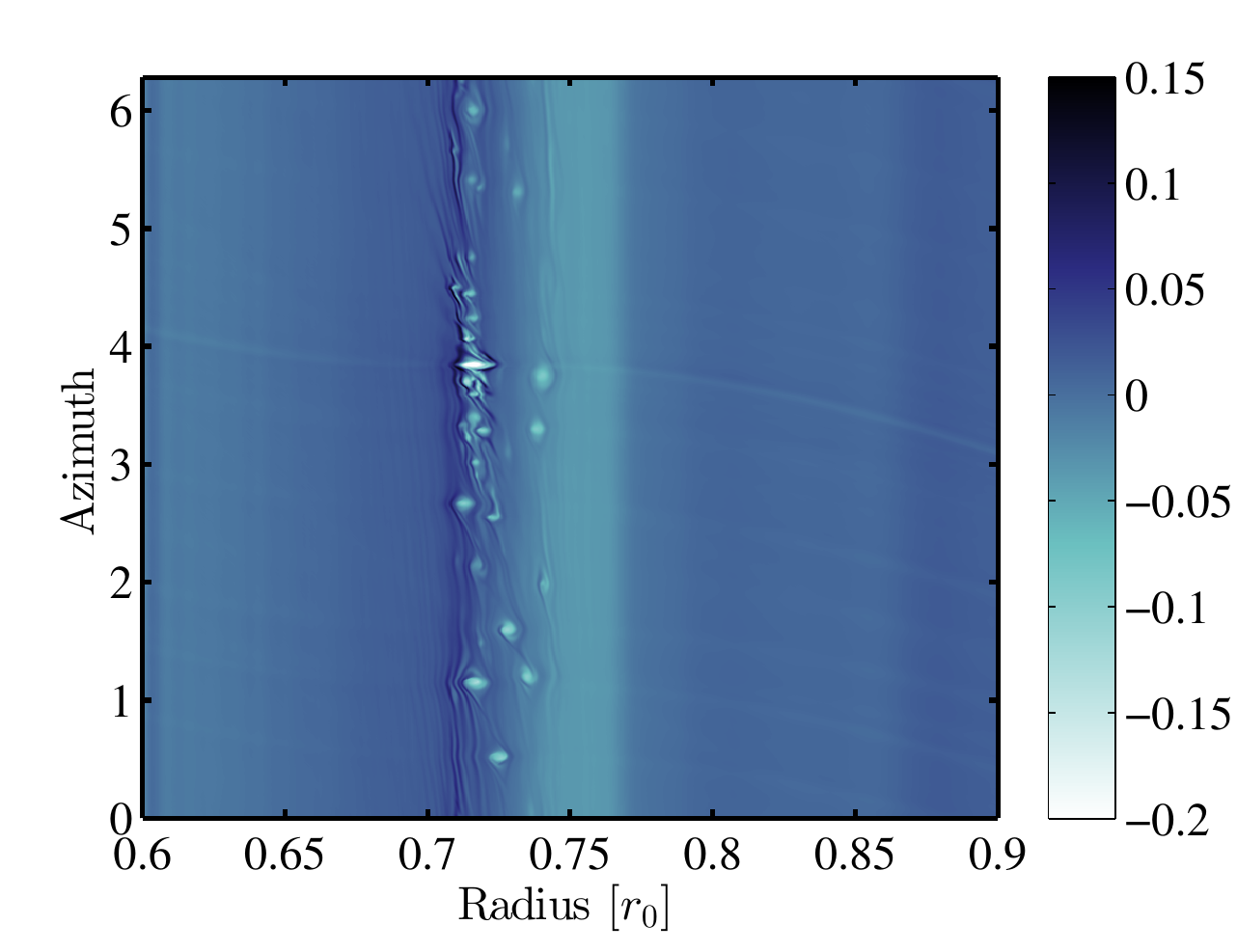}} &
	\imagetop{\includegraphics[height=5.5cm, trim=2mm 0mm 0mm 3mm, clip=true]{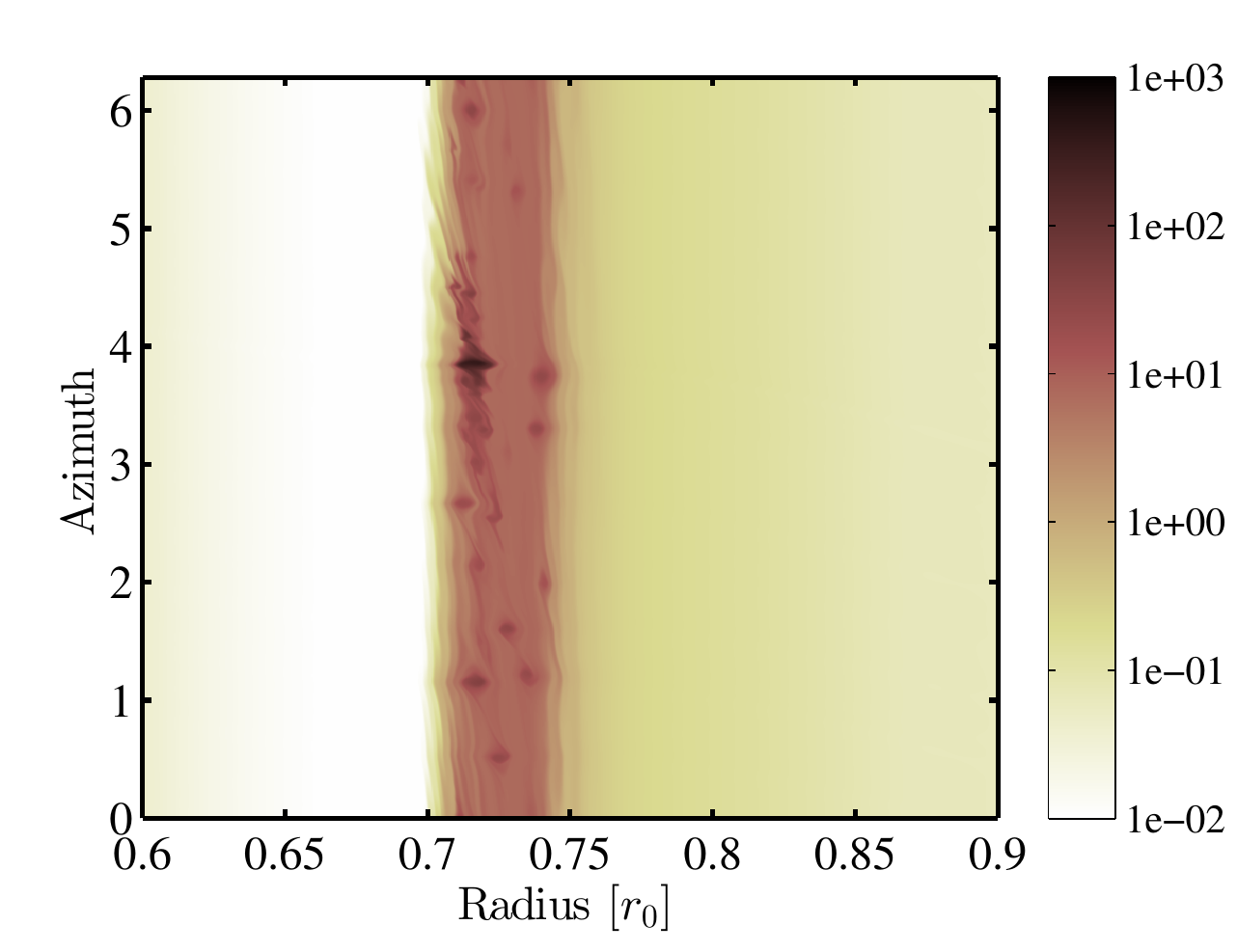}} &
	\scriptsize{ $\:$ \newline \newline $t=3750 \: rot$}

	\end{tabular}
	\caption{\label{Fig_Ring_Evo} Late evolution of the dusty ring for the high resolution additional run, with ${St= 4 \times 10^{-3}}$. From top to bottom: Snapshot at $t=3600$, $t=3675$, and $t=3750$ rotations respectively. On the left are reported the $(r, \: \theta)$ maps of the Rossby number and on the right the maps of the dust density, in a logarithmic color scale. The maps are zoomed in radially to reveal the details of the structures in the flow.}
      \end{center}
\end{figure*}

	After the capture phase and the vortex instability, which happen systematically for all Stokes numbers down to $10^{-3}$, the last phase of evolution of the dusty vortex is the formation of a turbulent dusty ring. These structures were depicted in \cite{Surville2016}, as a by-product of the vortex destruction. We will discuss here the particularities of such features for smaller grains, and their possible implications on the formation of planetesimals.

	During the description of the evolution of the different runs, Section \ref{Sect_Results}, we observed the formation of dust rings for the three largest Stokes numbers. The only case where the vortex survives up to the end of the simulation is with $St=10^{-3}$. We will discuss this exception during the next section. For the other dust populations, the formation of the dust ring happens after the unstable evolution of the vortex, at a characteristic time which is longer for smaller dust grains. Figure \ref{Fig_Ring_global} presents snapshots of the disk after $400$, $1400$, and $3000$ disk rotations, for the main runs with $St=4\times 10^{-2}$, $10^{-2}$, and $4\times 10^{-3}$, respectively. The choice of these times is motivated by the number of structures visible in the disk and in the dust rings. We also check that the vortex is fully dissipated in each case. 

	For the largest grains with $St=4\times 10^{-2}$ (first row), a chain of dusty eddies survives during a long period in the disk. The local dust-to-gas ratio is sustained to a few times unity, while smaller scale eddies and dense filaments are also present in the flow. On the left, the Rossby number map shows the large scale turbulence of the gas in this wide dust ring. As the effect of the drag is sufficiently fast for this Stokes number, estimated to the capture timescale i.e. around $30$ rotations, new eddies form frequently following the same process as the vortex instability, but on smaller scales. The only source of dissipation preventing high frequency turbulence, and eventually counteracting the drag force coupling is numerical diffusion. In this simulation, the dust ring is sustained during approximately $500$ rotations. 

	In the run with $St=10^{-2}$ (second row), the dust ring forms much later, which is in agreement with the self-similarity of the dust capture with respect to the Stokes number (see Section \ref{Sect_Linear_capture}). As the vortex had more time to migrate before being destroyed by the instability, the dust ring is established closer to the star, in the region ${0.9<r<0.96 \: r_0}$. Compared to the results with larger grains, the dust eddies are less individualized and detached from the main bump of dust, where the dust-to-gas ratio is around $50\%$. The number of dense eddies is reduced, but the Rossby number map reveals also the presence of smaller scale turbulence (on the left). The drag force effect onto the gas, which drives this turbulent dusty flow, is acting on longer time scale in this case. As a consequence, the numerical dissipation is more prominent, and counteracts the accumulation of solids in vortical regions. This accumulation is also physically slower, due the the reduction of the Stokes number. 

	There is a clear different behavior of the evolution resulting in the dust ring formation in this case, compared to the largest grain run. Rather than forming several high density clumps, the vortex evolution leads to only one big eddy which is a remnant of the original vortex containing material that was collected during the initial phase of dust concentration. The dust density in this clump is larger than in the ones obtained with ${St=4\times 10^{-2}}$, by a factor of a few. Over a few hundred orbits, the maximal dust-to-gas ratio in this clump is around $5$. 
	
	The last case is the result with ${St=4\times 10^{-3}}$ (bottom row). The dust ring is formed much later, after $\sim 3000$ disk rotations, and even more closer to the inner parts of the disk. The longer evolution of the disk and of the vortex is favorable for the numerical dissipation to act. Only a few small dusty eddies remain, with a much lower density than what was seen for larger grain sizes. The turbulence and vortical substructures in the dust ring are washed out by numerical diffusion, as seen on the Rossby number map (left). The longer timescale of action of the drag is also not favorable to sustain the generation of new small vortices. As in the case with ${St=10^{-2}}$, a solitary dusty eddy was formed during the phase of the vortex destruction, and survived during the dust ring formation. Despite the fewer number of eddies, the dust-to-gas ratio reached in the surviving eddy was >~5..
	
%-----------------------
\begin{figure}[t]
%% Figure 12
	\begin{center}
	\begin{tabular}{c}
	\scriptsize{Evolution of the dust-to-gas ratio} \\
	\includegraphics[height=6cm, trim=2mm 0mm 0mm 3mm, clip=true]{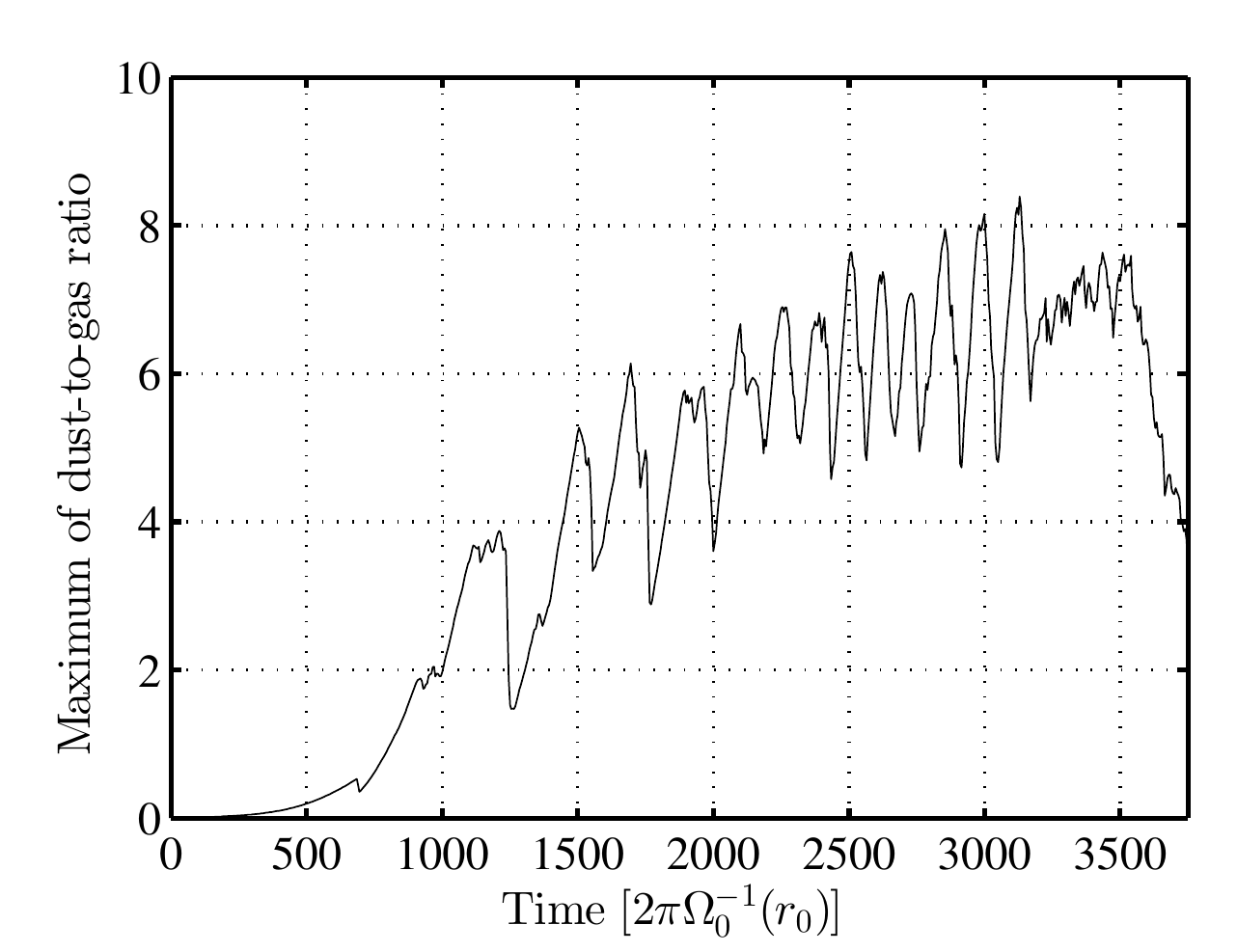} \\ \\
	\scriptsize{Dust-to-gas ratio at $t=3750\: rot$}  \\
	\includegraphics[height=6cm, trim=2mm 0mm 0mm 3mm, clip=true]{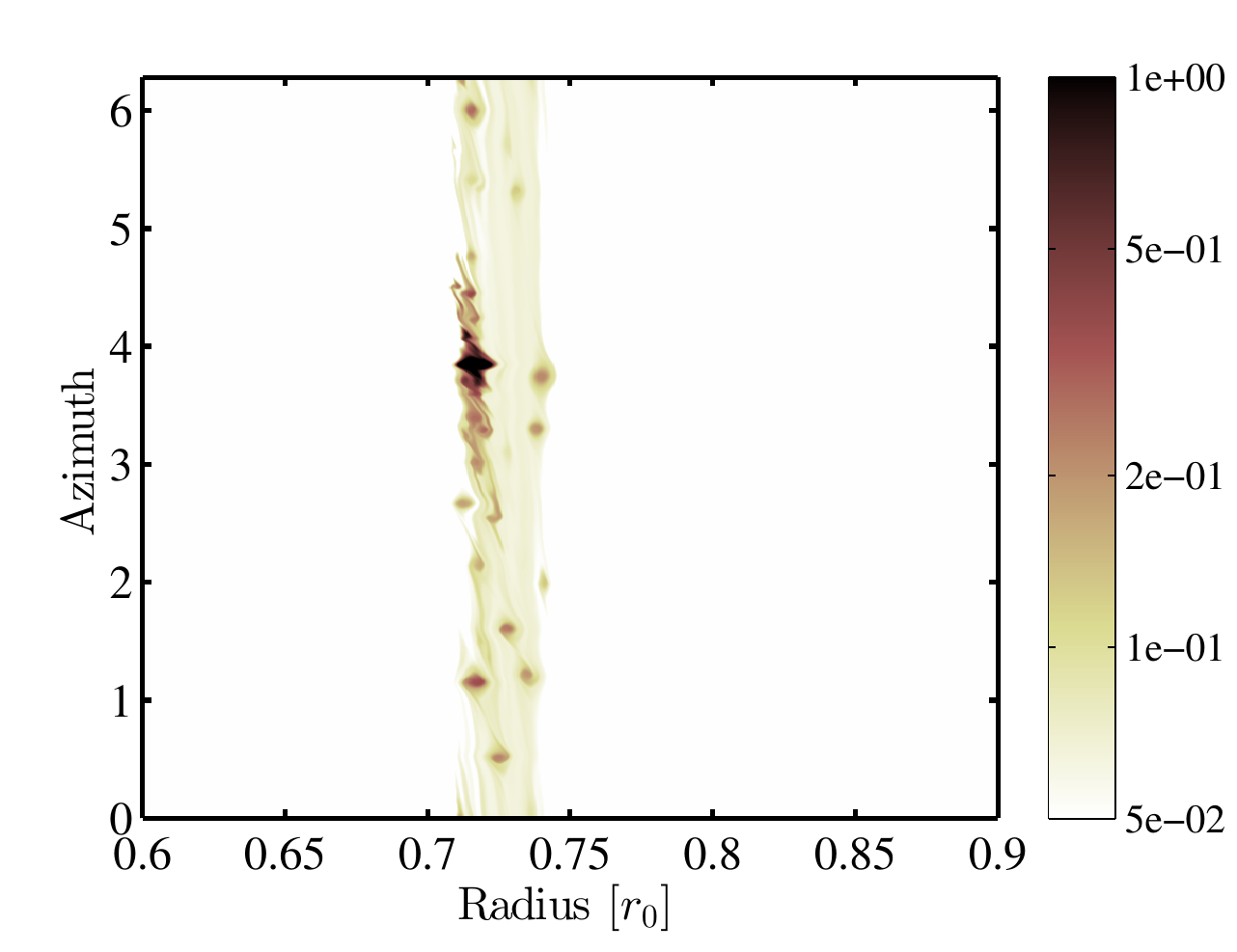} 
	\end{tabular}
	\caption{\label{Fig_Dust_gas_ratio_2k_4k}  The local dust-to-gas ratio for the high resolution run, with ${St= 4 \times 10^{-3}}$. Top: Evolution of the maximum of the local dust-to-gas ratio in the disk. After about $1500$ rotations of the disk, the highest value is much larger than unity, typically in the range $5-8$. Bottom: Map of the dust-to-gas ratio $\sigma_p / \sigma_g$ at the end of the run, i.e. $t=3750$ rotations.   }
      \end{center}
\end{figure}

	One of the outcomes of this comparison is that diffusion in the flow, numerical in our case, has a strong impact on the evolution of the dust ring. The use of a medium resolution in the main runs, $(N_r, \: N_\theta) = (1024, \: 2048)$, not only produces significant numerical diffusion, but also limits the resolved scale length of the flow. As the size of the dusty eddies in the dust ring seems to reduce when the Stokes number decreases, the convergence with resolution is crucial. We show Figure \ref{Fig_Ring_Evo} the evolution of the dust ring obtained in the high resolution run, with ${St= 4 \times 10^{-3}}$, and $(N_r, \: N_\theta) = (2048, \: 4096)$.
	
	When zooming in the radial dimension to reveal the details of the flow, one clearly sees that using higher resolution resolves much better the dusty eddies. A dozen of small scale dusty vortices survives over hundreds of disk rotations, with a local dust-to-gas ratio larger than $0.1$. The differentiation from the wide dusty bump is more prominent. If the resolved scale is smaller in this high resolution run, allowing to follow these small eddies, the sustain of turbulence by the drag instability, like with larger grains, is still difficult. In fact, the drag timescale is typically of order $\sim 1000$ orbits, which is of the order of the numerical diffusion timescale.
	
	The vortex remnant with the largest amount of dust, is stronger at higher resolution. At $t=3600$ disk rotations, top row, this dusty structure even excites density waves, visible in the Rossby number map (left). As time goes on, the size of this dusty eddy shrinks to smaller scales releasing some solid material to the dust ring in the form of smaller eddies. However, despite the decrease in the density, the dust-to-gas ratio stays much larger than unity, indicating a new interesting path to confine dust toward a gravitational instability. But this beyond the scope of this study.
	
	The evolution of the maximum dust-to-gas ratio during the high resolution run is presented Figure \ref{Fig_Dust_gas_ratio_2k_4k}, top, with a map obtained at $t=3750$ disk rotations, bottom. Even with a Stokes number of $4 \times 10^{-3}$, the maximum dust-to-gas ratio obtained during the evolution of the vortex is much larger than unity. During the first thousand orbits, the dust is accumulated inside the vortex, as described earlier. But during the unstable evolution of the vortex, for $1500<t<3200$ disk rotations, the dust confined in the vortex core reaches dust-to-gas ratios increasing in average from $4$ to $7$, with some peaks up to $8$. This period may be in favor for dust grain growth and gravitational instability, as the dust density is very large. On the other side, the unstable nature of the dusty vortex can counteract these processes.
	
	However, from $t=3200$ disk rotations to the end of the run, the dust ring is formed and the last compact eddy survives and shrinks, as explained previously. This period can be more favorable for the cited processes, because the dynamics of the dust in this big eddy is more quiet. The steady value of the dust-to-gas ratio around 7 sustained for $3200<t<3500$ is a signature of this aspect. Later on, the dust-to-gas ratio linearly decreases down to $4$ at the end of the run, while the size of the eddy shrinks. The state of the disk at the end of the run, bottom panel, shows that even if this extreme value is localized, the rest of the dust ring still corresponds to a dust enhancement larger than ten. In the smaller eddies, the dust-to-gas ratio is larger than $0.2$.

	The formation of a dust ring is a common outcome of the destruction of dusty vortices, and was observed in our runs for Stokes numbers down to $St=4\times 10^{-3}$. However, the impact of this structure on the dust dynamics and on possible paths towards planetesimal formation differs for different grain sizes. The larger grains, with $St>10^{-2}$ typically, sustain turbulence and generation of new dense eddies, as described in \cite{Surville2016}. It is the most active phase observed after the vortex instability. For this population, the dust ring evolution is the most favorable for grain growth and eventually gravitational collapse, because of the large dust densities, the small size of the eddies, and the confinement which could reduce collisional velocities. It is also a long lasting phase of the disk evolution, which provides chance for these processes to happen. Finally, a signature of these big grains is the large number of small dense eddies that survives, giving a possible path to form asteroid belts.

	The well-coupled grains, with $St<10^{-2}$, however, interact with the gas on timescales that are too long to sustain a turbulent flow in the dust ring. The evolution of the eddies is quiet, and sensitive to different forms of diffusion acting on timescales comparable to $1/St$ rotations. During the dust ring phase, the small solid material is passive, and evolve in a small number of eddies. Their density is not high enough to expect gravitational collapse to happen, however grain growth is possible due to small collisional velocities produced by the small amount of turbulence. Finally, a signature of these small grains is the isolated very dense eddy, remnant of the vortex destruction, that could produce a big size planetesimal if gravitational collapse starts.

%----------------------------------------------------
\subsection{ The crucial concept of timescale }
\label{Sect_Time_scale}

%-----------------------
\begin{figure*}
%% Figure 13
	\begin{center}
	\begin{tabular}{ccp{15mm}}
	\scriptsize{Rossby number} & \scriptsize{Dust density}  & \\
	\imagetop{\includegraphics[height=5.5cm, trim=2mm 0mm 0mm 3mm, clip=true]{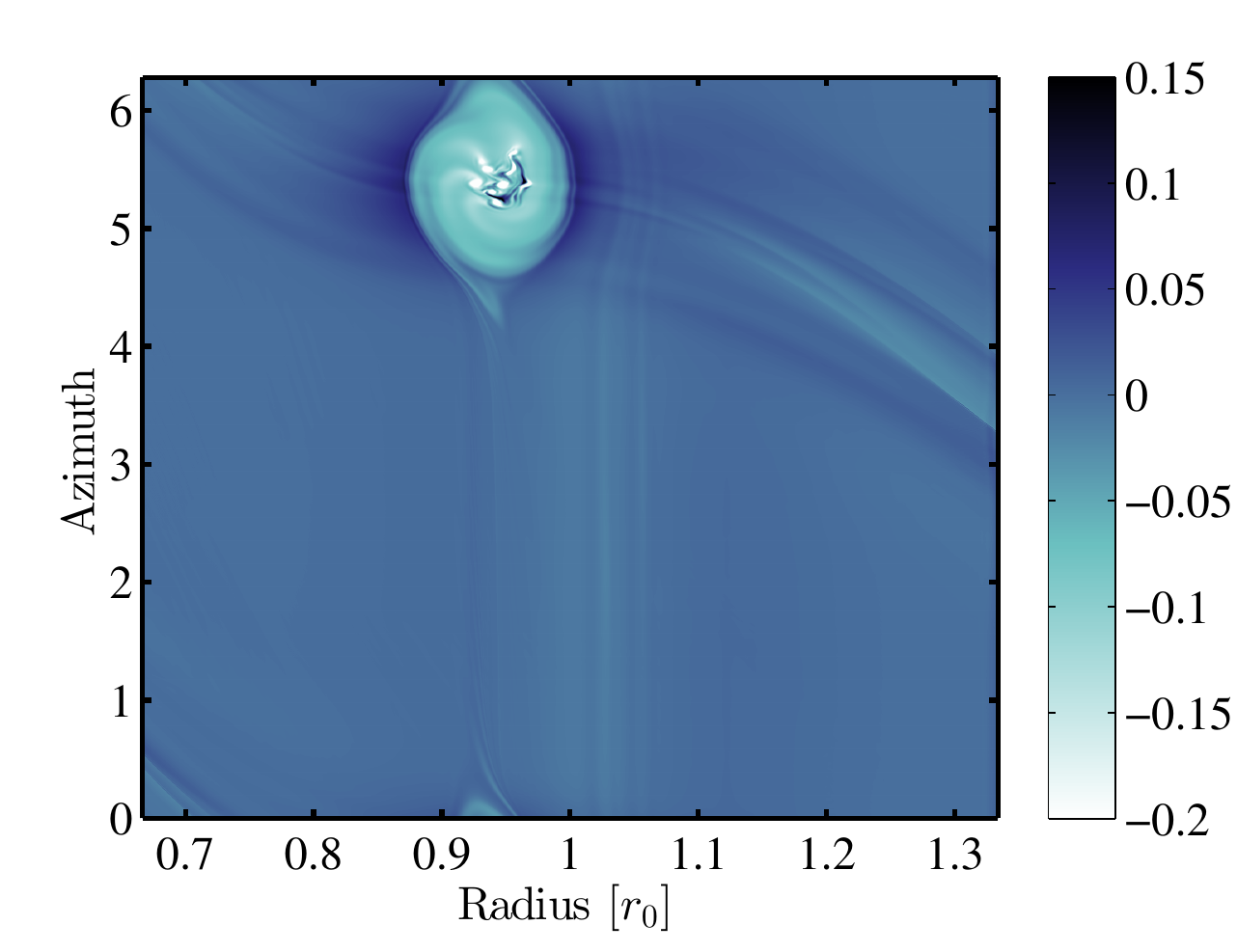}} &
	\imagetop{\includegraphics[height=5.5cm, trim=2mm 0mm 0mm 3mm, clip=true]{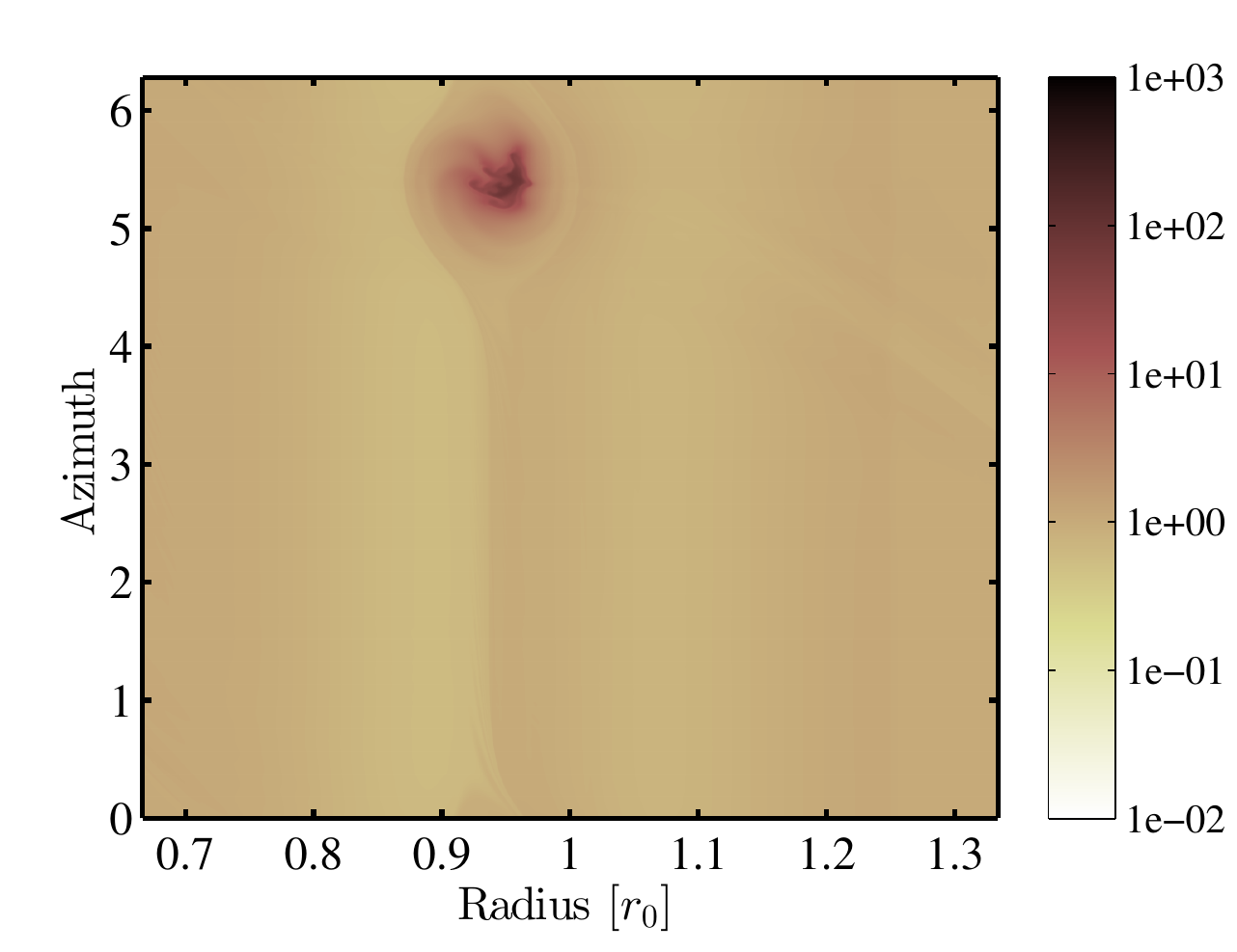}} &
	\scriptsize{ $\:$ \newline \newline $t=685 \: rot$} \\

	\imagetop{\includegraphics[height=5.5cm, trim=2mm 0mm 0mm 3mm, clip=true]{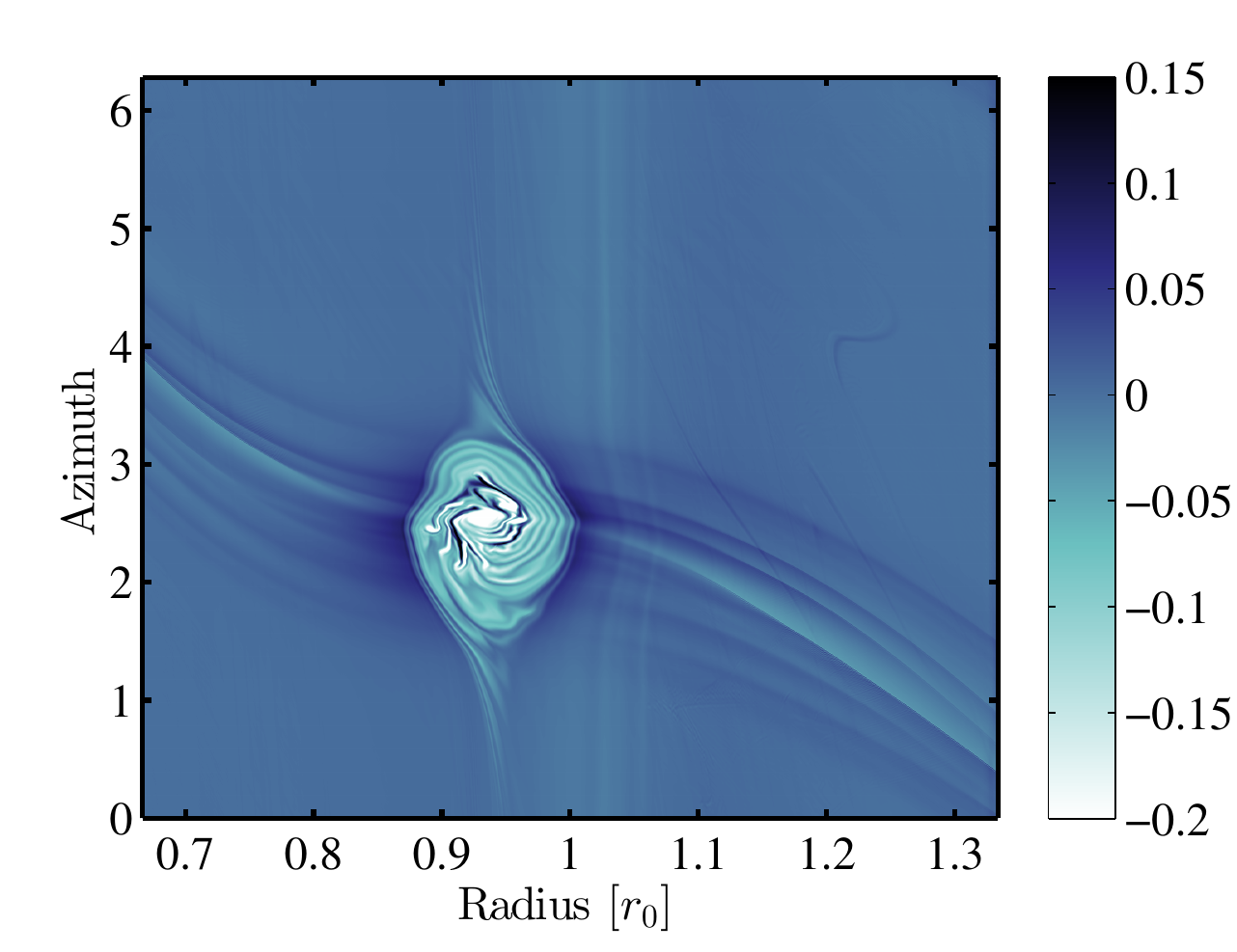}} &
	\imagetop{\includegraphics[height=5.5cm, trim=2mm 0mm 0mm 3mm, clip=true]{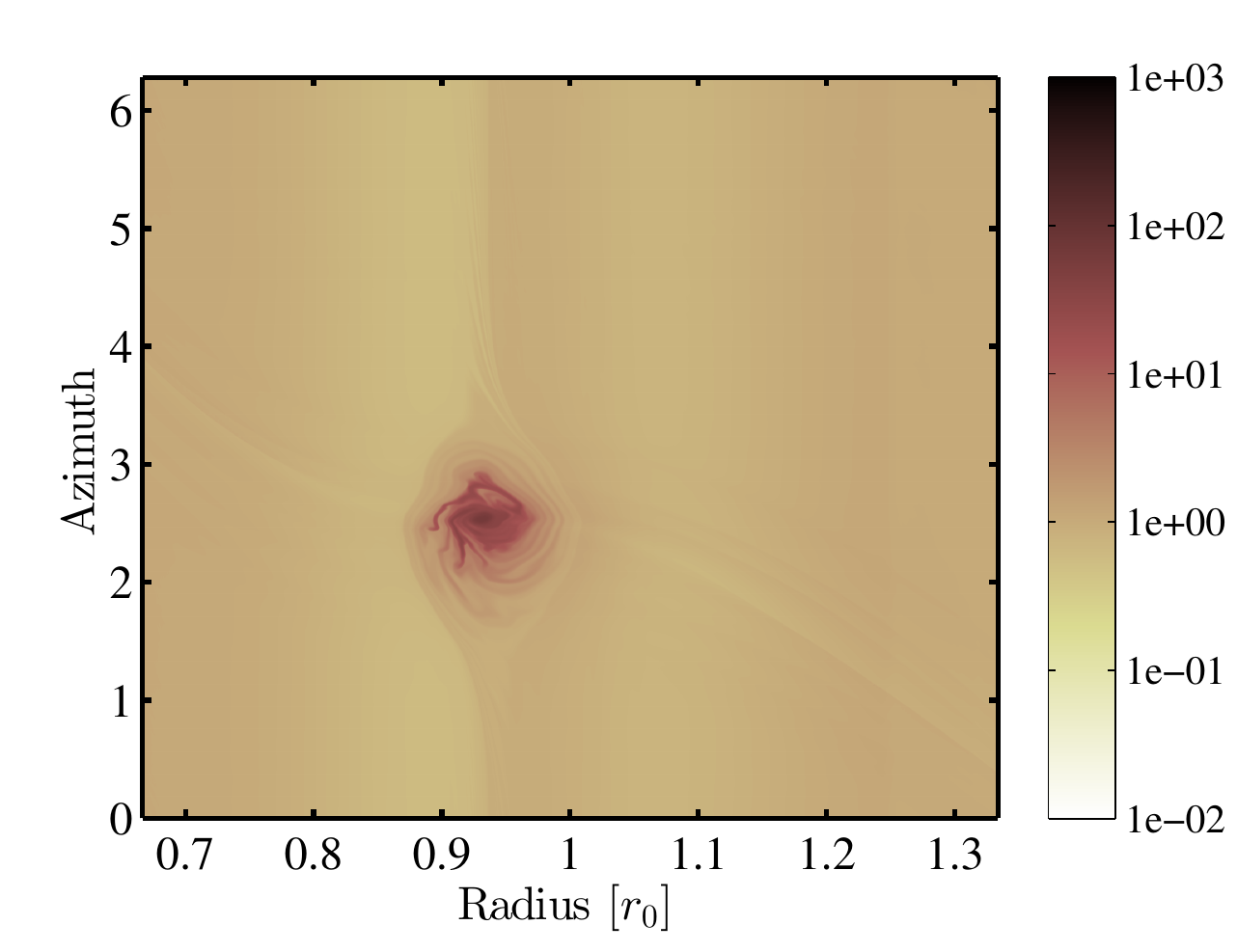}} &
	\scriptsize{ $\:$ \newline \newline $t=700 \: rot$} \\

	\imagetop{\includegraphics[height=5.5cm, trim=2mm 0mm 0mm 3mm, clip=true]{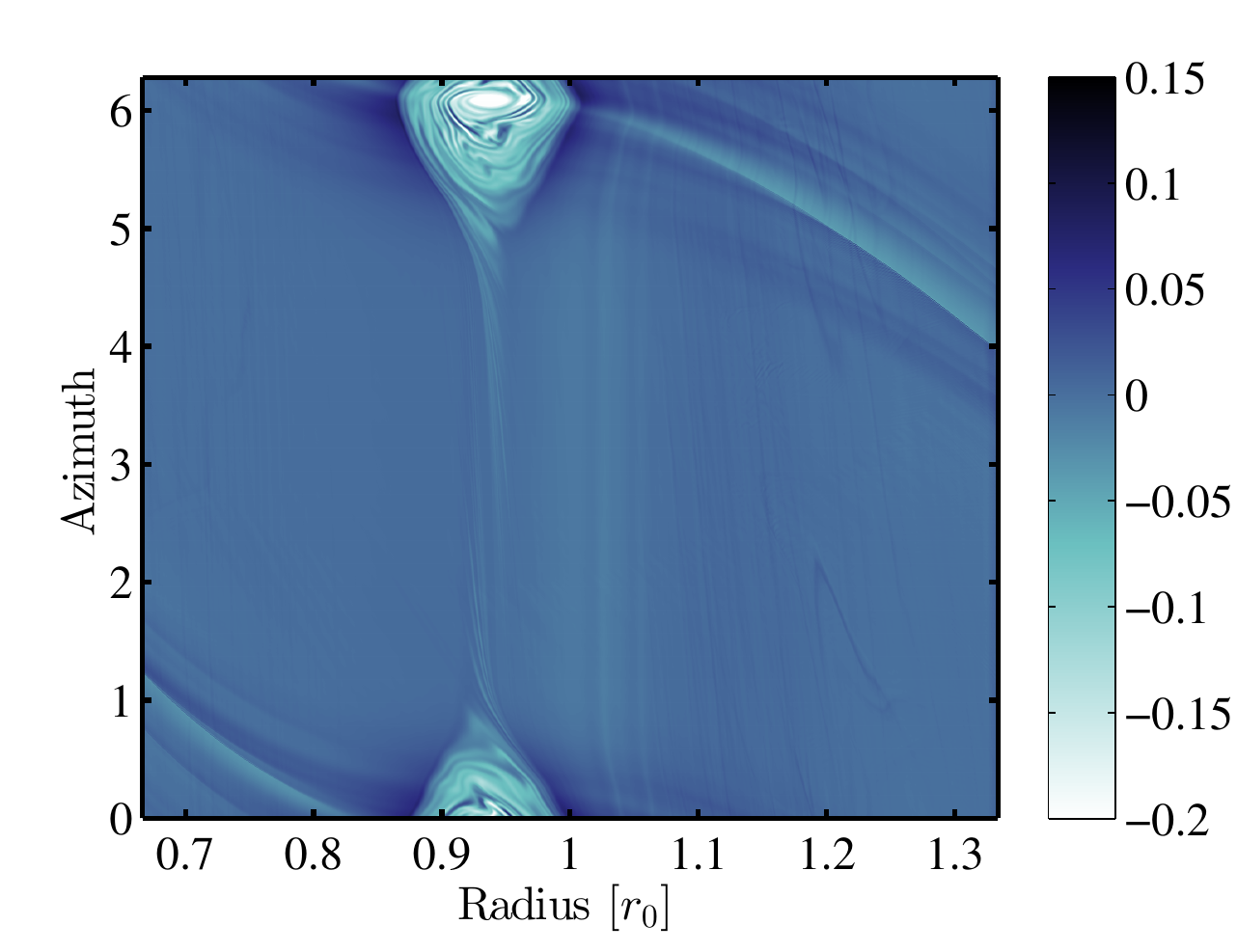}} &
	\imagetop{\includegraphics[height=5.5cm, trim=2mm 0mm 0mm 3mm, clip=true]{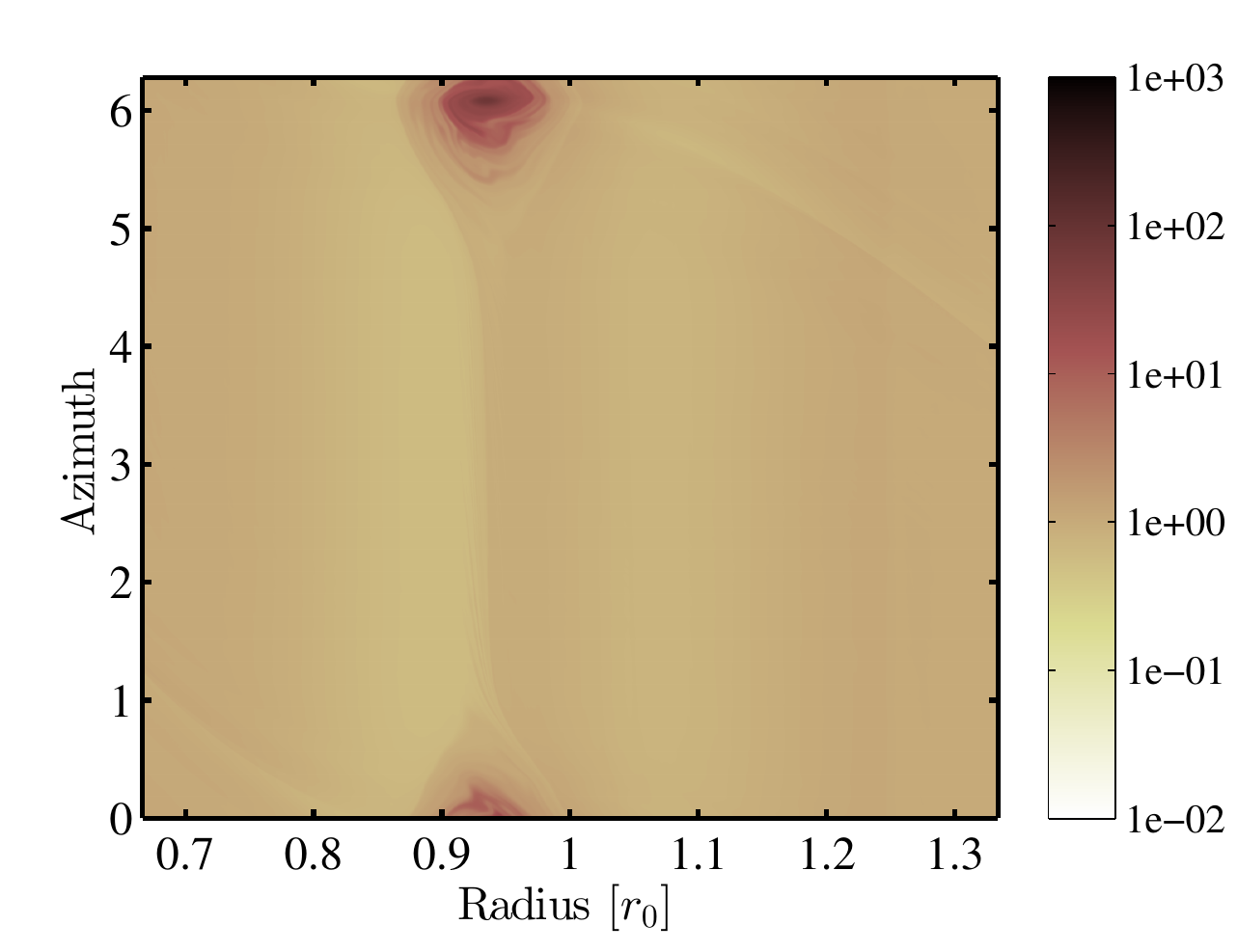}} &
	\scriptsize{ $\:$ \newline \newline $t=715 \: rot$} \\

	\imagetop{\includegraphics[height=5.5cm, trim=2mm 0mm 0mm 3mm, clip=true]{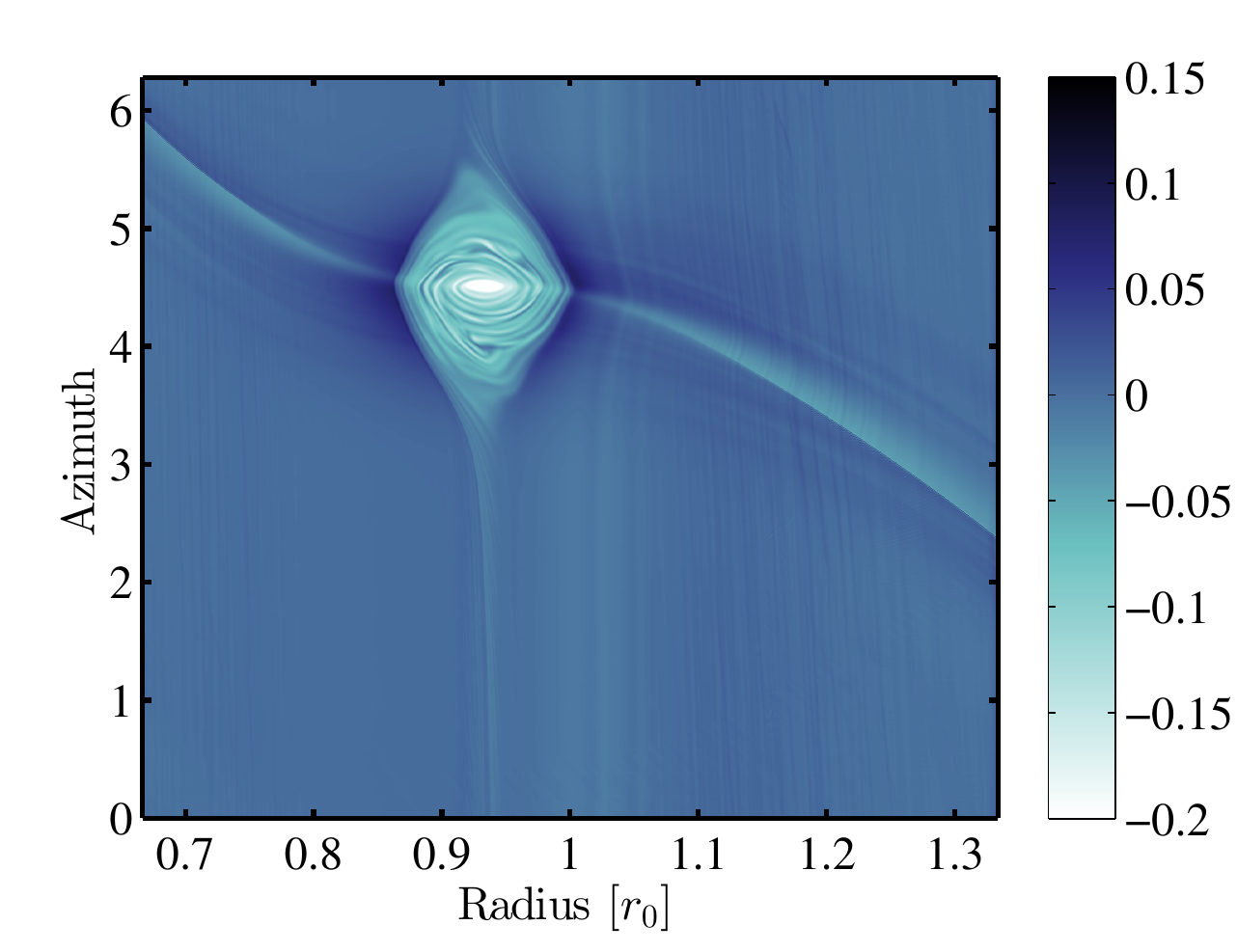}} &
	\imagetop{\includegraphics[height=5.5cm, trim=2mm 0mm 0mm 3mm, clip=true]{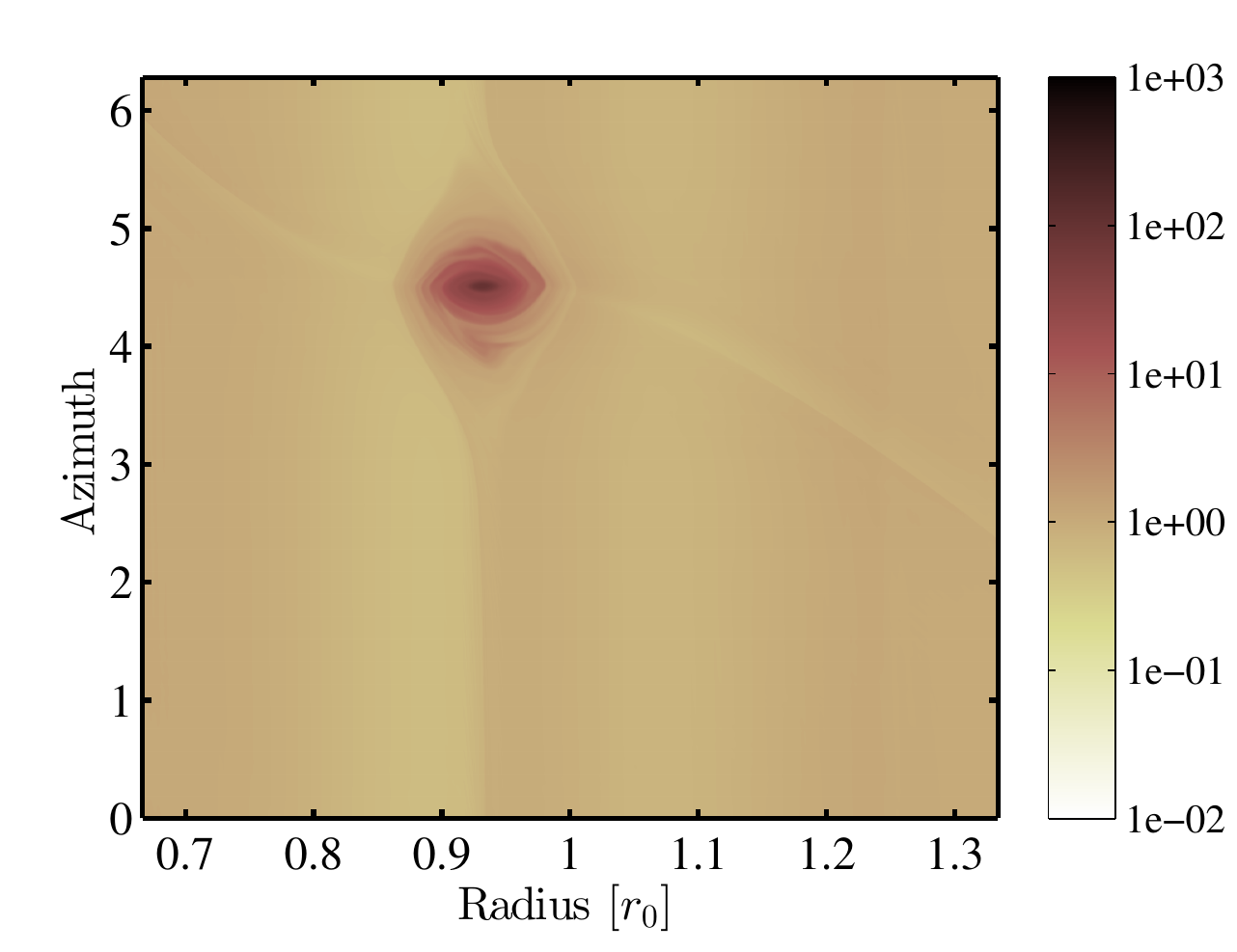}} &
	\scriptsize{ $\:$ \newline \newline $t=740 \: rot$} \\
	\end{tabular}
	\caption{\label{Fig_Vortex_insta_2k_4k} First vortex instability. We show the development of the instability after the linear capture observed in the high resolution run, for ${St=4\times 10^{-3}}$. From top to bottom: Snapshot at $t=685$, $t=700$, $t=715$, and $t=740$ rotations respectively. On the left are reported the $(r, \: \theta)$ maps of the Rossby number and on the right the maps of the dust density, in a logarithmic color scale. }
      \end{center}
\end{figure*}

	In Section \ref{Sect_Results}, we observe a repetition of instabilities in the disk when the Stokes number is smaller than $10^{-2}$ approximately. This phase lasts for a very long period, generating episodic events of strong mixing of gas and dust inside the vortex. To understand the origin of this effect, we analyze the $St=4 \times 10^{-3}$ high resolution run, just after the linear capture phase. Figure \ref{Fig_Vortex_insta_2k_4k} shows the evolution of the Rossby number of the gas and of the dust density at four instants between $t=685$ and $t=740$ disk rotations. First row, we observe the non linear saturation of the vortex instability, with the dust following the unstable vortex core. The range of values of the dust density is however restricted, with a high concentration at the vortex center. At $t=700$ rotations, second row, the instability has covered the whole vortex, and strong waves are excited from the inner parts, where the maximal vorticity is located. The mixing of the gas due to the vortical motion is visible as spiraling stripes of vorticity all over the vortex. This mixing is mirrored in the solid phase of the flow. The dust density map, on the right, reveals that the grains have spread out over a large part of the vortex, forming filaments that indicate a turbulent event. A dense core where the dust density is an order of magnitude larger than in the rest of the vortex appears, superimposed on a high vorticity core visible on the left panel.

	The two last rows, at $t=715$ and $t=740$ rotations respectively, illustrate the process of vortex relaxation. While the high vorticity core remains at the vortex center, embedding a dense clump of solids, the rest of the vortex reaches a smooth equilibrium. During this relaxation, the stripes of vorticity merge together toward the mean value of the vortex, $Ro\sim-0.13$. Because the coupling between gas and solids is strong, the dust follows this relaxation. On the density profile, on the right, the formation of an extended region, where the dust density is more than ten times larger than the initial value, is visible. During the vortex instability, and this relaxation process, the dust that was compacted at the vortex center during the capture is strongly mixed inside the whole vortex, and then redistributed into two structures: {\it{(i)}} a dense small core, where the density is comparable to the one reached at the end of the capture, and {\it{(ii)}} a wide homogeneous patch, covering almost all the vortex area, and containing a significant fraction of the captured solid material. This second structure is thus a reservoir of material which, after relaxation, continues to be pushed towards the vortex center, and which contributes to increase the mass of the inner dense core.

	The relaxation of the vortex is a hydrodynamical process that also happens without dust. In fact, it is observed during the first ten disk rotations when a Gaussian vortex model is used as an initial condition of the simulations. Vortices are in general in a equilibrium close to the geostrophic balance, as we discuss in \cite{Surville2015}. So when a disturbance displaces the vortex profile from this quasi equilibrium, without any physical process sustaining this disturbance, the vortex relaxes back to the smooth equilibrium profile. The typical timescale of this rearrangement is the circulation period inside the vortex, that is related to the vorticity of the vortex, i.e. $\sim 1/|Ro|$. This gives an estimate in the range $[5,10]$ orbital periods, for values of the Rossby number in the range ${[-0.2, -0.1]}$. This compares well with the results in Figure \ref{Fig_Vortex_insta_2k_4k}, where the vortex has strongly relaxed between $t=700$ and $t=715$ disk rotations.

	In the case of a dusty vortex, the drag changes the vorticity profile as time goes on. The timescale of effect on the vorticity is the same as the timescale of dust capture, because of the drag back reaction, i.e. $\sim 1/St$. As a results, if the stokes number of the dust grains is too small, then this timescale will be much larger than the relaxation of the vortex. Assuming the relaxation period to $10$ rotations, then an order of magnitude argument saying that the drag can be neglected when the timescale is ten times longer, i.e. $100$ rotations, gives a critical Stokes number of $10^{-2}$. This reasoning is in agreement with the numerical results, as we observe vortex relaxation after the first instability in the main run with $St=10^{-2}$ (see Section \ref{Sect_Results_1}).
	
	{{ After the relaxation, the capture of solids continues, until the vorticity profile in the vortex becomes unstable again. When the accumulation of dust is strong enough to trigger again the vortex instability, that the dust is distributed in a dense core at the vortex center, and an extended corona where the modes grow. Then the instability expends far from the vortex center, generating a chain of embedded vortices. The dust grains that were previously confined at the vortex center experience a strong mixing inside the vortex and are efficiently redistributed from the core to the corona. This effect can have a significant influence on the mechanisms of grain growth, in particular as concerns fragmentation and sticking, as these processes are fast, on timescales of the order of the local orbital period \citep{Birnstiel2012}.

	If the vortex can still relax, a high vorticity core forms at the center of the vortex. The area of this new core increases each time an instability event is triggered. The dust density inside the vortex increases a lot after each instability event. The vorticity at the vortex center also increases. As a consequence, each event of vortex instability reinforces the loading of dust in the vortex, reducing the timescale of the drag force, and also generates a less stable vorticity profile. As an outcome, the vortex cannot relax and an active dusty flow is sustained in its core, leading eventually to its disruption.}}

%-----------------------
\begin{figure}[t]
%% Figure 14
	\begin{center}
	\includegraphics[height=6.cm]{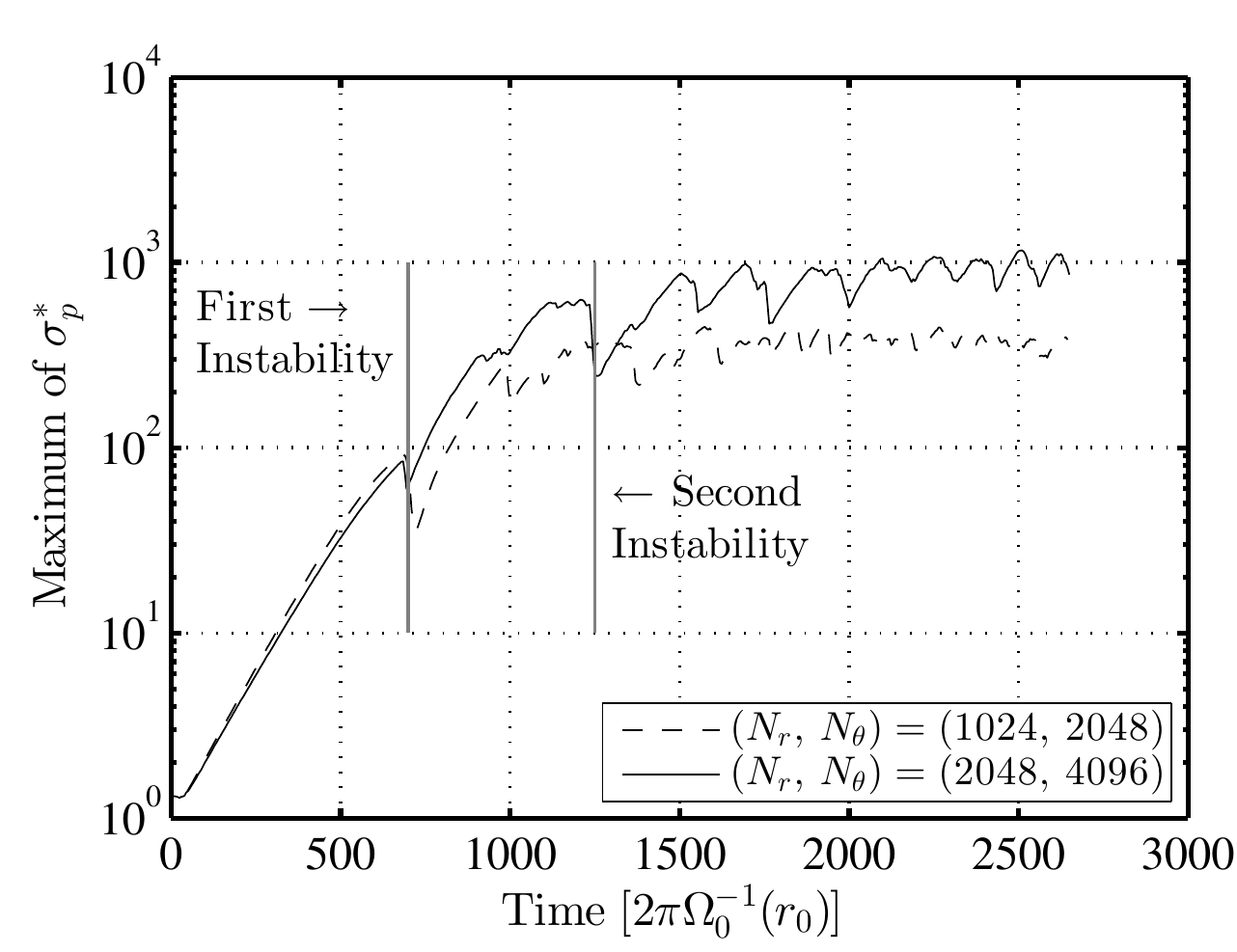}
	\caption{\label{Fig_Rho_p_max_compar} Identification of successive drag instabilities. We compare the evolution of the maximum of dust density inside the vortex for the main run and the high resolution run (dashed and solid lines respectively), for the same setup with ${St = 4 \times 10^{-3}}$. We identify the first and second main vortex instabilities, which produce fast reductions of the maximum of dust density.}
      \end{center}
\end{figure}
	
	{{The recurrence of phases of instability during the vortex evolution}} is a new aspect of the dust-vortex interaction, in the regime of well-coupled grains. It can be identified and traced by inspecting the evolution of the maximum dust density, as seen in Figure \ref{Fig_Rho_p_max_compar}. In the latter the time evolution of the maximum of the dust density (solid line) is shown for the high resolution run with ${St=4 \times 10^{-3}}$. From the initial time to $t=700$ disk rotations, the initial linear capture of dust is at play, with an exponential growth of the density, as discussed Section \ref{Sect_Results_1}. {{The first time the vortex instability grows is identified on the plot}}. There is a significant reduction of the maximum dust density, on a very short period, corresponding to the relaxation of the unstable vortex. Then a second phase of capture, with an exponential growth of the dust density, starts and lasts until ${t=1100}$ rotations. The growth rate is similar to the one of the first capture phase, because the Rossby number in the vortex has the same value, i.e. $Ro\sim -0.13$, in particular in the dust corona as discussed previously (see Figure \ref{Fig_Vortex_insta_2k_4k}).
	
	After a stabilization of the maximum density, {{the instability is triggered for a second time, at ${t=1200}$ rotations}}. The dust density is reduced by a factor two during the instability, because of the reorganization of the distribution of the solids inside the vortex, as described previously. Later on, several phases of instability happen until $t=2000$ disk rotations. Between these three additional event of instability, dust is accumulated at the same rate, until it reaches an enhancement of $10^3$ times the background value. In this case with ${St=4\times 10^{-3}}$, it takes almost $2000$ orbits and five instability phases to increase the maximum dust density up to $1000$ times the background value, and to prevent the vortex from relaxing.

	A comparison with the main run (dashed line), for the same grain size, gives similar results. The growth rates are comparable, but the times corresponding to the onset of the unstable phases differ slightly from the ones in the high resolution run, after the first instability. The maximum dust density obtained at the end of the sequence is also reduced to $400$ times the background value, and is more constant. This is due to the lack of spatial resolution, which filters out the high mode numbers of the instabilities. However, the same sequence of events and their related processes can be captured irrespective of resolution, lending strong support to our general findings.
	
%-----------------------
\begin{figure}[t]
%% Figure 15
	\begin{center}
	\begin{tabular}{l}
	\includegraphics[width=8.5cm, trim=4mm 5mm 6mm 0mm, clip=true]{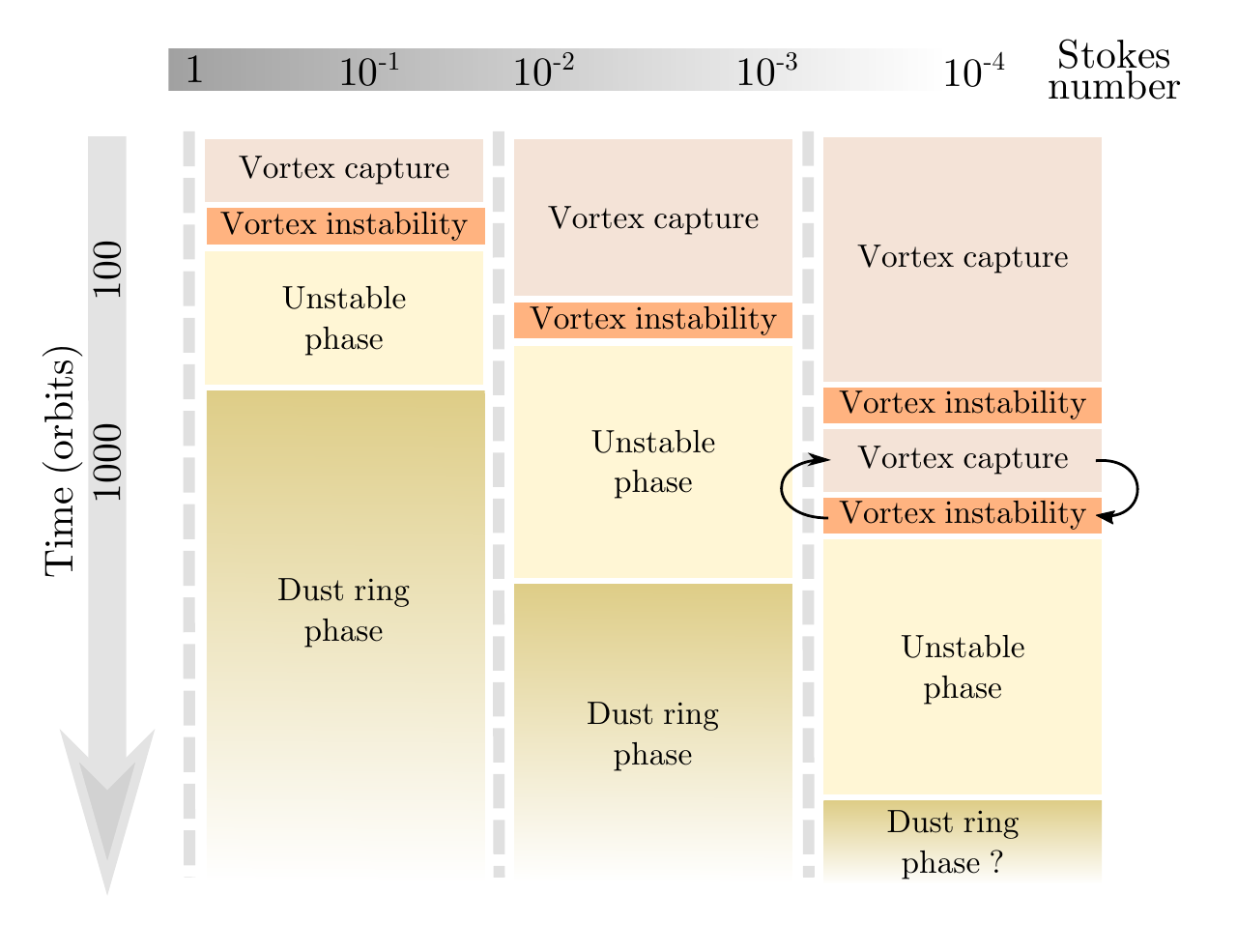}
	\end{tabular}
	\caption{\label{Fig_Small_stokes_cartoon}  Schematic view of the evolution of a dusty vortex as function of the grain size. The four main events during the evolution are identified with different colors. The time line on the left helps to visualize that the duration of the phases increases when the Stokes number reduces. The values of $St$ mentioned on the top bar are indicative, but we emphasize the critical difference between the $St>10^{-3}$ and $St<10^{-3}$ regimes. In the latter, the black arrows show the repetition of dust capture and vortex instabilities that we observe in this study. For too small dust grains, we could not obtain the formation of a dust ring before the end of the run. }
      \end{center}
\end{figure}

	The competition between several processes, which are vortex instability, relaxation of the vortex to a new quasi 
steady structure, dust drag effects during dust capture and dust ring phases, governs the evolution of dusty vortices. The 
relative importance of each effect depends on the timescale of the underlying physics, and changes during the whole 
evolution of the disk. In light of the results we obtained in this study, as well as in \cite{Surville2016}, we designed 
a diagram that summarizes the different evolutionary pathways of the vortex as function of the dust grain size (Figure 
\ref{Fig_Small_stokes_cartoon}).

	The main aim of this cartoon is to highlight the different evolutionary phases studied in this work, these being identified in different colors:
%-----------------------
\begin{itemize}
\item The dust capture phase in the vortex: this process happens on a timescale of ${\sim (\Omega_k St)^{-1}}$,
\item The vortex instability phase: this develops on a short timescale, of the order of ${\Omega_k^{-1}}$,
\item The unstable evolution phase: in competition with the drag, the typical timescale is the vortex relaxation ${\sim 
1-10 \times \Omega_k^{-1}}$,
\item The dust ring phase: the duration depends on the source of diffusion in the system (numerical, physical like viscous 
stress effects).
\end{itemize}
	
	For the largest dust grains, ${1<St<10^{-3}}$, that we can call {\it{weakly coupled grains}}, the four phases take place, and the system usually ends up to a long lived turbulent dust ring. The global amount of time necessary to cover the whole evolution is related to the drag timescale, and so is longer for smaller grains. However, the dust density enhancements to dust-to-gas ratios larger than unity is a common result weakly dependent on the Stokes number. The dust ring state is thus the most relevant for contributing in the process of planetesimal formation for these weakly coupled grains. 
	
	For the smallest dust grains, ${St<\sim10^{-3}}$, which we refer to as {\it{well-coupled grains}}, a sequence of vortex 
instabilities and dust capture, graphically indicated by the curved arrows in Figure \ref{Fig_Small_stokes_cartoon}, is necessary to really destabilized the vortex and eventually transition to the dust ring phase. There is a continuous transition between behavior at varying Stokes number, so that the number of subsequent instability episodes increases as the Stokes number 
decreases. Furthermore, if the drag timescale is too long compared to the relaxation timescale of the vortex, the formation of a dust ring is unlikely. During the long lived unstable phase, dust-to-gas ratios are larger than unity, and sustained during a very long period. It is thus the most relevant phase for contributing in the process of planetesimal formation for these well-coupled grains.

%%%%%%%%%%%%%%%%%%%%%%%%%%%%%%%%%%%%%%%%%%%%%%%%%%%%%
\section{ Conclusions }
\label{Sect_Conclusion}

	This study aims at clarifying the effect of small, well-coupled dust grains, on the evolution of large scale vortices in protoplanetary disks. Indeed, while we had shown, in our previous work, that grains with $St>10^{^-2}$ have a strong impact and produce turbulent dust rings with large dust-to-gas ratios, it was not clear if smaller grains can have similarly strong effects on disk evolution. In particular, the formation of long lived structures embedding a large amount of dust has key implications on the development of other phenomena such as the streaming instability, on grain growth models, and on the possibility of direct gravitational instability of the dust layer, which are all possible pathways leading to planetesimal formation.

	For a really wide range of Stokes numbers, corresponding to grains $<\sim$ mm $--$ $10$ cm in size, depending on distance from the star in the disk, large dust density concentrations, with dust-to-gas ratios in the range $1-10$, do arise one way or another. In one case because of dust capture and subsequent turbulence developing in the vortex when this is very long lived, as for the smallest grains. In the other case because a turbulent dust ring with multiple high dust concentration regions develops after the vortex dissipates relatively fast for the largest grains. Therefore the only required condition to lead to strong concentrations of dust is that a vortex is somehow generated in the disk at some point and depends weakly on how this evolves subsequently, this being the major different with previous works investigating vortices in dusty disks. However, when the vortex dissipates relatively fast and the dust ring develops afterward, which happens preferentially for larger Stokes numbers, a qualitative difference arises in that a larger area in the disk is covered with prominent dust concentrations owing to the global nature of the ring as opposed to the initial local nature of the vortex. More specifically we can summarize our findings as follows:

%-----------------------
\begin{itemize}
\item {{Because the drag force experienced by small dust grains is very strong, a specific implicit numerical scheme was implemented in the code RoSSBi to follow the dynamics of the disk accurately.}} This method is very accurate for any value of the Stokes number of the pressure-less dust fluid, and we use it to study the evolution a vortex model with dust characterized by Stokes numbers in the range ${4\times 10^{-2}<St< 10^{-3}}$.
\item Dust is systematically accumulated inside the vortex for every value of the Stokes number. The exponential growth rate of the dust density at the vortex center follows the analytical model developed in our previous study. The timescale depends on the grain size, but the density enhancement reached at the end, when the gas is perturbed by the drag force, is similar whatever the Stokes number, and produces dust-to-gas ratios of order unity. We highlight the existence of a generic capture process, characterized by a timescale scales with the Stokes number, which allows to compare disks with different dust sizes at an equivalent state of evolution albeit corresponding to intrinsically different physical timescales. 
\item The vortex instability is triggered at the end of the capture, whatever the grain size. The growth rates of the excited modes also depend weakly on the Stokes number. This effect is highly correlated to the similarity of the dust capture, and the concept of equivalent states of evolution, producing almost the same perturbation of the gas vorticity profile. 
\item A new phase of dusty vortex evolution is discovered for Stokes number in the order of $10^{-3}$ and below, where a succession of vortex instabilities followed by dust capture events happen. This period lasting several hundreds of disk orbits produces of strong mixing and a fast reorganization of the dust during the relaxation of the vortex. The dust-to-gas ratio increases progressively to high values, up to $8$ at the vortex center. The very dense core that forms, and survives even after eventual vortex dissipation could be a preferential place for planetesimal formation.
\item When the timescale of the drag effect is too long compared to the resistance and relaxation of the vortex, the formation of a dust ring becomes unlikely. In our case, the runs with $St= 10^{-3}$ where still in the unstable phase by the end of the simulations. Even when it forms, for $St<10^{-2}$, the dust ring produced by well-coupled grains have a fewer number of eddies than with larger Stokes numbers. Even if the dust-to-gas ratios are comparable, they suffer diffusion and their role in planetesimal formation is less promising than the unstable phase.
\end{itemize}

{{There are a number of caveats to our model that need to be pointed out. First, the fact that the simulations are still only two-dimensional, which could have a relevance both for the development of two-fluid turbulence during and after the vortex instability phase \citep[see discussion in ][]{Surville2016}. We find that the instability and the active flow in the dust ring or in the vortex are efficient sources of dust diffusion, for this class of non-turbulent compressible vortices.

	However, some other vortices, a class of incompressible vortices with closed streamlines, may be sensitive to 3D turbulence driven by the elliptical instability. Finally, $\alpha$-turbulence of the disk , driven by MRI, the SI, or the VSI, could act as additional source of diffusion of the dust density. We will investigate properly the competition between the diffusion of the two-fluid instability and other sources of turbulence in a full 3D study in an upcoming publication.}}

Related to this aspect, the only source of dissipation in our models is the numerical diffusion. However, we showed that we have enough resolution to capture the instability and the active flow in the dust ring in a converged manner regarding the resolution. Similarly to the SI, the instabilities we observe in our runs need a high resolution (and a low diffusion) to be captured. As a last argument, the $\alpha$ viscosity expected in the disk is as low as $10^{-4}-10^{-5}$, which correspond to diffusion timescales much longer than the duration of our runs.

Secondly, the absence of self-gravity of the dust, which could play a role as soon as the dust-to-gas ratio increases above unity locally, and may trigger gravitational collapse. Finally, and the fact that dust size is fixed in each individual simulation as we do not include a prescription for the coagulation and fragmentation of grains. Concerning this last point, it is important to note that, in addition to the mixing processes happening during the unstable evolution of the vortex, the dense dust eddy surviving can be a favorable place for grain growth processes. If the mean grain size increases, then the Stokes number of the dust will also increase, and the evolution path of the system will move to the left on the diagram shown in Figure \ref{Fig_Small_stokes_cartoon} even for grains with initially very small Stokes number. This means that destabilization of the vortex will be stronger and the probability to form a turbulent dust ring would increase. Since as the grains become larger evolution towards high concentrations is faster and most efficient owing to the formation of the turbulent ring, the implication would be that even very small dust grains, such as primordial dust, could evolve rapidly towards the formation of multiple dense eddies, and thus planetesimals. This hypothesis has to be confirmed with simulations including the different processes relevant to dust grain evolution. In a companion paper \citep{Tamfal2018}, we present the first results of a new sub-grid method that can indeed model coagulation and fragmentation of grains, thus including a variable grain size in the code RoSSBi. The new method will be applied soon to simulation setups analogous to those presented in this work.

%%%%%%%%%%%%%%%%%%%%%%%%%%%%%%%%%%%%%%%%%%%%%%%%%%%%%
\begin{acknowledgements}

	This work has been carried out within the frame of the National Center for Competence in Research {\it{PlanetS}} supported by the Swiss National Science Foundation (SNSF). The authors acknowledge the financial support of the SNSF. Numerical simulations were performed on the {\it{Piz Daint}} Cray XC50 system of the Swiss National Supercomputing Center (CSCS). C. Surville would like to thank M. Hutchison for his usefull comments and corrections to the draft.

\end{acknowledgements}

%%%%%%%%%%%%%%%%%%%%%%%%%%%%%%%%%%%%%%%%%%%%%%%%%%%%%
\bibliographystyle{aa}
\bibliography{Biblio}

\begin{thebibliography}{40}
\expandafter\ifx\csname natexlab\endcsname\relax\def\natexlab#1{#1}\fi

\bibitem[{Andrews {et~al.}(2016)Andrews, Wilner, Zhu, Birnstiel, Carpenter,
  P{\'{e}}rez, Bai, {\"{O}}berg, Hughes, Isella, \& Ricci}]{Andrews2016}
Andrews, S.~M., Wilner, D.~J., Zhu, Z., {et~al.} 2016, ApJ, 820, L40

\bibitem[{Barge {et~al.}(2017)Barge, Ricci, Carilli, \&
  Previn-Ratnasingam}]{Barge2017}
Barge, P., Ricci, L., Carilli, C.~L., \& Previn-Ratnasingam, R. 2017, A{\&}A,
  605, A122

\bibitem[{Birnstiel {et~al.}(2012)Birnstiel, Klahr, \&
  Ercolano}]{Birnstiel2012}
Birnstiel, T., Klahr, H., \& Ercolano, B. 2012, A{\&}A, 539, A148

\bibitem[{Booth {et~al.}(2015)Booth, Sijacki, \& Clarke}]{Booth2015}
Booth, R.~A., Sijacki, D., \& Clarke, C.~J. 2015, MNRAS, 452, 3932

\bibitem[{Brogan {et~al.}(2015)Brogan, P{\'{e}}rez, Hunter, Dent, Hales, Hills,
  Corder, Fomalont, Vlahakis, Asaki, Barkats, Hirota, Hodge, Impellizzeri,
  Kneissl, Liuzzo, Lucas, Marcelino, Matsushita, Nakanishi, Phillips, Richards,
  Toledo, Aladro, Broguiere, Cortes, Cortes, Espada, Galarza, Appadoo, Ramirez,
  Humphreys, Jung, Kameno, Laing, Leon, Marconi, Mignano, Nikolic, Nyman,
  Radiszcz, Remijan, Rod{\'{o}}n, Sawada, Takahashi, Tilanus, Vilaro, Watson,
  Wiklind, Akiyama, Chapillon, Monsalvo, Francesco, Gueth, Kawamura, Lee,
  Luong, Mangum, Pietu, Sanhueza, Saigo, Takakuwa, Ubach, van Kempen, Wootten,
  Carrizo, Francke, Gallardo, Garcia, Gonzalez, Hill, Kaminski, Kurono, Liu,
  Lopez, Morales, Plarre, Schieven, Testi, Videla, Villard, Andreani, Hibbard,
  \& Tatematsu}]{Partnership2015}
Brogan, C.~L., P{\'{e}}rez, L.~M., Hunter, T.~R., {et~al.} 2015, ApJ, 808, L3

\bibitem[{Bryden {et~al.}(2009)Bryden, Beichman, Carpenter, Rieke, Stapelfeldt,
  Werner, Tanner, Lawler, Wyatt, Trilling, Su, Blaylock, \&
  Stansberry}]{Bryden2009}
Bryden, G., Beichman, C.~A., Carpenter, J.~M., {et~al.} 2009, ApJ, 705, 1226

\bibitem[{Chatterjee \& Tan(2014)}]{Chatterjee2014}
Chatterjee, S. \& Tan, J.~C. 2014, ApJ, 780, 53

\bibitem[{Fu {et~al.}(2014)Fu, Li, Lubow, Li, \& Liang}]{Fu2014}
Fu, W., Li, H., Lubow, S., Li, S., \& Liang, E. 2014, ApJL, 795, L39

\bibitem[{Gonzalez {et~al.}(2012)Gonzalez, Pinte, Maddison, M{\'{e}}nard, \&
  Fouchet}]{Gonzalez2012}
Gonzalez, J.-F., Pinte, C., Maddison, S.~T., M{\'{e}}nard, F., \& Fouchet, L.
  2012, A{\&}A, 547, A58

\bibitem[{Gras-Vel{\'{a}}zquez \& Ray(2005)}]{Gras-Velazquez2005}
Gras-Vel{\'{a}}zquez, {\`{A}}. \& Ray, T.~P. 2005, A{\&}A, 443, 541

\bibitem[{Hopkins \& Squire(2017)}]{Hopkins2017}
Hopkins, P.~F. \& Squire, J. 2017, ArXiv e-prints, 1707.02997

\bibitem[{Jacquet {et~al.}(2011)Jacquet, Balbus, \& Latter}]{Jacquet2011}
Jacquet, E., Balbus, S., \& Latter, H. 2011, MNRAS, 415, 3591

\bibitem[{Johansen \& Klahr(2005)}]{Johansen2005}
Johansen, A. \& Klahr, H. 2005, ApJ, 634, 1353

\bibitem[{Johansen \& Youdin(2007)}]{Johansen2007}
Johansen, A. \& Youdin, A. 2007, ApJ, 662, 627

\bibitem[{Johansen {et~al.}(2012)Johansen, Youdin, \& Lithwick}]{Johansen2012}
Johansen, A., Youdin, A.~N., \& Lithwick, Y. 2012, A{\&}A, 537, A125

\bibitem[{Kowalik {et~al.}(2013)Kowalik, Hanasz, W{\'{o}}lta{\'{n}}ski, \&
  Gawryszczak}]{Kowalik2013}
Kowalik, K., Hanasz, M., W{\'{o}}lta{\'{n}}ski, D., \& Gawryszczak, A. 2013,
  MNRAS, 434, 1460

\bibitem[{Lambrechts {et~al.}(2014)Lambrechts, Johansen, \&
  Morbidelli}]{Lambrechts2014}
Lambrechts, M., Johansen, A., \& Morbidelli, A. 2014, A{\&}A, 572, A35

\bibitem[{{Lesur} \& {Papaloizou}(2009)}]{Lesur2009}
{Lesur}, G. \& {Papaloizou}, J. C.~B. 2009, A{\&}A, 498, 1

\bibitem[{Levison {et~al.}(2015)Levison, Kretke, \& Duncan}]{Levison2015}
Levison, H.~F., Kretke, K.~A., \& Duncan, M.~J. 2015, Nature, 524, 322

\bibitem[{Lin \& Youdin(2017)}]{Lin2017}
Lin, M.-K. \& Youdin, A.~N. 2017, ApJ, 849, 129

\bibitem[{Lyra \& Lin(2013)}]{Lyra2013}
Lyra, W. \& Lin, M.-K. 2013, ApJ, 775, 17

\bibitem[{{Meheut} {et~al.}(2010){Meheut}, {Casse}, {Varniere}, \&
  {Tagger}}]{Meheut2010}
{Meheut}, H., {Casse}, F., {Varniere}, P., \& {Tagger}, M. 2010, A{\&}A, 516,
  A31

\bibitem[{Meyer {et~al.}(2008)Meyer, Carpenter, Mamajek, Hillenbrand,
  Hollenbach, Moro-Martin, Kim, Silverstone, Najita, Hines, Pascucci, Stauffer,
  Bouwman, \& Backman}]{Meyer2008}
Meyer, M.~R., Carpenter, J.~M., Mamajek, E.~E., {et~al.} 2008, ApJL, 673, L181

\bibitem[{Paardekooper {et~al.}(2010)Paardekooper, Lesur, \&
  Papaloizou}]{Paardekooper2010}
Paardekooper, S.-J., Lesur, G., \& Papaloizou, J. C.~B. 2010, ApJ, 725, 146

\bibitem[{P{\'{e}}rez {et~al.}(2014)P{\'{e}}rez, Isella, Carpenter, \&
  Chandler}]{Perez2014}
P{\'{e}}rez, L.~M., Isella, A., Carpenter, J.~M., \& Chandler, C.~J. 2014,
  ApJL, 783, L13

\bibitem[{Sch{\"{a}}fer {et~al.}(2017)Sch{\"{a}}fer, Yang, \&
  Johansen}]{Schafer2017}
Sch{\"{a}}fer, U., Yang, C.-C., \& Johansen, A. 2017, A{\&}A, 597, A69

\bibitem[{Sierra {et~al.}(2017)Sierra, Lizano, \& Barge}]{Sierra2017}
Sierra, A., Lizano, S., \& Barge, P. 2017, ApJ, 850, 115

\bibitem[{Simon {et~al.}(2016)Simon, Armitage, Li, \& Youdin}]{Simon2016}
Simon, J.~B., Armitage, P.~J., Li, R., \& Youdin, A.~N. 2016, ApJ, 822, 55

\bibitem[{Smith {et~al.}(2008)Smith, Wyatt, \& Dent}]{Smith2008}
Smith, R., Wyatt, M.~C., \& Dent, W. R.~F. 2008, A{\&}A, 485, 897

\bibitem[{Squire \& Hopkins(2017)}]{Squire2017}
Squire, J. \& Hopkins, P. 2017, ArXiv e-prints, 1711.03975

\bibitem[{Surville \& Barge(2012)}]{Surville2012}
Surville, C. \& Barge, P. 2012, ArXiv e-prints, 1201.3257

\bibitem[{Surville \& Barge(2013)}]{Surville2013}
Surville, C. \& Barge, P. 2013, EPJ Web Conf., 46, 05002

\bibitem[{Surville \& Barge(2015)}]{Surville2015}
Surville, C. \& Barge, P. 2015, A{\&}A, 579, A100

\bibitem[{Surville {et~al.}(2016)Surville, Mayer, \& Lin}]{Surville2016}
Surville, C., Mayer, L., \& Lin, D. N.~C. 2016, ApJ, 831, 82

\bibitem[{{Tamfal} {et~al.}(2018){Tamfal}, {Dr{\c a}{\.z}kowska}, {Mayer}, \&
  {Surville}}]{Tamfal2018}
{Tamfal}, T., {Dr{\c a}{\.z}kowska}, J., {Mayer}, L., \& {Surville}, C. 2018,
  ApJ, 863, 97

\bibitem[{Yang \& Johansen(2016)}]{Yang2016}
Yang, C.-C. \& Johansen, A. 2016, ApJ Sup. Ser., 224, 39

\bibitem[{Youdin \& Goodman(2005)}]{Youdin2005}
Youdin, A. \& Goodman, J. 2005, ApJ, 620, 459

\bibitem[{Youdin \& Johansen(2007)}]{Youdin2007}
Youdin, A. \& Johansen, A. 2007, ApJ, 662, 613

\bibitem[{Youdin(2011)}]{Youdin2011}
Youdin, A.~N. 2011, ApJ, 731, 99

\bibitem[{Zhu {et~al.}(2014)Zhu, Stone, Rafikov, \& Bai}]{Zhu2014a}
Zhu, Z., Stone, J.~M., Rafikov, R.~R., \& Bai, X.-n. 2014, ApJ, 785, 122

\end{thebibliography}

\end{document}